\documentclass[reprint,longbibliography,amsmath,amssymb,aps,prd,floatfix]{revtex4-2}
\usepackage{graphicx}
\usepackage{array}[=2016-10-06]
\usepackage{dcolumn}
\usepackage{bm}
\usepackage[utf8]{inputenc}
\usepackage{xcolor}
\usepackage{booktabs}

\begin{document}
	
	\title{Anisotropic spin hydrodynamics in a magnetized medium: entropy-current formulation and minimal linear modes}

	\author{Yiwei Qiu$^{1}$}
	\email{qiuyw@mails.ccnu.edu.cn}
	\author{Duan She$^{2,1}$}
	\email{sheduan@hnas.ac.cn}
	\author{Defu Hou$^{1}$}
	\email{houdf@mail.ccnu.edu.cn}

	\affiliation{ $^{1}$ Institute of Particle Physics and Key Laboratory of Quark and Lepton Physics (MOE), Central China Normal University, Wuhan 430079, China }
	\affiliation{ $^{2}$ Institute of Physics, Henan Academy of Sciences, Zhengzhou 450046, China}
	
	\begin{abstract}
We formulate anisotropic relativistic spin hydrodynamics in a parity-even magnetized medium using an Israel--Stewart entropy-current construction.  Treating the magnetic field at leading order in the gradient expansion reduces the rotational symmetry to $SO(2)$ and separates spin susceptibilities and dissipative transport into longitudinal, transverse, and mixed channels.  A restricted quadratic entropy current yields transient equations for the retained dissipative stresses and spin fluxes.  We analyze a minimal linear closure around an unpolarized homogeneous equilibrium.  Magnetic-flux perturbations remain dynamical, and spin and magnetic fluctuations mix through the fluid variables without explicit spin-magnetization coupling.  Infrared coefficient equations and ultraviolet characteristic expansions organize the magnetosonic and Alfv\'en-like spectra together with spectator spin-flux poles.  The retained first-order boost sectors are unstable already at zero wave number, and the ultraviolet spectrum includes superlinear branches.  Positive effective inertias stabilize the zero-momentum gapped modes of the boost sector.  For the baseline parameters, finite relaxation times yield asymptotically linear propagation branches and finite relaxation poles, with no growing modes found on the sampled grid.  The characteristic speeds depend on the magnetic angle and the spin-transport ratios.  We determine their relation to the light cone.
	\end{abstract}
	
	\maketitle
	
	\section{Introduction}
\label{sec:introduction}

Relativistic heavy-ion collisions create matter with large orbital angular momentum, substantial vorticity, and intense electromagnetic fields.  Measurements of global and local hyperon polarization at RHIC, the LHC, and lower collision energies have established spin observables as quantitative probes of this rotational environment \cite{STAR2017,STAR2018,STAR2019Beam,ALICE2020,ALICESpinOrbit2020,STARXiOmega2021,STAR3GeV2021}.  The basic connection between orbital angular momentum and polarization was anticipated in early phenomenological studies of noncentral collisions \cite{LiangWang2005,BetzGyulassyTorrieri2007,BecattiniPiccininiRizzo2008}.  In parallel, event-by-event calculations show that the magnetic field has a preferred but fluctuating direction and can be comparable to other microscopic scales during the early evolution.  Medium response can prolong the field lifetime in an expanding viscous plasma \cite{HuangSheShiHuangLiao2023}.  These observations motivate a framework that treats rotation, spin, and magnetic flux together.

The macroscopic conversion between mechanical rotation and microscopic polarization is closely related to the Barnett and Einstein--de Haas phenomena \cite{Barnett1915,EinsteinDeHaas1915}.  Relativistic spin hydrodynamics implements this idea by adding a spin tensor and a conjugate spin potential to the usual hydrodynamic variables.  Reference~\cite{FlorkowskiKumarRyblewski2019} reviews the main formulations and phenomenological motivations.  Entropy-current, effective-action, torsional, and angular-momentum-based formulations have clarified how the antisymmetric part of the energy--momentum tensor mediates local spin--orbital exchange and how rotational transport controls spin relaxation \cite{HattoriEntropy2019,FukushimaPu2021,LiStephanovYee2021,SheHuangHouLiao2022,DaherCanonical2023,HongoEtAl2021,BecattiniDaherSheng2024,DaherShengWagnerBecattini2025,WangFangPu2021,GallegosGursoyYarom2021}.  Ideal and dissipative phenomenological realizations have also been developed for several flow backgrounds \cite{FlorkowskiFrimanJaiswalSperanza2018,FlorkowskiFrimanJaiswalRyblewskiSperanza2018,HidakaPuYang2018,FlorkowskiSperanzaBecattini2018,JiangFanSheYeZhang2025,SinghSophysRyblewski2021,SinghShokriRyblewski2021}.  In the present work we adopt a phenomenological spin tensor antisymmetric in its last two indices and count the spin chemical potential as $O(\partial)$.  In the equilibrium limit relevant to this representation, the spin potential is constrained by thermal vorticity, while the explicit nonequilibrium spin flux enters at higher order in the entropy analysis.

Microscopic approaches derive spin transport from Wigner functions, Kadanoff--Baym equations, nonlocal collision kernels, Chapman--Enskog expansions, and moment methods.  They relate macroscopic spin variables to kinetic degrees of freedom and make the dependence on pseudo-gauge choice, matching conditions, and collision dynamics explicit \cite{WeickgenanntShengSperanzaWangRischke2019,BhaduryEtAl2021PLB,WeickgenanntEtAl2021PRL,ShiGaleJeon2021,SperanzaWeickgenannt2021,BhaduryEtAl2021PRD,BhaduryBhattJaiswalKumar2021,PengZhangShengWang2021,ShengEtAl2021,ShengWangRischke2022,HuChapmanEnskog2022,HuLinearMode2022,FangPuYang2022,WangSpinEvolution2022}.  Different pseudo-gauges redistribute local energy--momentum and spin densities and can affect constitutive identifications away from global equilibrium.  Statistical and thermodynamic analyses of rotating systems help specify the hydrodynamic representation used in a calculation \cite{MontenegroTintiTorrieri2017,MontenegroTintiTorrieriSound2017,MontenegroTorrieri2019,MontenegroTorrieri2020,BecattiniPiccinini2008,BecattiniTinti2010,BecattiniTinti2013,FlorkowskiHontarenko2025,BecattiniFlorkowskiSperanza2019,HuKubo2021,Buzzegoli2022,DasFlorkowskiRyblewskiSingh2021}.

Dissipation imposes additional consistency requirements.  First-order spin-hydrodynamic models may contain unstable or acausal modes, and the outcome depends on the frame, the spin equation of state, the background spin density, and the retained dissipative channels \cite{DaherDasRyblewski2023,SarwarEtAl2023,BiswasBoost2023,XieWangYangPu2023,Weickgenannt2023,LuZhongFeng2026}.  Entropy-current and Kubo formulations introduce second-order spin variables and finite relaxation times \cite{BiswasSecondOrder2023,DaherSecondOrder2024,SheQiuHou2025,SheQiuJiangHou2025,BhaduryNovelRTA2025}.  The spin susceptibilities and finite-momentum spectrum must also satisfy independent stability conditions \cite{DaherSpinEOS2024,RenYangWangPu2024}.  This situation parallels the broader development of causal relativistic dissipative hydrodynamics, from the M\"uller--Israel--Stewart theory to kinetic moment expansions and modern first-order formulations \cite{Muller1967,Israel1976,IsraelStewart1979,HiscockLindblom1983,HiscockLindblom1985,HiscockLindblom1987,KoideDenicolMotaKodama2007,DenicolKoideRischke2010,PuKoideRischke2010,DenicolNiemiMolnarRischke2012,BaierEtAl2008,Kovtun2019,BemficaDisconziNoronha2019,HoultKovtun2020,BritoDenicol2020,FloerchingerGrossi2018,RochaWagnerDenicolNoronhaRischke2024,AbbasiDavodyTahery2024}.  Density-frame evolution, which is first order in time and uses no auxiliary relaxation fields, has also been benchmarked against QCD kinetic theory and transient hydrodynamics \cite{BhambureEtAl2025DensityFrame}.  The instability analyzed below concerns the specified Landau-type truncation.  Nonnegative entropy production alone does not guarantee hyperbolicity, causality, or stability at finite wave number.

The magnetic sector supplies an independent source of anisotropy \cite{AbbasiDavody2016}.  Modern RMHD can be organized around energy--momentum conservation and magnetic-flux conservation, the latter associated with a magnetic one-form symmetry \cite{HernandezKovtun2017,GrozdanovHofmanIqbal2017,ArmasJain2019,ArmasJain2020,GrozdanovPoovuttikul2019,HongoHattori2021,HattoriHongoHuang2022}.  Higher-form formulations have classified nondissipative second-order transport at strong field and clarified the lifetime of screened electric-flux excitations \cite{BenenowskiPoovuttikul2019,PoovuttikulRajagopal2021}.  One-form effective actions also construct dissipative magnetic-flux dynamics without an entropy-current ansatz \cite{VardhanEtAl2024,VardhanEtAl2025}.  These symmetry- and action-based approaches and the entropy-current construction used below constrain overlapping physics from different starting points.  In the strong-field counting used here, the magnetic field is an $O(\partial^0)$ variable and reduces the local rotational symmetry from $SO(3)$ to the $SO(2)$ subgroup about the field direction.  The ideal pressure and the dissipative stresses consequently split into longitudinal, transverse, and mixed channels.  Kinetic, phenomenological, and transient studies have developed anisotropic first- and higher-order RMHD constitutive relations and have shown that characteristic speeds and stability bounds depend on both the propagation angle and combinations of transport coefficients and relaxation times \cite{DenicolEtAl2018,DenicolMolnarNiemiRischke2019,PandaDashBiswasRoy2021,BiswasDashHaquePuRoy2020,FangHattoriHu2024,HoultKovtun2025,MolnarRischke2025,KushwahDenicol2024,ArmasCamilloni2022,QiuSheHou2026}.  Expanding-flow MHD studies describe the evolution of nonresistive and resistive plasmas \cite{SheJiangHouYang2019,MoghaddamAlbericoSheKordAzadegan2020,JiangLiuHuZhengShe2025,JingSheJiang2026}.

Spin magnetohydrodynamics combines these two sources of structure.  Kinetic theory has exhibited relativistic spin--magnetic responses and analogues of the Barnett and Einstein--de Haas effects \cite{BhadurySpinMHD2022}, while entropy-current and wave analyses have shown that spin, vorticity, and magnetic anisotropy can generate a richer constitutive and mode structure \cite{KiamariSadooghiJafari2024,FangHattoriHu2025}.  The present work focuses on anisotropic spin hydrodynamics in a magnetized medium, coupled to magnetic-flux dynamics.  The residual symmetry about the equilibrium field direction $b^\mu$ permits separate longitudinal and transverse spin susceptibilities.  It also decomposes the antisymmetric stress and the rank-three nonequilibrium spin flux into inequivalent $SO(2)$ scalar, vector, symmetric-traceless, and antisymmetric representations.  These coefficients are treated as independent phenomenological data unless a microscopic calculation relates them.  The minimal spectral model uses an unpolarized equilibrium and a constant-permeability equation of state, without an explicit spin contribution to magnetization.  Its vector cross-transport coefficients are set to zero, so spin and magnetic perturbations mix through the fluid variables.  Magnetic-flux perturbations remain dynamical; the background field specifies the equilibrium direction, not an externally fixed magnetic sector.  Explicit magnetic polarization and its feedback on spin transport lie beyond this closure.

We develop an entropy-current formulation through second order.  Adiabaticity fixes the ideal Euler relation and magnetic pressure anisotropy; entropy production and Onsager--Casimir reciprocity constrain the first-order transport matrices.  A restricted quadratic entropy current then promotes the retained stresses and spin-flux amplitudes to transient variables.  The additional structures omitted from this closure are specified in Sec.~\ref{sec:second-order-entropy}.

The first-order spin-MHD formulation of Fang, Hattori, and Hu uses a totally antisymmetric spin current and identifies anisotropic waves and a nonuniform transverse-angle limit \cite{FangHattoriHu2024FirstOrder,FangHattoriHu2025}.  Our spin tensor is antisymmetric only in its last two indices, allowing independent boost-spin densities and dissipative spin-flux channels.  Building also on the second-order entropy-current construction for an unmagnetized spin fluid \cite{BiswasSecondOrder2023}, we construct transient equations for each irreducible $SO(2)$ channel.  Its additional degrees of freedom have identifiable consequences: the retained boost sectors generate a zero-momentum instability in the first-order truncation, while the split spin flux supplies separate propagating characteristic families.  We derive the effective-inertia thresholds that stabilize the gapped modes of the boost sector and resolve the spin-flux part of the principal spectrum analytically.

We then linearize with the magnetic field along the $z$ direction and the wave vector in the $x$--$z$ plane, imposing the magnetic constraint $\mathbf k\cdot\delta\mathbf B=0$ before forming the determinants.  Ordinary and Newton--Puiseux expansions characterize the small- and large-$k$ roots and are compared with numerical roots.  In the generic first-order magnetosonic determinant, the $K$ and $q^\mu$ channels produce a fractional $\omega\sim k^{4/3}$ family alongside parabolic, linear, finite, and vanishing roots.  These channels also generate unstable boost-sector poles.  For the minimal second-order closure and parameters examined here, the ultraviolet spectrum instead consists of linear and finite branches, including exact spin-flux spectators.  The associated characteristic speeds provide a necessary causality condition for each mode.

The paper is organized as follows.  Section~\ref{sec:ideal-entropy} discusses the ideal anisotropic thermodynamics and spin-tensor structure.  Section~\ref{sec:first-order-entropy} develops the first-order dynamics, entropy production, constitutive matrices, positivity conditions, and Onsager--Casimir relations.  Section~\ref{sec:second-order-entropy} introduces the restricted transient entropy-current extension, and Sec.~\ref{sec:linearized-blocks} gives the linearized equations and block matrices.  Sections~\ref{sec:ideal-modular-analysis}--\ref{sec:second-order-modular-analysis} analyze the ideal, first-order, and second-order spectra.  Section~\ref{sec:numerical-causality-analysis} presents the characteristic speeds and their relation to causality, and Sec.~\ref{sec:summary} summarizes the results and outlook.  The appendices collect the tensor decompositions, matching conditions, entropy-current expressions, explicit matrix entries, and asymptotic coefficient equations.

Throughout this work we use natural units $c=\hbar=k_B=1$ and the Minkowski metric $g_{\mu\nu}=\mathrm{diag}(1,-1,-1,-1)$.  The fluid velocity obeys $u^\mu u_\mu=1$.  We denote the magnetic-field magnitude by $B\equiv\sqrt{-B^\mu B_\mu}$ and its unit direction by $b^\mu\equiv B^\mu/B$, so that $b^\mu b_\mu=-1$ and $u^\mu b_\mu=0$.  The projectors orthogonal to $u^\mu$ and to both $u^\mu$ and $b^\mu$ are $\Delta^{\mu\nu}\equiv g^{\mu\nu}-u^\mu u^\nu$ and $\Xi^{\mu\nu}\equiv\Delta^{\mu\nu}+b^\mu b^\nu$, respectively.  In the two-dimensional subspace projected by $\Xi^{\mu\nu}$, the symmetric-traceless projector is $\Xi^{\mu\nu\alpha\beta}\equiv\tfrac12(\Xi^{\mu\alpha}\Xi^{\nu\beta}+\Xi^{\mu\beta}\Xi^{\nu\alpha}-\Xi^{\mu\nu}\Xi^{\alpha\beta})$, and $\epsilon_\perp^{\mu\nu}\equiv\epsilon^{\mu\nu\alpha\beta}u_\alpha b_\beta$.  We choose $\epsilon_{0123}=+1$, so that $\epsilon^{0123}=-1$ and $\epsilon_\perp^{xy}=+1$ for a field along $+z$.  Derivatives along the flow, along the magnetic direction, and in the transverse plane are denoted by $D\equiv u^\mu\partial_\mu$, $\widetilde D\equiv-b^\mu\partial_\mu$, and $\widetilde\nabla^\mu\equiv\Xi^{\mu\nu}\partial_\nu$.

Repeated antisymmetric index pairs are summed over both orderings.  Thus $\omega_{\mu\nu}S^{\mu\nu}=2\sum_{\mu<\nu}\omega_{\mu\nu}S^{\mu\nu}$, with no additional factor $1/2$ in the thermodynamic relations below.

The leading-order magnetic-field counting treats $u^\mu$, $T$, $B^\mu$, $p_\perp$, $p_\parallel$, $s$, and $S^{\mu\nu}$ as $O(\partial^0)$, while $\omega_{\mu\nu}$ and the dissipative corrections are $O(\partial)$.  Hence the antisymmetric force $\partial_{[\mu}\beta_{\nu]}+2\beta\omega_{\mu\nu}$ is first order, whereas $Q^\mu=O(\partial^2)$ and $\partial_\mu Q^\mu=O(\partial^3)$.  This is a macroscopic gradient counting, not a microscopic expansion in $\hbar$, and it does not require magnetic energy to dominate the matter enthalpy.

\section{Ideal anisotropic spin fluid and magnetic-flux dynamics}
	\label{sec:ideal-entropy}
	
	The ideal entropy analysis fixes the thermodynamic identities and tensor basis used
	in the dissipative theory, following the entropy-current approach to RMHD and spin
	hydrodynamics \cite{HattoriHongoHuang2022,BiswasSecondOrder2023}.
	Appendix~\ref{app:tensor-decompositions} gives the tensor decompositions, and
	Appendix~\ref{app:matching-conditions} describes the matching conditions.
	
	The residual $SO(2)$ symmetry splits tensor responses into longitudinal, transverse,
	and mixed channels.  In the spin sector, an asymmetric energy--momentum tensor at dissipative order allows spin and orbital angular momentum
	to exchange locally \cite{FlorkowskiKumarRyblewski2019,XieWangYangPu2023}.  Magnetic pressure anisotropy is constrained by thermodynamics, while spin enters through angular-momentum balance and the Gibbs relation.
	
	The governing conservation and balance equations are
	\begin{equation}
		\partial_\mu T^{\mu\nu}=0,
		\qquad
		\partial_\mu \tilde F^{\mu\nu}=0,
		\qquad
		\partial_\lambda S^{\lambda\mu\nu}=-2T^{[\mu\nu]}.
		\label{eq:eom-main}
	\end{equation}
	The rank-three spin tensor is antisymmetric in its last two indices,
	$S^{\lambda\mu\nu}=-S^{\lambda\nu\mu}$, and its projection along the fluid velocity defines
	the hydrodynamic spin density, $S^{\mu\nu}\equiv u_\lambda S^{\lambda\mu\nu}$.  The
	remaining nonequilibrium spin flux is transverse to the flow in its first index,
	\begin{equation}
		S^{\lambda\mu\nu}
		=u^\lambda S^{\mu\nu}
		+
		\phi^{\langle\lambda\rangle\mu\nu},
		\qquad u_\lambda \phi^{\langle\lambda\rangle\mu\nu}=0 .
		\label{eq:spin-decomp}
	\end{equation}
	Thus $S^{\mu\nu}$ is the sole ideal spin-density variable, whereas
	$\phi^{\langle\lambda\rangle\mu\nu}$ describes dissipative spin transport relative to the
	local fluid rest frame.
	
	At ideal order we take
	\begin{equation}
		T_{(0)}^{\mu\nu}
		=
		\epsilon\,u^\mu u^\nu
		-
		p_\perp \Xi^{\mu\nu}
		+
		p_\parallel b^\mu b^\nu,
		\label{eq:T0}
	\end{equation}
	\begin{equation}
		\tilde F_{(0)}^{\mu\nu}
		=
		B^\mu u^\nu-B^\nu u^\mu,
		\qquad
		B^\mu=B\,b^\mu,
		\label{eq:F0}
	\end{equation}
	\begin{equation}
		S_{(0)}^{\lambda\mu\nu}
		=u^\lambda S^{\mu\nu}.
		\label{eq:S0}
	\end{equation}\vspace{-1pt}
	The ideal energy--momentum tensor is symmetric, $T_{(0)}^{[\mu\nu]}=0$; consequently,
	spin and orbital angular momentum do not exchange locally at zeroth derivative order.
	
	The local first law is
	\begin{equation}
		d\epsilon
		=
		T\,ds
		+
		\omega_{\mu\nu}\,dS^{\mu\nu}
		+
		H_\mu\,dB^\mu .
		\label{eq:firstlaw}
	\end{equation}
	The conjugate magnetic field is chosen flow-transverse, $u_\mu H^\mu=0$.
	Here ``ideal'' labels the constitutive tensors in Eqs.~\eqref{eq:T0}--\eqref{eq:S0}.  Because $\omega_{\mu\nu}=O(\partial)$ in the adopted counting, the spin term in Eq.~\eqref{eq:firstlaw} is retained to the order needed by the subsequent entropy analysis; it does not introduce an additional zeroth-order anisotropic stress.  At this stage $p_\perp$, $p_\parallel$, and $H^\mu$ are regarded as local thermodynamic coefficient functions.  Requiring the ideal entropy current to be adiabatic fixes their algebraic relations below.  Generalized thermodynamic relations for perfect spin hydrodynamics and their relation to dissipative constructions have been discussed in Refs.~\cite{FlorkowskiHontarenko2025,BecattiniDaherSheng2024}.
	
	The ideal entropy current is taken as
	\begin{equation}
		s^\mu_{(0)}=s\,u^\mu.
		\label{eq:s0}
	\end{equation}
	Its divergence is
	\begin{equation}
		\partial_\mu s^\mu_{(0)}=Ds+s\theta,
		\qquad
		\theta\equiv \partial_\mu u^\mu .
	\end{equation}
	
	Projecting the ideal energy--momentum and magnetic-flux equations along $u^\mu$ and
	$H^\mu$, respectively, gives
	\begin{equation}
		D\epsilon
		+
		(\epsilon+p_\perp)\theta
		+
		(p_\perp-p_\parallel)b^\mu b^\nu\partial_\mu u_\nu
		=
		0,
		\label{eq:ideal-energy}
	\end{equation}
	\begin{equation}
		- H_\mu D B^\mu
		-(B\!\cdot\!H)\theta
		+
		B^\mu H^\nu\partial_\mu u_\nu
		=
		0,
		\label{eq:ideal-magnetic}
	\end{equation}
	while the ideal spin-balance equation reduces to
	\begin{equation}
		D S^{\mu\nu}
		+\theta\,S^{\mu\nu}
		=0.
		\label{eq:ideal-spin}
	\end{equation}
	
	Using Eqs.~\eqref{eq:firstlaw}, \eqref{eq:ideal-energy}, \eqref{eq:ideal-magnetic},
	and \eqref{eq:ideal-spin}, the divergence of the ideal entropy current becomes
	\begin{widetext}
		\begin{equation}
			\partial_\mu s^\mu_{(0)}
			=
			\beta
			\Big(
			Ts-\epsilon-p_\perp+B^\mu H_\mu+\omega_{\mu\nu}S^{\mu\nu}
			\Big)\theta
			-
			\beta\,b^\mu
			\Big[
			(p_\perp-p_\parallel)b^\nu+BH^\nu
			\Big]
			\partial_\mu u_\nu .
			\label{eq:ideal-entropy}
		\end{equation}
	\end{widetext}
	Requiring this divergence to vanish independently for the scalar expansion and anisotropic strain structures yields
	\begin{equation}
		Ts+\omega_{\mu\nu}S^{\mu\nu}+B^\mu H_\mu
		=
		\epsilon+p_\perp ,
		\label{eq:euler}
	\end{equation}
	\begin{equation}
		(p_\perp-p_\parallel)b^\nu+BH^\nu=0 .
		\label{eq:ideal-pressure-anisotropy}
	\end{equation}
	
	Equation~\eqref{eq:euler} is the Euler-type thermodynamic identity selected by the
	ideal entropy analysis.  Together with the local first law, it implies the Gibbs--Duhem
	relation $dp_\perp=s\,dT+S^{\mu\nu}d\omega_{\mu\nu}+B^\mu dH_\mu$.
	Equation~\eqref{eq:ideal-pressure-anisotropy} gives the corresponding magnetic pressure
	anisotropy.  With the convention $H^\mu=-H b^\mu$, it reduces to
	$p_\perp-p_\parallel=BH=B^\mu H_\mu$.  Together with the residual $SO(2)$ symmetry,
	this relation motivates separate longitudinal and transverse response coefficients in the
	subsequent dissipative theory, including $\zeta_\perp$, $\zeta_\parallel$,
	$\rho_\perp$, $\rho_\parallel$, $\eta_\perp$, and $\eta_\parallel$.
	
	\section{First-order derivative corrections and entropy-current analysis}
	\label{sec:first-order-entropy}
	
	We insert the dissipative decompositions into the conservation laws, identify the
	thermodynamic forces in the Navier--Stokes entropy production, and constrain the
	resulting $SO(2)$ transport matrices by the local second law and Onsager--Casimir reciprocity.
	
	The decomposition details and the matching procedure are collected in
	Appendix~\ref{app:tensor-decompositions} and
	Appendix~\ref{app:matching-conditions}.
	We work in a Landau-type frame adapted to RMHD: perturbative shifts of the energy density
	and magnetic-field magnitude are absorbed into the local equilibrium variables, and the
	redundant heat and magnetic-frame components are removed.  In this matching choice,
	$h^\mu$ and $g^\mu$ are set to zero.
	The matched first-order constitutive structures are then
	\begin{equation}
		T^{\mu\nu}_{(1)}
		=
		-\Pi_\perp \Xi^{\mu\nu}
		+\Pi_\parallel b^\mu b^\nu
		+2f^{(\mu}b^{\nu)}
		+\pi_{\perp}^{\mu\nu}
		+T_{(1)}^{[\mu\nu]},
		\label{eq:T1}
	\end{equation}
	\begin{equation}
		\tilde F^{\mu\nu}_{(1)}
		=
		2b^{[\mu}\ell^{\nu]}
		+
		m^{\mu\nu},
		\label{eq:F1}
	\end{equation}
	and
	\begin{equation}
		S^{\lambda\mu\nu}_{(1)}=\phi^{\langle\lambda\rangle\mu\nu}.
		\label{eq:S1}
	\end{equation}
	The antisymmetric part of the first-order energy-momentum tensor is written as
	\begin{equation}
		T_{(1)}^{[\mu\nu]}
		=
		K\,u^{[\mu}b^{\nu]}
		+
		2u^{[\mu}q^{\nu]}
		+
		2b^{[\mu}r^{\nu]}
		+
		\phi^{\mu\nu},
		\label{eq:T1anti}
	\end{equation}
	with
	\begin{equation}
		u_\mu q^\mu=b_\mu q^\mu=0,
		\qquad
		u_\mu r^\mu=b_\mu r^\mu=0.
	\end{equation}
	Here $\phi^{\mu\nu}$ denotes the two-index transverse antisymmetric stress; it is distinct from the rank-three nonequilibrium spin flux $\phi^{\langle\lambda\rangle\mu\nu}$ in Eq.~\eqref{eq:spin-decomp}.
	The explicit first-order spin flux is decomposed as
	\begin{equation}
		\begin{aligned}
			\phi^{\langle\lambda\rangle\mu\nu}={}&
			2b^{\lambda}u^{[\mu}b^{\nu]}\Lambda_{\parallel}
			-2u^{[\mu}b^{\nu]}\Lambda_{\perp}^{\lambda}
			\\
			&-2b^{\lambda}u^{[\mu}\Phi^{\nu]}
			+u^{[\mu}\Xi^{\nu]\lambda}M_{(t)}
			\\
			&+2u^{[\mu}M_{(s)}^{\nu]\lambda}
			-2u^{[\mu}M_{(a)}^{\nu]\lambda}
			\\
			&+2b^{\lambda}b^{[\mu}W^{\nu]}
			-b^{[\mu}\Xi^{\nu]\lambda}N_{(t)}
			\\
			&-2b^{[\mu}N_{(s)}^{\nu]\lambda}
			+2b^{[\mu}N_{(a)}^{\nu]\lambda}
			\\
			&-b^{\lambda}\Omega_{\parallel}^{\mu\nu}
			+\Omega^{\lambda\mu\nu} .
		\end{aligned}
		\label{eq:spinflux}
	\end{equation}
	Appendix~\ref{app:tensor-decompositions} defines the projected scalar, vector, and tensor amplitudes in Eq.~\eqref{eq:spinflux} and gives the rank-three decomposition.
	
	The first-order conservation laws follow from inserting
	Eqs.~\eqref{eq:T1}--\eqref{eq:spinflux} into Eq.~\eqref{eq:eom-main}.
	The projected energy equation becomes
	\begin{equation}
		\begin{aligned}
			0={}&
			D\epsilon
			+
			(\epsilon+p_\perp)\theta
			\\
			&+(p_\perp-p_\parallel)b^\mu b^\nu\partial_\mu u_\nu
			+\Pi_\perp\theta_\perp
			+\Pi_\parallel\theta_\parallel
			\\
			&-(b^\mu f^\nu+b^\nu f^\mu)\partial_\mu u_\nu
			-\pi_{\perp}^{\mu\nu}\partial_\mu u_\nu
			\\
			&+u_\nu\partial_\mu T_{(1)}^{[\mu\nu]},
		\end{aligned}
		\label{eq:first-energy}
	\end{equation}
	where
	\begin{equation}
		\theta_\perp\equiv \Xi^{\mu\nu}\partial_\mu u_\nu,
		\qquad
		\theta_\parallel\equiv -b^\mu b^\nu\partial_\mu u_\nu.
	\end{equation}
	The longitudinal and transverse momentum equations read
	\begin{equation}
		(\epsilon+p_\perp)b_\nu D u^\nu
		+\tilde D p_\parallel
		+(p_\perp-p_\parallel)\partial_\mu b^\mu
		+b_\nu\partial_\mu T_{(1)}^{\mu\nu}=0,
		\label{eq:first-momentum-par}
	\end{equation}
	\begin{equation}
		\begin{aligned}
		&(\epsilon+p_\perp)\Xi^\rho{}_{\nu} D u^\nu
		-\tilde\nabla^\rho p_\perp
		\\
		&\quad+(p_\perp-p_\parallel)\Xi^\rho{}_{\nu}\tilde D b^\nu
		+\Xi^\rho{}_{\nu}\partial_\mu T_{(1)}^{\mu\nu}=0.
		\end{aligned}
		\label{eq:first-momentum-perp}
	\end{equation}
	The magnetic projections yield
	\begin{equation}
		- H_\mu D B^\mu
		-(B\!\cdot\!H)\theta
		+
		B^\mu H^\nu\partial_\mu u_\nu
		+
		H_\nu\partial_\mu \tilde F^{\mu\nu}_{(1)}=0,
		\label{eq:first-magnetic-scalar}
	\end{equation}
	\begin{equation}
		-\,B\,\Xi^\rho{}_{\nu}\left(D b^\nu+\tilde D u^\nu\right)
		+
		\Xi^\rho{}_{\nu}\partial_\mu \tilde F^{\mu\nu}_{(1)}=0.
		\label{eq:first-magnetic-transverse}
	\end{equation}
	The first-order spin-balance equation is
	\begin{equation}
		\partial_\lambda
		\left(
		u^\lambda S^{\mu\nu}
		+
		\phi^{\langle\lambda\rangle\mu\nu}
		\right)
		=
		-2T_{(1)}^{[\mu\nu]}.
		\label{eq:first-spinbalance}
	\end{equation}
	
	For the entropy analysis, we use the nonequilibrium entropy current.  Related first-order entropy-current classifications and transport-coefficient analyses are discussed in Refs.~\cite{BecattiniDaherSheng2024,DaherShengWagnerBecattini2025}.
	\begin{equation}
		\begin{aligned}
			s^\mu_{\rm NS}={}&s^\mu_{(0)}
			+\beta_\nu T^{\mu\nu}_{(1)}
			+\mathcal H_\nu \tilde F^{\mu\nu}_{(1)}
			\\
			&-\beta\,\omega_{\alpha\beta}
			\phi^{\langle\mu\rangle\alpha\beta},
			\qquad
			\beta^\mu=\beta u^\mu,
			\quad
			\mathcal H_\mu=\beta H_\mu .
		\end{aligned}
		\label{eq:sNS}
	\end{equation}
	Its divergence can be written as
	\begin{equation}
		\begin{aligned}
			\partial_\mu s^\mu_{\rm NS}={}&
			T_{(1)}^{\mu\nu}\partial_\mu\beta_\nu
			+\tilde F_{(1)}^{\mu\nu}\partial_\mu\mathcal H_\nu
			\\
			&-\phi^{\langle\lambda\rangle\mu\nu}
			\partial_\lambda(\beta\omega_{\mu\nu})
			+2\beta\,\omega_{\mu\nu}T_{(1)}^{[\mu\nu]} .
		\end{aligned}
		\label{eq:sNSdiv-compact}
	\end{equation}
	The first two corrections in Eq.~\eqref{eq:sNS} are linear in the dissipative currents, whereas the final spin-flux term is already quadratic in the adopted counting.  Specifically, the spin chemical potential and the matched spin flux satisfy
	$\omega_{\mu\nu}\sim O(\partial)$ and
	$\phi^{\langle\lambda\rangle\mu\nu}\sim O(\partial)$.
	Consequently,
	$\phi^{\langle\lambda\rangle\mu\nu}\partial_\lambda(\beta\omega_{\mu\nu})$
	is of order $O(\partial^3)$ and does not enter the Navier--Stokes entropy
	production.  It is retained in Eq.~\eqref{eq:sNSdiv-compact} to identify
	the spin-flux term needed in the second-order entropy-current analysis.  The first-order entropy production is therefore
	\begin{equation}
		\begin{aligned}
			\partial_\mu s^\mu_{\rm NS}={}&
			T_{(1)}^{\mu\nu}\partial_\mu\beta_\nu
			+\tilde F_{(1)}^{\mu\nu}\partial_\mu\mathcal H_\nu
			\\
			&+2\beta\,\omega_{\mu\nu}T_{(1)}^{[\mu\nu]}
			+O(\partial^3).
		\end{aligned}
		\label{eq:sNSdiv-NSorder}
	\end{equation}
	The last term follows from the first-order spin-balance equation
	\eqref{eq:first-spinbalance}; it is the only explicit spin-dependent contribution that remains in the Navier--Stokes entropy production.
	
	Substituting the irreducible decompositions of $T^{\mu\nu}_{(1)}$
	and $\tilde F^{\mu\nu}_{(1)}$ into Eq.~\eqref{eq:sNSdiv-NSorder}, the
	Navier--Stokes entropy production becomes
	\begin{equation}
		\begin{aligned}
			\partial_\mu s^\mu_{\rm NS}={}&
			-\beta \Pi_\perp\theta_\perp
			-\beta \Pi_\parallel\theta_\parallel
			+2f^\mu b^{(\alpha}\Xi^{\beta)}{}_{\mu}
			\partial_\alpha\beta_\beta
			\\
			&+\beta\pi_\perp^{\mu\nu}
			\Xi_{\mu\nu}{}^{\alpha\beta}\partial_\alpha u_\beta
			+2\ell^\mu b^{[\alpha}\Xi^{\beta]}{}_{\mu}
			\partial_\alpha\mathcal H_\beta
			\\
			&+m_{\mu\nu}\Xi^{\mu[\alpha}\Xi^{\beta]\nu}
			\partial_\alpha\mathcal H_\beta
			\\
			&+K u^\mu b^\nu
			\big(\partial_{[\mu}\beta_{\nu]}+2\beta\omega_{\mu\nu}\big)
			\\
			&+2q^\rho u^\mu\Xi^\nu{}_{\rho}
			\big(\partial_{[\mu}\beta_{\nu]}+2\beta\omega_{\mu\nu}\big)
			\\
			&+2r^\rho b^\mu\Xi^\nu{}_{\rho}
			\big(\partial_{[\mu}\beta_{\nu]}+2\beta\omega_{\mu\nu}\big)
			\\
			&+\phi_{\mu\nu}\Xi^\mu{}_{\rho}\Xi^\nu{}_{\sigma}
			\big(\partial^{[\rho}\beta^{\sigma]}
			+2\beta\omega^{\rho\sigma}\big)
			+O(\partial^3).
		\end{aligned}
		\label{eq:sNSdiv-full}
	\end{equation}
	The spin-flux amplitudes in Eq.~\eqref{eq:spinflux} are therefore absent from this
	Navier--Stokes quadratic form; their transport constraints enter at second order.
	
	Equation~\eqref{eq:sNSdiv-full} also fixes the sign convention for the transport
	coefficients.  Bulk stresses respond with the usual minus sign to expansion, while the
	symmetric-traceless and transverse antisymmetric tensor channels have positive quadratic
	norms in the local rest frame.  The vector forces are spacelike.  Their constitutive signs are chosen so that
	the corresponding transport matrices enter the entropy production as
	positive-semidefinite quadratic forms.  Within the retained ansatz, cross couplings are included only between forces with the same tensorial character and compatible discrete-symmetry properties, in accordance with Curie's principle and Onsager--Casimir reciprocity in a background magnetic field.
	
	We organize the resulting constitutive relations by irreducible $SO(2)$ sector.
	For the magnetic transport coefficients, we define
	\begin{equation}
		\rho_a\equiv\beta_0\bar\rho_a,
		\qquad
		\kappa_{AB}\equiv\sqrt{\beta_0}\,\bar\kappa_{AB},
		\label{eq:magnetic-transport-rescaling}
	\end{equation}
	where $\beta_0\equiv1/T_0$ is the equilibrium inverse temperature,
	$a=\parallel,\perp$, and $AB=\ell q,q\ell,m\phi,\phi m$.
	
	The scalar sector is
	\begin{equation}
		\begin{aligned}
			\Pi_{\perp}&=-\zeta_{\perp}\theta_{\perp}
			-\zeta_{\times}\theta_{\parallel},
			\\
			\Pi_{\parallel}&=-\zeta_{\parallel}\theta_{\parallel}
			-\zeta_{\times}^{\prime}\theta_{\perp},
			\\
			\label{eq:constitutive-scalar}
			K&=\lambda_K u^\mu b^\nu
			\big(\partial_{[\mu}\beta_{\nu]}
			+2\beta\omega_{\mu\nu}\big).
		\end{aligned}
	\end{equation}
	
	For the axial-vector sector,
	\begin{equation}
		\begin{aligned}
			f^\mu={}&-2\eta_{\parallel}
			b^{(\alpha}\Xi^{\beta)\mu}\partial_\alpha\beta_\beta
			\\
			&-\kappa_{fr}b^{\alpha}\Xi^{\mu\beta}
			\big(\partial_{[\alpha}\beta_{\beta]}
			+2\beta\omega_{\alpha\beta}\big),
			\\
			r^\mu={}&-\kappa_{rf}
			b^{(\alpha}\Xi^{\beta)\mu}\partial_\alpha\beta_\beta
			\\
			&-\lambda_r b^{\alpha}\Xi^{\mu\beta}
			\big(\partial_{[\alpha}\beta_{\beta]}
			+2\beta\omega_{\alpha\beta}\big).
		\end{aligned}
		\label{eq:constitutive-axial}
	\end{equation}
	
	For the polar-vector sector,
	\begin{equation}
		\begin{aligned}
			\ell^\mu={}&-\frac{2\rho_{\parallel}}{\beta_0}
			b^{[\alpha}\Xi^{\beta]\mu}\partial_\alpha\mathcal H_\beta
			\\
			&-\frac{\kappa_{\ell q}}{\sqrt{\beta_0}}u^{\alpha}\Xi^{\mu\beta}
			\big(\partial_{[\alpha}\beta_{\beta]}
			+2\beta\omega_{\alpha\beta}\big),
			\\
			q^\mu={}&-\frac{\kappa_{q\ell}}{\sqrt{\beta_0}}
			b^{[\alpha}\Xi^{\beta]\mu}\partial_\alpha\mathcal H_\beta
			\\
			&-\lambda_q u^{\alpha}\Xi^{\mu\beta}
			\big(\partial_{[\alpha}\beta_{\beta]}
			+2\beta\omega_{\alpha\beta}\big).
		\end{aligned}
		\label{eq:constitutive-polar}
	\end{equation}
	
	For the transverse symmetric-traceless tensor sector,
	\begin{equation}
		\pi_{\perp}^{\mu\nu}
		=2\eta_{\perp}\Xi^{\mu\nu\alpha\beta}
		\partial_\alpha u_\beta.
		\label{eq:pi-irred}
	\end{equation}
	
	The correction $m^{\mu\nu}$ belongs to the pseudotensor $\tilde F^{\mu\nu}$,
	whereas $\phi^{\mu\nu}$ belongs to the ordinary tensor $T^{\mu\nu}$.
	Their opposite spatial parity gives $\kappa_{m\phi}=\kappa_{\phi m}=0$
	in a parity-even equilibrium.  Their constitutive relations are
	\begin{equation}
		\begin{aligned}
			m^{\mu\nu}={}&\frac{2\rho_\perp}{\beta_0}
			\Xi^{\mu[\alpha}\Xi^{\beta]\nu}
			\partial_\alpha\mathcal H_\beta,
			\\
			\phi^{\mu\nu}={}&\lambda_\phi\Xi^\mu{}_{\rho}\Xi^\nu{}_{\sigma}
			\big(\partial^{[\rho}\beta^{\sigma]}
			+2\beta\omega^{\rho\sigma}\big).
		\end{aligned}
		\label{eq:constitutive-AT}
	\end{equation}
	
	Eqs.~\eqref{eq:constitutive-scalar}--\eqref{eq:constitutive-AT} constitute the retained Navier--Stokes constitutive set under the adopted power counting.

	The coupled first-order constitutive relations decompose into independent irreducible
	sectors.  Substituting them into Eq.~\eqref{eq:sNSdiv-full} gives a set of quadratic forms
	whose explicit matrices are collected in Appendix~\ref{app:first-order-entropy-matrices}.
	For arbitrary first-order gradients, the local second law requires the corresponding
	symmetric transport matrices to be positive semidefinite.  We therefore obtain
		\begin{align}
			&\zeta_\perp\ge 0,
			\qquad
			\zeta_\parallel\ge 0,
			\qquad
			4\zeta_\perp\zeta_\parallel-(\zeta_\times+\zeta_\times')^2\ge 0,
			\notag\\
			&\lambda_K\ge 0,
			\label{eq:scalar-positive-main}
			\\
			&\eta_\parallel\ge 0,
			\qquad
			\lambda_r\ge 0,
			\qquad
			8\eta_\parallel\lambda_r-(\kappa_{fr}+\kappa_{rf})^2\ge 0,
			\label{eq:axial-positive-main}
			\\
			&\rho_\parallel\ge0,
			\qquad
			\lambda_q\ge0,
			\qquad
			8\rho_\parallel\lambda_q-(\kappa_{\ell q}+\kappa_{q\ell})^2\ge0,
			\label{eq:polar-positive-main}
			\\
			&\eta_\perp\ge0,
			\label{eq:st-positive-main}
			\\
			&\rho_\perp\ge0,
			\qquad
			\lambda_\phi\ge0 .
			\label{eq:at-positive-main}
		\end{align}
		Assuming microscopic time-reversal invariance and the conventional parities of the paired forces, Onsager--Casimir reciprocity further relates the cross coefficients under magnetic-field reversal \cite{HattoriHongoHuang2022,RochaWagnerDenicolNoronhaRischke2024}:
		\begin{equation}
			\begin{aligned}
				\zeta_\times(B)&=\zeta_\times'(-B),
				&\qquad
				\kappa_{fr}(B)&=\kappa_{rf}(-B),
				\\
				\kappa_{\ell q}(B)&=\kappa_{q\ell}(-B).
			\end{aligned}
			\label{eq:onsager-main}
		\end{equation}
	Within the stated time-reversal assumptions, these relations complete the first-order entropy analysis at the level needed below.  The
	transport coefficients associated with the split spin-flux channels in
	Eq.~\eqref{eq:spinflux} are not constrained at Navier--Stokes order; their sign constraints
	enter the second-order analysis.

	\section{Second-order anisotropic spin hydrodynamics from the entropy current}
	\label{sec:second-order-entropy}
	
	The retained Navier--Stokes theory has an instantaneous dissipative response,
	which can lead to acausality or instability.  In first-order theories with
	antisymmetric stress, the stability constraints also depend on the spin equation of state
	\cite{XieWangYangPu2023,Weickgenannt2023}.  We therefore consider an Israel--Stewart-type
	construction in which dissipative currents and spin-flux variables are promoted to
	independent transient fields.  This follows the entropy-current strategy used in
	second-order spin hydrodynamics \cite{BiswasSecondOrder2023}, including
	the additional anisotropic tensor splitting required by RMHD\@.  In particular, the magnetic
	axis resolves the spin flux into inequivalent $SO(2)$ scalar, vector, symmetric-traceless, and
	antisymmetric channels, which may carry distinct relaxation scales and cross responses.
	Related spin-hydrodynamic analyses show how finite relaxation replaces pathological
	first-order short-wavelength behavior with a transient spectrum.  Its causality and
	stability depend on the transport data and the background state
	\cite{DaherSecondOrder2024,LuZhongFeng2026}.
	
	The Navier--Stokes analysis used Eq.~\eqref{eq:sNSdiv-NSorder}, where the spin-flux contribution $-\phi^{\langle\lambda\rangle\mu\nu}\partial_\lambda(\beta\omega_{\mu\nu})$ was counted as $O(\partial^3)$ and therefore omitted from the first-order entropy production.  In the Israel--Stewart analysis it is retained because it is of the same order as the divergence of the quadratic entropy correction.  We therefore start
	from
	\begin{equation}
		\partial_\mu s^\mu_{\rm IS}
		=
		\left[\partial_\mu s^\mu_{\rm NS}\right]_{O(\partial^2)}
		-
		\phi^{\langle\lambda\rangle\mu\nu}
		\partial_\lambda(\beta\omega_{\mu\nu})
		+\partial_\mu Q^\mu .
		\label{eq:sISdiv-start}
	\end{equation}
	Here $\left[\partial_\mu s^\mu_{\rm NS}\right]_{O(\partial^2)}$ denotes the $O(\partial^2)$ entropy production displayed in Eq.~\eqref{eq:sNSdiv-full}, before the Navier--Stokes constitutive relations are substituted.
	
	We split the second-order entropy-current correction into a flow-aligned entropy-density
	correction and a flow-transverse entropy flux,
	\begin{equation}
		Q^\mu=u^\mu Q_u+Q_\perp^\mu,
		\qquad
		u_\mu Q_\perp^\mu=0 .
		\label{eq:Q-split}
	\end{equation}
	The transient closure considered here sets the vector transport
	cross coefficients $\kappa$ and the transverse entropy-flux coefficients $b_i,c_i$
	to zero.  The anisotropic bulk cross response is retained with
	$\zeta_\times=\zeta_\times'$ and a common bulk relaxation time
	$\tau_{\Pi_\perp}=\tau_{\Pi_\parallel}\equiv\tau_\Pi$.
	For the flow-aligned part, we retain the quadratic form
	\begin{equation}
		\begin{aligned}
			Q_u={}&
			\frac{a_\perp}{2}\Pi_\perp^2
			+\frac{a_\parallel}{2}\Pi_\parallel^2
			+a_\times\Pi_\perp\Pi_\parallel
			+\frac{a_K}{2}K^2
			\\
			&+\frac{a_f}{2}f^\lambda f_\lambda
			+\frac{a_\ell}{2}\ell^\lambda\ell_\lambda
			+\frac{a_q}{2}q^\lambda q_\lambda
			+\frac{a_r}{2}r^\lambda r_\lambda
			\\
			&+\frac{a_\pi}{2}\pi_\perp^{\lambda\nu}\pi_{\perp\lambda\nu}
			+\frac{a_m}{2}m^{\lambda\nu}m_{\lambda\nu}
			+\frac{a_\phi}{2}\phi^{\lambda\nu}\phi_{\lambda\nu}
			\\
			&+\frac{a_{\Lambda\parallel}}{2}\Lambda_\parallel^2
			+\frac{a_{\Lambda\perp}}{2}
			\Lambda_\perp^\lambda\Lambda_{\perp\lambda}
			+\frac{a_\Phi}{2}\Phi^\lambda\Phi_\lambda
			\\
			&+\frac{a_W}{2}W^\lambda W_\lambda
			+\frac{a_{Mt}}{2}M_{(t)}^2
			+\frac{a_{Nt}}{2}N_{(t)}^2
			\\
			&+\frac{a_{Ms}}{2}M_{(s)}^{\lambda\nu}M_{(s)\lambda\nu}
			+\frac{a_{Ns}}{2}N_{(s)}^{\lambda\nu}N_{(s)\lambda\nu}
			\\
			&+\frac{a_{Ma}}{2}M_{(a)}^{\lambda\nu}M_{(a)\lambda\nu}
			+\frac{a_{Na}}{2}N_{(a)}^{\lambda\nu}N_{(a)\lambda\nu}
			\\
			&+\frac{a_{\Omega\parallel}}{2}
			\Omega_\parallel^{\lambda\nu}\Omega_{\parallel\lambda\nu}
			+\frac{a_\Omega}{2}\Omega^{\rho\lambda\nu}\Omega_{\rho\lambda\nu} .
		\end{aligned}
		\label{eq:Qu}
	\end{equation}
	This closure retains a subset of the allowed second-order entropy-current
	structures.  A general construction would also contain further off-diagonal quadratic terms,
	longitudinal entropy flux along the magnetic direction, Hall-like transverse structures, and
	additional spin-flux couplings.  A nonlocal statistical-operator derivation also produces second-order couplings between different tensor ranks \cite{SheQiuJiangHou2025}, beyond the restricted entropy-current ansatz used here.  A representative symmetry-allowed transverse quadratic flux is
	\begin{equation}
		\begin{aligned}
			Q_\perp^\mu={}&
			b_1\Pi_\perp\ell^\mu+b_2\Pi_\perp q^\mu
			+b_3\Pi_\parallel\ell^\mu+b_4\Pi_\parallel q^\mu
			\\
			&+b_5Kf^\mu+b_6Kr^\mu
			+c_1\pi_\perp^{\mu\nu}\ell_\nu
			+c_2\pi_\perp^{\mu\nu}q_\nu
			\\
			&+c_3m^{\mu\nu}f_\nu+c_4m^{\mu\nu}r_\nu .
		\end{aligned}
		\label{eq:Qperp}
	\end{equation}
	The omitted $b_i,c_i$ terms would generate cross-gradient relaxation couplings
	without changing the Navier--Stokes limit.
	
	Within the quadratic near-equilibrium ansatz, the local entropy density is required
	to have a maximum at equilibrium.  The bulk entropy matrix is negative definite;
	the remaining diagonal coefficients have the signs
	\begin{equation}
		\begin{aligned}
			&a_\perp,a_\parallel,a_K,a_\pi,a_m,a_\phi,
			a_{\Lambda\parallel},a_{Mt},a_{Nt},
			\\
			&a_{Ms},a_{Ns},a_{Ma},a_{Na},a_{\Omega\parallel}\le0,
			\\
			&a_f,a_\ell,a_q,a_r,a_{\Lambda\perp},a_\Phi,a_W,a_\Omega\ge0 .
		\end{aligned}
		\label{eq:a-signs}
	\end{equation}
	
	The coefficients multiplying spacelike vector norms and the fully transverse rank-three
	norm have the opposite sign because those contractions are negative in the local rest
	frame.  The bulk cross term additionally requires $a_\perp a_\parallel-a_\times^2>0$.
	
	After reorganizing $\partial_\mu Q^\mu$ by irreducible channels, we write
	\begin{equation}
		\begin{aligned}
			\partial_\mu Q^\mu={}&
			\Pi_\perp\mathcal A_\perp
			+\Pi_\parallel\mathcal A_\parallel
			+K\mathcal A_K
			\\
			&+2f_\mu\mathcal B_f^\mu
			+2\ell_\mu\mathcal B_\ell^\mu
			+2q_\mu\mathcal B_q^\mu
			+2r_\mu\mathcal B_r^\mu
			\\
			&+\beta\pi_{\perp\mu\nu}\mathcal C_\pi^{\mu\nu}
			+m_{\mu\nu}\mathcal C_m^{\mu\nu}
			+\phi_{\mu\nu}\mathcal C_\phi^{\mu\nu}
			\\
			&+\Lambda_\parallel\mathcal A_{\Lambda\parallel}
			+2\Lambda_{\perp\mu}\mathcal B_{\Lambda\perp}^\mu
			+2\Phi_\mu\mathcal B_\Phi^\mu
			+2W_\mu\mathcal B_W^\mu
			\\
			&+M_{(t)}\mathcal A_{Mt}+N_{(t)}\mathcal A_{Nt}
			\\
			&+2M_{(s)\mu\nu}\mathcal D_{Ms}^{\mu\nu}
			+2N_{(s)\mu\nu}\mathcal D_{Ns}^{\mu\nu}
			\\
			&+2M_{(a)\mu\nu}\mathcal E_{Ma}^{\mu\nu}
			+2N_{(a)\mu\nu}\mathcal E_{Na}^{\mu\nu}
			\\
			&+\Omega_{\parallel\mu\nu}\mathcal E_{\Omega\parallel}^{\mu\nu}
			+\Omega_{\lambda\mu\nu}\mathcal G_\Omega^{\lambda\mu\nu} .
		\end{aligned}
		\label{eq:dQ-grouped}
	\end{equation}
	The bulk matrix, diagonal pieces, and transverse entropy flux are described in
	Appendix~\ref{app:second-order-entropy-contributions}.
	
	Substituting Eq.~\eqref{eq:dQ-grouped} into Eq.~\eqref{eq:sISdiv-start} gives
	\begin{widetext}
		\begin{equation}
			\begin{aligned}
				\partial_\mu s^\mu_{\rm IS}={}&
				-\beta\Pi_\perp(\theta_\perp-T\mathcal A_\perp)
				-\beta\Pi_\parallel(\theta_\parallel-T\mathcal A_\parallel)
				+K\left[u^\mu b^\nu\Big(\partial_{[\mu}\beta_{\nu]}+2\beta\omega_{\mu\nu}\Big)+\mathcal A_K\right]
				+2f^\mu\left[b^{(\alpha}\Xi^{\beta)}{}_{\mu}\partial_\alpha\beta_\beta+\mathcal B_{f\mu}\right]
				\\
				&+2\ell^\mu\left[b^{[\alpha}\Xi^{\beta]}{}_{\mu}\partial_\alpha\mathcal H_\beta+\mathcal B_{\ell\mu}\right]
				+2q^\mu\left[u^\alpha\Xi^\beta{}_{\mu}
				\Big(\partial_{[\alpha}\beta_{\beta]}+2\beta\omega_{\alpha\beta}\Big)+\mathcal B_{q\mu}\right]
				+2r^\mu\left[b^\alpha\Xi^\beta{}_{\mu}
				\Big(\partial_{[\alpha}\beta_{\beta]}+2\beta\omega_{\alpha\beta}\Big)+\mathcal B_{r\mu}\right]
				\\
				&+\beta\pi_\perp^{\mu\nu}\left[\Xi_{\mu\nu}{}^{\alpha\beta}\partial_\alpha u_\beta+\mathcal C_{\pi\mu\nu}\right]
				+m_{\mu\nu}\left[\Xi^{\mu[\alpha}\Xi^{\beta]\nu}\partial_\alpha\mathcal H_\beta+\mathcal C_m^{\mu\nu}\right]
				+\phi_{\mu\nu}\left[\Xi^\mu{}_{\rho}\Xi^\nu{}_{\sigma}
				\Big(\partial^{[\rho}\beta^{\sigma]}+2\beta\omega^{\rho\sigma}\Big)+\mathcal C_\phi^{\mu\nu}\right]
				\\
				&+\Lambda_\parallel\left[-2b^\rho u^{[\mu}b^{\nu]}\partial_\rho(\beta\omega_{\mu\nu})+\mathcal A_{\Lambda\parallel}\right]
				+2\Lambda_\perp^\rho\left[u^{[\mu}b^{\nu]}\partial_\rho(\beta\omega_{\mu\nu})+\mathcal B_{\Lambda\perp\rho}\right]
				+2\Phi^\sigma\left[b^\rho u^{[\mu}\Xi^{\nu]}{}_{\sigma}\partial_\rho(\beta\omega_{\mu\nu})+\mathcal B_{\Phi\sigma}\right]
				\\
				&+2W^\sigma\left[-b^\rho b^{[\mu}\Xi^{\nu]}{}_{\sigma}\partial_\rho(\beta\omega_{\mu\nu})+\mathcal B_{W\sigma}\right]
				+M_{(t)}\left[-u^{[\mu}\Xi^{\nu]\rho}\partial_\rho(\beta\omega_{\mu\nu})+\mathcal A_{Mt}\right]
				+N_{(t)}\left[b^{[\mu}\Xi^{\nu]\rho}\partial_\rho(\beta\omega_{\mu\nu})+\mathcal A_{Nt}\right]
				\\
				&+2M_{(s)}^{\rho\sigma}\left[-u^{[\mu}\Xi^{\nu]}{}_{\sigma}\partial_\rho(\beta\omega_{\mu\nu})+\mathcal D_{Ms\rho\sigma}\right]
				+2N_{(s)}^{\rho\sigma}\left[b^{[\mu}\Xi^{\nu]}{}_{\sigma}\partial_\rho(\beta\omega_{\mu\nu})+\mathcal D_{Ns\rho\sigma}\right]
				\\
				&+2M_{(a)}^{\rho\sigma}\left[-u^{[\mu}\Xi^{\nu]}{}_{\sigma}\partial_\rho(\beta\omega_{\mu\nu})+\mathcal E_{Ma\rho\sigma}\right]
				+2N_{(a)}^{\rho\sigma}\left[b^{[\mu}\Xi^{\nu]}{}_{\sigma}\partial_\rho(\beta\omega_{\mu\nu})+\mathcal E_{Na\rho\sigma}\right]
				\\
				&+\Omega_\parallel^{\mu\nu}\left[b^\rho\partial_\rho(\beta\omega_{\mu\nu})+\mathcal E_{\Omega\parallel\mu\nu}\right]
				+\Omega^{\rho\mu\nu}\left[-\partial_\rho(\beta\omega_{\mu\nu})+\mathcal G_{\Omega\rho\mu\nu}\right].
			\end{aligned}
			\label{eq:sISdiv-full}
		\end{equation}
	\end{widetext}
	
	In Eq.~\eqref{eq:sISdiv-full}, the divergence of $Q^\mu$ supplies the
	comoving derivatives of the dissipative currents.  Separating these terms from
	the remaining second-order gradients gives relaxation equations.  This is the
	M\"uller--Israel--Stewart mechanism: the currents approach their
	Navier--Stokes values over a finite relaxation time.  In the bulk sector define
	\begin{widetext}
	\begin{equation}
		\begin{gathered}
			\Pi=(\Pi_\perp,\Pi_\parallel)^{\mathsf T},
			\qquad
			\mathsf Z=
			\begin{pmatrix}\zeta_\perp&\zeta_\times\\
			\zeta_\times&\zeta_\parallel\end{pmatrix},\\
			A_\Pi=
			\begin{pmatrix}a_\perp&a_\times\\
			a_\times&a_\parallel\end{pmatrix}
			=-\beta\tau_\Pi\mathsf Z^{-1},
			\qquad \mathsf Z>0 .
		\end{gathered}
		\label{eq:bulk-entropy-matrix}
	\end{equation}
	Thus $Q_{u,\Pi}=\Pi^{\mathsf T}A_\Pi\Pi/2$,
	$-T\mathsf Z A_\Pi=\tau_\Pi\mathsf I$, and the bulk entropy production is
	$\beta\Pi^{\mathsf T}\mathsf Z^{-1}\Pi\geq0$.
	The matrix construction retains the RMHD bulk cross response
	\cite{QiuSheHou2026} without introducing additional linear transient couplings.
	We define
		\begin{equation}
			\begin{gathered}
				T\mathcal A_\Pi
				=-\tau_\Pi\mathsf Z^{-1}D\Pi
				+\widehat{\mathcal A}_\Pi,
				\qquad
				\mathcal A_K=-\frac{\tau_K}{\lambda_K}DK+\widehat{\mathcal A}_K,
				\\
				\mathcal B_f^\mu=\frac{\tau_f}{2\eta_\parallel}\Xi^\mu{}_{\nu}Df^\nu+\widehat{\mathcal B}_f^\mu,
				\qquad
				\mathcal B_\ell^\mu=\frac{\beta_0\tau_\ell}{2\rho_\parallel}\Xi^\mu{}_{\nu}D\ell^\nu+\widehat{\mathcal B}_\ell^\mu,
				\\
				\mathcal B_q^\mu=\frac{\tau_q}{\lambda_q}\Xi^\mu{}_{\nu}Dq^\nu+\widehat{\mathcal B}_q^\mu,
				\qquad
				\mathcal B_r^\mu=\frac{\tau_r}{\lambda_r}\Xi^\mu{}_{\nu}Dr^\nu+\widehat{\mathcal B}_r^\mu,
				\\
				\mathcal C_\pi^{\mu\nu}=-\frac{\tau_\pi}{2\eta_\perp}\Xi^{\mu\nu}{}_{\alpha\beta}D\pi_\perp^{\alpha\beta}+\widehat{\mathcal C}_\pi^{\mu\nu},
				\qquad
				\mathcal C_m^{\mu\nu}=-\frac{\beta_0\tau_m}{2\rho_\perp}\Xi^{[\mu}{}_{\alpha}\Xi^{\nu]}{}_{\beta}Dm^{\alpha\beta}+\widehat{\mathcal C}_m^{\mu\nu},
				\\
				\mathcal C_\phi^{\mu\nu}=-\frac{\tau_\phi}{\lambda_\phi}\Xi^{[\mu}{}_{\alpha}\Xi^{\nu]}{}_{\beta}D\phi^{\alpha\beta}+\widehat{\mathcal C}_\phi^{\mu\nu}.
			\end{gathered}
			\label{eq:ordinary-source-relaxation-defs}
		\end{equation}
		Here $\mathcal A_\Pi=(\mathcal A_\perp,\mathcal A_\parallel)^{\mathsf T}$.
		The bulk relaxation time is fixed by Eq.~\eqref{eq:bulk-entropy-matrix};
		the remaining relaxation times are
		\begin{equation}
			\begin{gathered}
				\tau_K\equiv -\lambda_K a_K,
				\\
				\tau_f\equiv \eta_\parallel a_f,
				\qquad
				\tau_\ell\equiv \frac{\rho_\parallel}{\beta_0}a_\ell,
				\qquad
				\tau_q\equiv \frac12\lambda_q a_q,
				\qquad
				\tau_r\equiv \frac12\lambda_r a_r,
				\\
				\tau_\pi\equiv -2T\eta_\perp a_\pi,
				\qquad
				\tau_m\equiv -\frac{2\rho_\perp}{\beta_0}a_m,
				\qquad
				\tau_\phi\equiv -\lambda_\phi a_\phi .
			\end{gathered}
			\label{eq:ordinary-relaxation-times}
		\end{equation}
		The ordinary magnetohydrodynamic and antisymmetric-stress channels then obey
		\begin{equation}
			\begin{aligned}
				\tau_{\Pi_\perp}D\Pi_\perp+\Pi_\perp={}&
				-\zeta_\perp\theta_\perp-\zeta_\times\theta_\parallel
				+\zeta_\perp\widehat{\mathcal A}_\perp+\zeta_\times\widehat{\mathcal A}_\parallel,
				\\
				\tau_{\Pi_\parallel}D\Pi_\parallel+\Pi_\parallel={}&
				-\zeta_\parallel\theta_\parallel-\zeta_\times\theta_\perp
				+\zeta_\times\widehat{\mathcal A}_\perp+\zeta_\parallel\widehat{\mathcal A}_\parallel,
				\\
				\tau_KDK+K={}&
				\lambda_K u^\mu b^\nu\Big(\partial_{[\mu}\beta_{\nu]}+2\beta\omega_{\mu\nu}\Big)
				+\lambda_K\widehat{\mathcal A}_K,
				\\
				\tau_f\Xi^\mu{}_{\nu}Df^\nu+f^\mu={}&
				-2\eta_\parallel b^{(\alpha}\Xi^{\beta)\mu}\partial_\alpha\beta_\beta
				-2\eta_\parallel\widehat{\mathcal B}_f^\mu,
				\\
				\tau_r\Xi^\mu{}_{\nu}Dr^\nu+r^\mu={}&
				-\lambda_r b^\alpha\Xi^{\mu\beta}\Big(\partial_{[\alpha}\beta_{\beta]}+2\beta\omega_{\alpha\beta}\Big)
				-\lambda_r\widehat{\mathcal B}_r^\mu,
				\\
				\tau_\ell\Xi^\mu{}_{\nu}D\ell^\nu+\ell^\mu={}&
				-\frac{2\rho_\parallel}{\beta_0}b^{[\alpha}\Xi^{\beta]\mu}\partial_\alpha\mathcal H_\beta
				-\frac{2\rho_\parallel}{\beta_0}\widehat{\mathcal B}_\ell^\mu,
				\\
				\tau_q\Xi^\mu{}_{\nu}Dq^\nu+q^\mu={}&
				-\lambda_q u^\alpha\Xi^{\mu\beta}\Big(\partial_{[\alpha}\beta_{\beta]}+2\beta\omega_{\alpha\beta}\Big)
				-\lambda_q\widehat{\mathcal B}_q^\mu,
				\\
				\tau_\pi\Xi^{\mu\nu}{}_{\alpha\beta}D\pi_\perp^{\alpha\beta}+\pi_\perp^{\mu\nu}={}&
				2\eta_\perp\Xi^{\mu\nu\alpha\beta}\partial_\alpha u_\beta
				+2\eta_\perp\widehat{\mathcal C}_\pi^{\mu\nu},
				\\
				\tau_m\Xi^{[\mu}{}_{\alpha}\Xi^{\nu]}{}_{\beta}Dm^{\alpha\beta}+m^{\mu\nu}={}&
				\frac{2\rho_\perp}{\beta_0}\Xi^{\mu[\alpha}\Xi^{\beta]\nu}\partial_\alpha\mathcal H_\beta
				+\frac{2\rho_\perp}{\beta_0}\widehat{\mathcal C}_m^{\mu\nu},
				\\
				\tau_\phi\Xi^{[\mu}{}_{\alpha}\Xi^{\nu]}{}_{\beta}D\phi^{\alpha\beta}+\phi^{\mu\nu}={}&
				\lambda_\phi\Xi^\mu{}_{\rho}\Xi^\nu{}_{\sigma}
				\Big(\partial^{[\rho}\beta^{\sigma]}+2\beta\omega^{\rho\sigma}\Big)
				+\lambda_\phi\widehat{\mathcal C}_\phi^{\mu\nu} .
			\end{aligned}
			\label{eq:second-ordinary-constitutive}
		\end{equation}
		
		For the split spin-flux channels, we separate the comoving derivatives as
		
		\begin{equation}
			\begin{gathered}
				\mathcal A_{\Lambda\parallel}=-\frac{\tau_{\Lambda\parallel}}{\chi_{\Lambda\parallel}}D\Lambda_\parallel+\widehat{\mathcal A}_{\Lambda\parallel},
				\qquad
				\mathcal B_{\Lambda\perp}^\mu=\frac{\tau_{\Lambda\perp}}{\chi_{\Lambda\perp}}\Xi^\mu{}_{\nu}D\Lambda_\perp^\nu+\widehat{\mathcal B}_{\Lambda\perp}^\mu,
				\\
				\mathcal B_\Phi^\mu=\frac{\tau_\Phi}{\chi_\Phi}\Xi^\mu{}_{\nu}D\Phi^\nu+\widehat{\mathcal B}_\Phi^\mu,
				\qquad
				\mathcal B_W^\mu=\frac{\tau_W}{\chi_W}\Xi^\mu{}_{\nu}DW^\nu+\widehat{\mathcal B}_W^\mu,
				\\
				\mathcal A_{Mt}=-\frac{\tau_{Mt}}{\chi_{Mt}}DM_{(t)}+\widehat{\mathcal A}_{Mt},
				\qquad
				\mathcal A_{Nt}=-\frac{\tau_{Nt}}{\chi_{Nt}}DN_{(t)}+\widehat{\mathcal A}_{Nt},
				\\
				\mathcal D_{Ms}^{\mu\nu}=-\frac{\tau_{Ms}}{\chi_{Ms}}\Xi^{\mu\nu}{}_{\alpha\beta}DM_{(s)}^{\alpha\beta}+\widehat{\mathcal D}_{Ms}^{\mu\nu},
				\qquad
				\mathcal D_{Ns}^{\mu\nu}=-\frac{\tau_{Ns}}{\chi_{Ns}}\Xi^{\mu\nu}{}_{\alpha\beta}DN_{(s)}^{\alpha\beta}+\widehat{\mathcal D}_{Ns}^{\mu\nu},
				\\
				\mathcal E_{Ma}^{\mu\nu}=-\frac{\tau_{Ma}}{\chi_{Ma}}\Xi^{[\mu}{}_{\alpha}\Xi^{\nu]}{}_{\beta}DM_{(a)}^{\alpha\beta}+\widehat{\mathcal E}_{Ma}^{\mu\nu},
				\qquad
				\mathcal E_{Na}^{\mu\nu}=-\frac{\tau_{Na}}{\chi_{Na}}\Xi^{[\mu}{}_{\alpha}\Xi^{\nu]}{}_{\beta}DN_{(a)}^{\alpha\beta}+\widehat{\mathcal E}_{Na}^{\mu\nu},
				\\
				\mathcal E_{\Omega\parallel}^{\mu\nu}=-\frac{\tau_{\Omega\parallel}}{\chi_{\Omega\parallel}}\Xi^{[\mu}{}_{\alpha}\Xi^{\nu]}{}_{\beta}D\Omega_\parallel^{\alpha\beta}+\widehat{\mathcal E}_{\Omega\parallel}^{\mu\nu},
				\qquad
				\mathcal G_\Omega^{\lambda\mu\nu}=\frac{\tau_\Omega}{\chi_\Omega}\Xi^\lambda{}_{\rho}\Xi^{[\mu}{}_{\alpha}\Xi^{\nu]}{}_{\beta}D\Omega^{\rho\alpha\beta}+\widehat{\mathcal G}_\Omega^{\lambda\mu\nu}.
			\end{gathered}
			\label{eq:spinflux-source-relaxation-defs}
		\end{equation}
		The corresponding relaxation times are defined as
		\begin{equation}
			\begin{gathered}
				\tau_{\Lambda\parallel}\equiv -\chi_{\Lambda\parallel}a_{\Lambda\parallel},
				\qquad
				\tau_{\Lambda\perp}\equiv \frac12\chi_{\Lambda\perp}a_{\Lambda\perp},
				\qquad
				\tau_\Phi\equiv \frac12\chi_\Phi a_\Phi,
				\qquad
				\tau_W\equiv \frac12\chi_W a_W,
				\\
				\tau_{Mt}\equiv -\chi_{Mt}a_{Mt},
				\qquad
				\tau_{Nt}\equiv -\chi_{Nt}a_{Nt},
				\qquad
				\tau_{Ms}\equiv -\frac12\chi_{Ms}a_{Ms},
				\qquad
				\tau_{Ns}\equiv -\frac12\chi_{Ns}a_{Ns},
				\\
				\tau_{Ma}\equiv -\frac12\chi_{Ma}a_{Ma},
				\qquad
				\tau_{Na}\equiv -\frac12\chi_{Na}a_{Na},
				\qquad
				\tau_{\Omega\parallel}\equiv -\chi_{\Omega\parallel}a_{\Omega\parallel},
				\qquad
				\tau_\Omega\equiv \chi_\Omega a_\Omega .
			\end{gathered}
			\label{eq:spinflux-relaxation-times}
		\end{equation}
		The split spin-flux channels obey
		\begin{equation}
			\begin{aligned}
				\tau_{\Lambda\parallel}D\Lambda_\parallel+\Lambda_\parallel={}&
				\chi_{\Lambda\parallel}\left[-2b^\rho u^{[\mu}b^{\nu]}\partial_\rho(\beta\omega_{\mu\nu})+\widehat{\mathcal A}_{\Lambda\parallel}\right],
				\\
				\tau_{\Lambda\perp}\Xi^\rho{}_{\sigma}D\Lambda_\perp^\sigma+\Lambda_\perp^\rho={}&
				-\chi_{\Lambda\perp}\left[\Xi^\rho{}_{\eta}u^{[\mu}b^{\nu]}\partial^\eta(\beta\omega_{\mu\nu})+\widehat{\mathcal B}_{\Lambda\perp}^{\rho}\right],
				\\
				\tau_\Phi\Xi^\sigma{}_{\rho}D\Phi^\rho+\Phi^\sigma={}&
				-\chi_\Phi\left[b^\rho u^{[\mu}\Xi^{\nu]\sigma}\partial_\rho(\beta\omega_{\mu\nu})+\widehat{\mathcal B}_\Phi^{\sigma}\right],
				\\
				\tau_W\Xi^\sigma{}_{\rho}DW^\rho+W^\sigma={}&
				-\chi_W\left[-b^\rho b^{[\mu}\Xi^{\nu]\sigma}\partial_\rho(\beta\omega_{\mu\nu})+\widehat{\mathcal B}_W^{\sigma}\right],
				\\
				\tau_{Mt}DM_{(t)}+M_{(t)}={}&
				\chi_{Mt}\big[-u^{[\mu}\Xi^{\nu]\rho}\partial_\rho(\beta\omega_{\mu\nu})+\widehat{\mathcal A}_{Mt}\big],
				\\
				\tau_{Nt}DN_{(t)}+N_{(t)}={}&
				\chi_{Nt}\big[b^{[\mu}\Xi^{\nu]\rho}\partial_\rho(\beta\omega_{\mu\nu})+\widehat{\mathcal A}_{Nt}\big],
				\\
				\tau_{Ms}\Xi^{\rho\sigma}{}_{\alpha\beta}DM_{(s)}^{\alpha\beta}+M_{(s)}^{\rho\sigma}={}&
				\chi_{Ms}\left[-\Xi^{\rho\sigma}{}_{\eta\zeta}u^{[\mu}\Xi^{\nu]\zeta}\partial^\eta(\beta\omega_{\mu\nu})+\widehat{\mathcal D}_{Ms}^{\rho\sigma}\right],
				\\
				\tau_{Ns}\Xi^{\rho\sigma}{}_{\alpha\beta}DN_{(s)}^{\alpha\beta}+N_{(s)}^{\rho\sigma}={}&
				\chi_{Ns}\left[\Xi^{\rho\sigma}{}_{\eta\zeta}b^{[\mu}\Xi^{\nu]\zeta}\partial^\eta(\beta\omega_{\mu\nu})+\widehat{\mathcal D}_{Ns}^{\rho\sigma}\right],
				\\
				\tau_{Ma}\Xi^{[\rho}{}_{\alpha}\Xi^{\sigma]}{}_{\beta}DM_{(a)}^{\alpha\beta}+M_{(a)}^{\rho\sigma}={}&
				\chi_{Ma}\left[-\Xi^{[\rho}{}_{\eta}\Xi^{\sigma]}{}_{\zeta}u^{[\mu}\Xi^{\nu]\zeta}\partial^\eta(\beta\omega_{\mu\nu})+\widehat{\mathcal E}_{Ma}^{\rho\sigma}\right],
				\\
				\tau_{Na}\Xi^{[\rho}{}_{\alpha}\Xi^{\sigma]}{}_{\beta}DN_{(a)}^{\alpha\beta}+N_{(a)}^{\rho\sigma}={}&
				\chi_{Na}\left[\Xi^{[\rho}{}_{\eta}\Xi^{\sigma]}{}_{\zeta}b^{[\mu}\Xi^{\nu]\zeta}\partial^\eta(\beta\omega_{\mu\nu})+\widehat{\mathcal E}_{Na}^{\rho\sigma}\right],
				\\
				\tau_{\Omega\parallel}\Xi^{[\mu}{}_{\alpha}\Xi^{\nu]}{}_{\beta}D\Omega_\parallel^{\alpha\beta}+\Omega_\parallel^{\mu\nu}={}&
				\chi_{\Omega\parallel}\left[\Xi^{[\mu}{}_{\alpha}\Xi^{\nu]}{}_{\beta}b^\rho\partial_\rho(\beta\omega^{\alpha\beta})+\widehat{\mathcal E}_{\Omega\parallel}^{\mu\nu}\right],
				\\
				\tau_\Omega\Xi^\lambda{}_{\rho}\Xi^{[\mu}{}_{\alpha}\Xi^{\nu]}{}_{\beta}D\Omega^{\rho\alpha\beta}+\Omega^{\lambda\mu\nu}={}&
				\chi_\Omega\Xi^\lambda{}_{\rho}\Xi^{[\mu}{}_{\alpha}\Xi^{\nu]}{}_{\beta}\partial^\rho(\beta\omega^{\alpha\beta})-\chi_\Omega\widehat{\mathcal G}_\Omega^{\lambda\mu\nu} .
			\end{aligned}
			\label{eq:second-spinflux-constitutive}
		\end{equation}
	\end{widetext}
	The hatted sources in Eqs.~\eqref{eq:second-ordinary-constitutive} and
	\eqref{eq:second-spinflux-constitutive} collect coefficient-gradient,
	projector-derivative, and nonlinear terms.  With $b_i=c_i=0$, they vanish
	in the homogeneous linearization.
	
	Within this minimal transient closure, nonnegative entropy production imposes the first-order transport constraints in
	Eqs.~\eqref{eq:scalar-positive-main}--\eqref{eq:at-positive-main} and
	\begin{equation}
		\begin{gathered}
			\chi_{\Lambda\parallel},\chi_{Mt},\chi_{Nt},\chi_{Ms},\chi_{Ns},\chi_{Ma},\chi_{Na},\chi_{\Omega\parallel}\ge0,
			\\
			\chi_{\Lambda\perp},\chi_\Phi,\chi_W,\chi_\Omega\ge0 .
		\end{gathered}
		\label{eq:second-positive-chi}
	\end{equation}
	The finite-relaxation model also adopts the entropy-density signs in Eq.~\eqref{eq:a-signs}
	and strictly positive relaxation times.  Source definitions containing inverse transport
	coefficients apply to retained channels with strictly positive coefficients.
	The two lines group the positive transport amplitudes according to the sign of the corresponding rest-frame quadratic norm; the explicit signs in the relaxation equations have been chosen consistently with those norms.
	
	\section{Linearized equations and block-matrix formulation}
	\label{sec:linearized-blocks}
	
	Residual axial symmetry separates linear perturbations into magnetosonic and
	Alfv\'en-like sectors.  Dissipation adds gapped spin-relaxation channels and,
	at second order, independent spin-flux poles.  Their coupling to the fluid and
	magnetic variables depends on the propagation angle.
	
	\subsection{Spin equation of state and susceptibility closure}
	
	For the linear analysis we specialize to an unpolarized homogeneous equilibrium,
	\begin{equation}
		\bar S^{\mu\nu}=0,
		\qquad
		\bar\omega_{\mu\nu}=0,
	\end{equation}
	and eliminate the spin chemical-potential perturbations through the anisotropic linear
	spin equation of state,
	\begin{equation}
		\delta S^{0z}=\chi_b^{\parallel}\delta\omega_{0z},
		\qquad
		\delta S^{0i}=\chi_b^{\perp}\delta\omega_{0i},
	\end{equation}
	\begin{equation}
		\delta S^{zi}=\chi_s^{\parallel}\delta\omega_{zi},
		\qquad
		\delta S^{xy}=\chi_s^{\perp}\delta\omega_{xy},
	\end{equation}
	where $i=x,y$ runs over the directions transverse to the background magnetic field.  Equivalently,
	\begin{equation}
		\delta(\beta\omega_{0z})=
		\Gamma_{b\parallel}\delta S^{0z},
		\qquad
		\delta(\beta\omega_{0i})=
		\Gamma_{b\perp}\delta S^{0i},
	\end{equation}
	\begin{equation}
		\delta(\beta\omega_{zi})=
		\Gamma_{s\parallel}\delta S^{zi},
		\qquad
		\delta(\beta\omega_{xy})=
		\Gamma_{s\perp}\delta S^{xy},
	\end{equation}
	with
	\begin{equation}
		\begin{aligned}
			\Gamma_{b\parallel}&\equiv \frac{\beta_0}{\chi_b^\parallel},
			&\qquad
			\Gamma_{b\perp}&\equiv \frac{\beta_0}{\chi_b^\perp},
			\\
			\Gamma_{s\parallel}&\equiv \frac{\beta_0}{\chi_s^\parallel},
			&\qquad
			\Gamma_{s\perp}&\equiv \frac{\beta_0}{\chi_s^\perp}.
		\end{aligned}
		\label{eq:matrix-shorthands}
	\end{equation}
	We therefore take the spin-density perturbations
	\begin{equation}
		\delta S^{0z},\quad \delta S^{0i},\quad
		\delta S^{zi},\quad \delta S^{xy}
	\end{equation}
	as independent variables.
	
	\subsection{Kinematics and component conventions}
	
	We work in the local rest frame with the background magnetic field along the $z$
	direction.  The amplitudes $u_a,B_a,f_a,\ell_a,q_a,r_a$ denote the covariant
	spatial components, e.g., $\delta u_a=-\delta u^a$ and $\delta B_a=-\delta B^a$.
	For the split spin flux, $\Lambda_{\perp a},\Phi_a,W_a,\Omega_a$ are component
	labels for the contravariant amplitudes
	$\Lambda_\perp^a,\Phi^a,W^a,\Omega^a$.
	Transverse component sums and componentwise symmetric-traceless projections use $\delta_{ij}$, while $\delta S^{0i}$ and $\delta S^{zi}$ retain their contravariant spin-density indices.  Thus $\delta_{jl}\delta S^{0l}$ denotes a Euclidean component contraction, not Lorentz lowering of the index $l$.  We use
	\begin{equation}
		k_\perp^i=(k_x,k_y),
		\qquad
		k_\perp^i\delta u_i=k_x\delta u_x+k_y\delta u_y,
		\qquad
		\epsilon_\perp^{xy}=1 .
	\end{equation}
		Plane waves are written as
	\begin{equation}
		\delta\mathcal O(t,\mathbf x)=
		\delta\mathcal O(\omega,\mathbf k)
		\exp(-\mathrm i\omega t+\mathrm i\mathbf k\cdot\mathbf x),
	\end{equation}
	so that
	\begin{equation}
		\partial_t\rightarrow -\mathrm i\omega,
		\qquad
		\partial_i\rightarrow \mathrm i k_{\perp i},
		\qquad
			\partial_z\rightarrow \mathrm i k_z .
		\end{equation}
		The homogeneous Maxwell equation contains both induction equations and a constraint.
		Its projection along the equilibrium flow velocity is
		\begin{equation}
			u_\nu\partial_\mu\widetilde F^{\mu\nu}=0
			\quad\Longrightarrow\quad
			\mathrm i k_\perp^i\delta B_i+\mathrm i k_z\delta B_z=0 .
			\label{eq:magnetic-gauss-linear}
		\end{equation}
		Equation~\eqref{eq:magnetic-gauss-linear} constrains the initial data to the
		divergence-free subspace and is preserved by the induction equations.  The projections
		along $b^\mu$ and $\Xi^{\mu\nu}$ supply the magnetic evolution equations.  We impose it before
		forming every characteristic determinant below.
		For the mode analysis we close the matter sector by a barotropic linear response and do not introduce an additional conserved charge.  We therefore use
	\begin{equation}
		\begin{aligned}
			\delta p&=c_s^2\delta\varepsilon,
			&\qquad
			\delta\beta&=\beta_\varepsilon\delta\varepsilon,
			\\
			h&\equiv \epsilon+p_\perp,
			&\qquad
			\eta_f&\equiv \beta_0\eta_\parallel .
		\end{aligned}
	\end{equation}
	For the constant-permeability equation of state used here,
	\begin{equation}
		\begin{aligned}
			\epsilon&=\varepsilon+\frac{B^2}{2\mu_m},\\
			p_\perp&=p+\frac{B^2}{2\mu_m},
			\qquad p_\parallel=p-\frac{B^2}{2\mu_m}.
		\end{aligned}
		\label{eq:matter-magnetic-split}
	\end{equation}
	The transverse and longitudinal momentum densities therefore involve different
	equilibrium inertias,
	\begin{equation}
		\begin{aligned}
			h&=\epsilon+p_\perp=\varepsilon+p+\frac{B^2}{\mu_m},\\
			h_\parallel&\equiv\epsilon+p_\parallel
			=h-\frac{B^2}{\mu_m}=h(1-v_A^2).
		\end{aligned}
		\label{eq:directional-enthalpies}
	\end{equation}
	Here $v_A^2=B^2/(\mu_m h)$.  Linearizing Eq.~\eqref{eq:T0}, with
	$\delta b^0=\delta u^z$, gives $\delta T_{(0)}^{0z}=h_\parallel\delta u^z$
	and $\delta T_{(0)}^{0i}=h\delta u^i$ for $i=x,y$.
	The magnetic anisotropic stress thus cancels the magnetic contribution to the
	longitudinal inertia.  In the unpolarized background, the spin term in the Euler
	relation has no linear variation and does not alter these equilibrium coefficients.
	This directional inertia also appears in first-order RMHD and spin-MHD
	\cite{FangHattoriHu2024,FangHattoriHu2025}.
	In the zero-charge background, $h_\parallel=T_0s$.
	The Gibbs--Duhem relation $dp=s\,dT$ then fixes
	\begin{equation}
		\beta_\varepsilon=-\frac{\beta_0c_s^2}{h_\parallel}.
		\label{eq:thermodynamic-beta-response}
	\end{equation}
	For constant $\mu_m$ and $H^\mu=-B^\mu/\mu_m$, the magnetic response is
	\begin{equation}
		\begin{aligned}
			\delta\mathcal H_i&=-\frac{\beta_0}{\mu_m}\delta B_i,\\
			\delta\mathcal H_z&=-\frac{\beta_0}{\mu_m}\delta B_z
			+\frac{B}{\mu_m}\beta_\varepsilon\delta\varepsilon .
		\end{aligned}
		\label{eq:linear-magnetic-response}
	\end{equation}
	The temperature term follows from $\delta(\beta H_\mu)=
	\beta_0\delta H_\mu+H_{0\mu}\delta\beta$; it is not an independent susceptibility.
	With $\rho_a=\beta_0\bar\rho_a$, the resistive rows therefore involve
	$\rho_a/\mu_m$ and $\rho_a B\beta_\varepsilon/(\beta_0\mu_m)$.
	Here $\delta\varepsilon$ and $\delta p$ denote matter-sector perturbations, whereas $\delta\epsilon$, $\delta p_\perp$, and $\delta p_\parallel$ denote the corresponding total quantities.  With our index conventions, they are related by
	\begin{equation}
		\begin{aligned}
			\delta\epsilon&=
			\delta\varepsilon-\frac{B}{\mu_m}\delta B_z,
			\\
			\delta p_\perp&=
			\delta p-\frac{B}{\mu_m}\delta B_z,
			\\
			\delta p_\parallel&=
			\delta p+\frac{B}{\mu_m}\delta B_z .
		\end{aligned}
	\end{equation}
	Equivalently, the energy equation can be written in terms of $\delta\epsilon$;
	then its time derivative is simply $-\mathrm i\omega\delta\epsilon$.
	Using $\delta\varepsilon$ as the independent amplitude introduces the magnetic
	time-derivative term below and requires the same change of variables in the
	pressure and temperature responses.
	The transverse symmetric-traceless projection is
	\begin{equation}
		k_{\perp\langle i}\delta u_{j\rangle}
		=\frac12\left(k_{\perp i}\delta u_j+k_{\perp j}\delta u_i
		-\delta_{ij}k_\perp^l\delta u_l\right).
	\end{equation}
	For the split spin-flux variables we use the component definitions
	\begin{equation}
		\begin{aligned}
			M^{ij}&=-\frac12\delta^{ij}M_{(t)}+M_{(s)}^{ij}+M_{(a)}^{ij},
			\\
			N^{ij}&=-\frac12\delta^{ij}N_{(t)}+N_{(s)}^{ij}+N_{(a)}^{ij}.
		\end{aligned}
	\end{equation}
	Here
	\begin{equation}
		\begin{aligned}
			M_{(a)}^{ij}&=\epsilon_\perp^{ij}M_{(a)},
			&\qquad
			N_{(a)}^{ij}&=\epsilon_\perp^{ij}N_{(a)},
			\\
			\Omega_\parallel^{ij}&=\epsilon_\perp^{ij}\Omega_\parallel,
			&\qquad
			\Omega^{kij}&=\epsilon_\perp^{ij}\Omega^k .
		\end{aligned}
	\end{equation}
	These definitions realize Eq.~\eqref{eq:spinflux} componentwise:
	$M_{(t,s,a)}$ and $N_{(t,s,a)}$ are the trace, symmetric-traceless, and
	antisymmetric parts of the nonequilibrium spin flux.
	
	\subsection{Zeroth-order linearization}
	
	At ideal order all dissipative currents and the nonequilibrium spin flux are set to zero,
	\begin{equation}
		\begin{aligned}
			\Pi_\perp=\Pi_\parallel=f_i=\ell_i=\pi_\perp^{ij}=m=0,
			\\
			K=q_i=r_i=\phi=0,
			\qquad
			\phi^{\langle\lambda\rangle\mu\nu}=0 .
		\end{aligned}
	\end{equation}
	The ideal energy equation is
	\begin{equation}
		0=-\mathrm i\omega\delta\varepsilon
		+\mathrm i\omega\frac{B}{\mu_m}\delta B_z
		-\mathrm ih(k_\perp^i\delta u_i+k_z\delta u_z)
		+\mathrm i\frac{B^2}{\mu_m}k_z\delta u_z .
	\end{equation}
	The longitudinal momentum equation is
	\begin{equation}
		0=-\mathrm i\omega h_\parallel\delta u_z-\mathrm i c_s^2 k_z\delta\varepsilon .
	\end{equation}
	The transverse momentum equation is
	\begin{equation}
		0=\mathrm i\omega h\delta u_i
		+\mathrm i k_{\perp i}\left(c_s^2\delta\varepsilon-\frac{B}{\mu_m}\delta B_z\right)
		+\mathrm i\frac{B}{\mu_m}k_z\delta B_i .
	\end{equation}
		The two displayed induction components are
		\begin{equation}
			0=\mathrm i\omega\delta B_z-\mathrm iB k_\perp^i\delta u_i,
			\qquad
			0=-\mathrm i\omega\delta B_i-\mathrm iB k_z\delta u_i .
		\end{equation}
		They are supplemented by the non-evolutionary constraint
		Eq.~\eqref{eq:magnetic-gauss-linear}; only one linear combination of the $x$ and $z$
		induction equations is independent in the magnetosonic plane.
	The spin densities are passive at ideal order,
	\begin{equation}
		\begin{aligned}
			0&=-\mathrm i\omega\delta S^{0z},
			&\qquad
			0&=-\mathrm i\omega\delta S^{0i},
			\\
			0&=-\mathrm i\omega\delta S^{zi},
			&\qquad
			0&=-\mathrm i\omega\delta S^{xy}.
		\end{aligned}
	\end{equation}
	
	\subsection{First-order linearization}
	
	At Navier--Stokes order the retained dissipative currents are algebraic auxiliary
	variables.  The entropy analysis does not fix algebraic spin-flux responses at
	this order, so the first-order spectral truncation sets these amplitudes to zero:
	\begin{equation}
		\begin{aligned}
			0&=\Lambda_\parallel=\Lambda_{\perp i}
			=\Phi_i=W_i=M_{(t)}=N_{(t)}
			\\
			&=M_{(s)}^{ij}=N_{(s)}^{ij}
			=M_{(a)}=N_{(a)}=\Omega_\parallel=\Omega_i .
		\end{aligned}
	\end{equation}
	The conservation equations are
	\begin{equation}
		\begin{aligned}
			0={}&-\mathrm i\omega\delta\varepsilon
			+\mathrm i\omega\frac{B}{\mu_m}\delta B_z
			\\
			&-\mathrm ih(k_\perp^i\delta u_i+k_z\delta u_z)
			+\mathrm i\frac{B^2}{\mu_m}k_z\delta u_z
			\\
			&+\mathrm i k_\perp^i\delta q_i
			-\frac{\mathrm i}{2}k_z\delta K,
		\end{aligned}
	\end{equation}
	\begin{equation}
		\begin{aligned}
			0={}&-\mathrm i\omega h_\parallel\delta u_z
			-\mathrm i k_z(c_s^2\delta\varepsilon+\delta\Pi_\parallel)
			\\
			&+\mathrm i k_\perp^i\delta f_i
			+\frac{\mathrm i\omega}{2}\delta K
			-\mathrm i k_\perp^i\delta r_i,
		\end{aligned}
	\end{equation}
	The longitudinal time derivative is
	$-\mathrm i\omega(h_\parallel\delta u_z-\delta K/2)$.
	The antisymmetric spin stress contributes the independent $K$ amplitude;
	it does not replace $h_\parallel$ by $h$.
	\begin{equation}
		\begin{aligned}
			0={}&\mathrm i\omega h\delta u_i
			+\mathrm i k_{\perp i}\left(c_s^2\delta\varepsilon
			-\frac{B}{\mu_m}\delta B_z+\delta\Pi_\perp\right)
			\\
			&+\mathrm i\frac{B}{\mu_m}k_z\delta B_i
			-\mathrm i k_z\delta f_i
			+\mathrm i k_{\perp j}\delta\pi_\perp^{ji}
			\\
			&+\mathrm i\omega\delta q_i
			-\mathrm i k_z\delta r_i
			-\mathrm i\epsilon_{\perp ij}k_\perp^j\delta\phi .
		\end{aligned}
	\end{equation}
	\begin{equation}
		\begin{aligned}
			0={}&\mathrm i\omega\delta B_z
			-\mathrm iB k_\perp^i\delta u_i
			-\mathrm i k_\perp^i\delta\ell_i .
		\end{aligned}
	\end{equation}
		\begin{equation}
			\begin{aligned}
				0={}&-\mathrm i\omega\delta B_i
			-\mathrm iB k_z\delta u_i
			\\
			&-\mathrm i k_z\delta\ell_i
			-\mathrm i\epsilon_{\perp ij}k_\perp^j\delta m .
			\end{aligned}
		\end{equation}
		At dissipative order the flow projection of the homogeneous Maxwell equation remains
		Eq.~\eqref{eq:magnetic-gauss-linear}.  The $\ell_i$ terms modify the induction dynamics
		but not the magnetic Gauss constraint.
		The first-order spin-balance equations are obtained by dropping the spin-flux divergence,
	\begin{equation}
		\begin{aligned}
			0&=-\mathrm i\omega\delta S^{0z}+\delta K,\\
			0&=-\mathrm i\omega\delta S^{0i}-2\delta q_i,
		\end{aligned}
	\end{equation}
	\begin{equation}
		\begin{aligned}
			0&=-\mathrm i\omega\delta S^{zi}-2\delta r_i,\\
			0&=-\mathrm i\omega\delta S^{xy}+2\delta\phi .
		\end{aligned}
	\end{equation}
	The algebraic Navier--Stokes rows are
	\begin{equation}
		\begin{aligned}
			0={}&\delta\Pi_\perp-\mathrm i\zeta_\perp k_\perp^i\delta u_i
			-\mathrm i\zeta_\times k_z\delta u_z,\\
			0={}&\delta\Pi_\parallel-\mathrm i\zeta_\parallel k_z\delta u_z
			-\mathrm i\zeta_\times' k_\perp^i\delta u_i .
		\end{aligned}
	\end{equation}
	\begin{equation}
		\begin{aligned}
			0={}&\delta K-\lambda_K\Big[-\frac{\mathrm i}{2}
			(\omega\beta_0\delta u_z+k_z\beta_\varepsilon\delta\varepsilon)\\
			&\hspace{5.2em}+2\Gamma_{b\parallel}\delta S^{0z}\Big] .
		\end{aligned}
	\end{equation}
	\begin{equation}
		\begin{aligned}
			0={}&\delta f_i+\mathrm i\eta_f(k_{\perp i}\delta u_z+k_z\delta u_i)\\
			&+\kappa_{fr}\Big[\frac{\mathrm i\beta_0}{2}(k_z\delta u_i-k_{\perp i}\delta u_z)
			+2\Gamma_{s\parallel}\delta S^{zi}\Big] .
		\end{aligned}
	\end{equation}
	\begin{equation}
		\begin{aligned}
			0={}&\delta r_i+\frac{\mathrm i\kappa_{rf}}{2T_0}(k_{\perp i}\delta u_z+k_z\delta u_i)\\
			&+\lambda_r\Big[\frac{\mathrm i\beta_0}{2}(k_z\delta u_i-k_{\perp i}\delta u_z)
			+2\Gamma_{s\parallel}\delta S^{zi}\Big] .
		\end{aligned}
	\end{equation}
	\begin{equation}
		\begin{aligned}
			0={}&\delta\ell_i-\frac{\mathrm i\rho_\parallel}{\mu_m}
			\left(k_z\delta B_i-k_{\perp i}\delta B_z
			+\frac{B}{\beta_0}k_{\perp i}\beta_\varepsilon\delta\varepsilon\right)\\
			&+\frac{\kappa_{\ell q}}{\sqrt{\beta_0}}\Big[-\frac{\mathrm i}{2}(\omega\beta_0\delta u_i
			+k_{\perp i}\beta_\varepsilon\delta\varepsilon)+2\Gamma_{b\perp}\delta S^{0i}\Big] .
		\end{aligned}
	\end{equation}
	\begin{equation}
		\begin{aligned}
			0={}&\delta q_i-\frac{\mathrm i\sqrt{\beta_0}\kappa_{q\ell}}{2\mu_m}
			\left(k_z\delta B_i-k_{\perp i}\delta B_z
			+\frac{B}{\beta_0}k_{\perp i}\beta_\varepsilon\delta\varepsilon\right)\\
			&+\lambda_q\Big[-\frac{\mathrm i}{2}(\omega\beta_0\delta u_i
			+k_{\perp i}\beta_\varepsilon\delta\varepsilon)+2\Gamma_{b\perp}\delta S^{0i}\Big] .
		\end{aligned}
	\end{equation}
	\begin{equation}
		0=\delta\pi_{\perp ij}-2\mathrm i\eta_\perp k_{\perp\langle i}\delta u_{j\rangle},
	\end{equation}
	\begin{equation}
		0=\delta m+\frac{\mathrm i\rho_\perp}{\mu_m}\epsilon_\perp^{ij}k_{\perp i}\delta B_j .
	\end{equation}
	\begin{equation}
		\begin{aligned}
			0={}&\delta\phi-\lambda_\phi\Big[\frac{\mathrm i\beta_0}{2}\epsilon_\perp^{ij}k_{\perp i}\delta u_j
			+2\Gamma_{s\perp}\delta S^{xy}\Big] .
		\end{aligned}
	\end{equation}
	
	\subsection{Second-order minimal linearization}
	
	The minimal second-order system keeps the conservation equations above and restores the
	spin-flux divergence in the spin-balance equations.  Each algebraic
	constitutive row becomes a relaxation row.  The linearized spin-balance equations are
	\begin{equation}
		\begin{aligned}
			0={}&-\mathrm i\omega\delta S^{0z}+\mathrm i k_z\delta\Lambda_\parallel\\
			&-\mathrm i k_\perp^i\delta\Lambda_{\perp i}+\delta K,
		\end{aligned}
	\end{equation}
	\begin{equation}
		\begin{aligned}
			0={}&-\mathrm i\omega\delta S^{0i}-\mathrm i k_z\delta\Phi_i\\
			&+\mathrm i k_{\perp j}\delta M^{ji}-2\delta q_i,
		\end{aligned}
	\end{equation}
	\begin{equation}
		\begin{aligned}
			0={}&-\mathrm i\omega\delta S^{zi}+\mathrm i k_z\delta W_i\\
			&-\mathrm i k_{\perp j}\delta N^{ji}-2\delta r_i,
		\end{aligned}
	\end{equation}
	\begin{equation}
		\begin{aligned}
			0={}&-\mathrm i\omega\delta S^{xy}-\mathrm i k_z\delta\Omega_\parallel\\
			&+\mathrm i k_\perp^i\delta\Omega_i+2\delta\phi .
		\end{aligned}
	\end{equation}
	Within the minimal truncation, $\kappa=b_i=c_i=0$ and the bulk sector obeys
	Eq.~\eqref{eq:bulk-entropy-matrix}.  The hatted sources vanish in the homogeneous
	linearization.  The ordinary constitutive rows consequently acquire the factor
	$1-\mathrm i\omega\tau_X$ for each current:
	\begin{equation}
		\begin{aligned}
			0={}&(1-\mathrm i\omega\tau_{\Pi_\perp})\delta\Pi_\perp
			-\mathrm i\zeta_\perp k_\perp^i\delta u_i\\
			&-\mathrm i\zeta_\times k_z\delta u_z,
		\end{aligned}
	\end{equation}
	\begin{equation}
		\begin{aligned}
			0={}&(1-\mathrm i\omega\tau_{\Pi_\parallel})\delta\Pi_\parallel
			-\mathrm i\zeta_\parallel k_z\delta u_z\\
			&-\mathrm i\zeta_\times k_\perp^i\delta u_i,
		\end{aligned}
	\end{equation}
	\begin{equation}
		\begin{aligned}
			0={}&(1-\mathrm i\omega\tau_K)\delta K
			-\lambda_K\Big[-\frac{\mathrm i}{2}(\omega\beta_0\delta u_z\\
			&\hspace{7.5em}+k_z\beta_\varepsilon\delta\varepsilon)
			+2\Gamma_{b\parallel}\delta S^{0z}\Big] .
		\end{aligned}
	\end{equation}
	\begin{equation}
		\begin{aligned}
			0={}&(1-\mathrm i\omega\tau_f)\delta f_i
			+\mathrm i\eta_f(k_{\perp i}\delta u_z+k_z\delta u_i) .
		\end{aligned}
	\end{equation}
	\begin{equation}
		\begin{aligned}
			0={}&(1-\mathrm i\omega\tau_r)\delta r_i
			+\lambda_r\Big[\frac{\mathrm i\beta_0}{2}(k_z\delta u_i-k_{\perp i}\delta u_z)\\
			&+2\Gamma_{s\parallel}\delta S^{zi}\Big] .
		\end{aligned}
	\end{equation}
	\begin{equation}
		\begin{aligned}
			0={}&(1-\mathrm i\omega\tau_\ell)\delta\ell_i
			-\frac{\mathrm i\rho_\parallel}{\mu_m}
			\Big[k_z\delta B_i-k_{\perp i}\delta B_z\\
			&+\frac{B}{\beta_0}k_{\perp i}\beta_\varepsilon\delta\varepsilon\Big] .
		\end{aligned}
	\end{equation}
	\begin{equation}
		\begin{aligned}
			0={}&(1-\mathrm i\omega\tau_q)\delta q_i
			+\lambda_q\Big[-\frac{\mathrm i}{2}(\omega\beta_0\delta u_i
			+k_{\perp i}\beta_\varepsilon\delta\varepsilon)\\
			&\hspace{6.1em}+2\Gamma_{b\perp}\delta S^{0i}\Big] .
		\end{aligned}
	\end{equation}
	\begin{equation}
		\begin{aligned}
			0={}&(1-\mathrm i\omega\tau_\pi)\delta\pi_{\perp ij}\\
			&-2\mathrm i\eta_\perp k_{\perp\langle i}\delta u_{j\rangle},
		\end{aligned}
	\end{equation}
	\begin{equation}
		\begin{aligned}
			0={}&(1-\mathrm i\omega\tau_m)\delta m
			+\frac{\mathrm i\rho_\perp}{\mu_m}\epsilon_\perp^{ij}k_{\perp i}\delta B_j .
		\end{aligned}
	\end{equation}
	\begin{equation}
		\begin{aligned}
			0={}&(1-\mathrm i\omega\tau_\phi)\delta\phi
			-\lambda_\phi\Big[\frac{\mathrm i\beta_0}{2}\epsilon_\perp^{ij}k_{\perp i}\delta u_j
			+2\Gamma_{s\perp}\delta S^{xy}\Big] .
		\end{aligned}
	\end{equation}
	The split spin-flux relaxation equations are diagonal in the minimal approximation:
	\begin{equation}
		\begin{aligned}
			0={}&(1-\mathrm i\omega\tau_{\Lambda\parallel})\delta\Lambda_\parallel\\
			&+2\mathrm i\chi_{\Lambda\parallel}k_z\Gamma_{b\parallel}\delta S^{0z},
		\end{aligned}
	\end{equation}
	\begin{equation}
		\begin{aligned}
			0={}&(1-\mathrm i\omega\tau_{\Lambda\perp})\delta\Lambda_{\perp i}\\
			&-\mathrm i\chi_{\Lambda\perp}k_{\perp i}\Gamma_{b\parallel}\delta S^{0z},
		\end{aligned}
	\end{equation}
	\begin{equation}
		\begin{aligned}
			0={}&(1-\mathrm i\omega\tau_{\Phi})\delta\Phi_i
			-\mathrm i\chi_{\Phi}k_z\Gamma_{b\perp}\delta S^{0i},\\
			0={}&(1-\mathrm i\omega\tau_{W})\delta W_i
			+\mathrm i\chi_{W}k_z\Gamma_{s\parallel}\delta S^{zi}.
		\end{aligned}
	\end{equation}
	\begin{equation}
		\begin{aligned}
			0={}&(1-\mathrm i\omega\tau_{Mt})\delta M_{(t)}
			-\mathrm i\chi_{Mt}k_\perp^i\Gamma_{b\perp}\delta S^{0i},\\
			0={}&(1-\mathrm i\omega\tau_{Nt})\delta N_{(t)}
			+\mathrm i\chi_{Nt}k_\perp^i\Gamma_{s\parallel}\delta S^{zi}.
		\end{aligned}
	\end{equation}
	\begin{equation}
		\begin{aligned}
			0={}&(1-\mathrm i\omega\tau_{Ms})\delta M_{(s)ij}\\
			&+\mathrm i\chi_{Ms}k_{\perp\langle i}\delta_{j\rangle l}\Gamma_{b\perp}\delta S^{0l},
		\end{aligned}
	\end{equation}
	\begin{equation}
		\begin{aligned}
			0={}&(1-\mathrm i\omega\tau_{Ns})\delta N_{(s)ij}\\
			&-\mathrm i\chi_{Ns}k_{\perp\langle i}\delta_{j\rangle l}\Gamma_{s\parallel}\delta S^{zl},
		\end{aligned}
	\end{equation}
	With the scalar amplitudes defined above, the antisymmetric projectors in
	Eq.~\eqref{eq:second-spinflux-constitutive} supply a factor $1/2$ in the next two rows.
	The coefficients $\chi_{Ma}$ and $\chi_{Na}$ retain the covariant normalization used in
	Table~\ref{tab:numerical-parameters}.
	\begin{equation}
		\begin{aligned}
			0={}&(1-\mathrm i\omega\tau_{Ma})\delta M_{(a)}\\
			&+\frac{\mathrm i\chi_{Ma}}{2}\epsilon_\perp^{ij}k_{\perp i}\Gamma_{b\perp}\delta S^{0j},
		\end{aligned}
		\label{eq:Ma-component-relaxation}
	\end{equation}
	\begin{equation}
		\begin{aligned}
			0={}&(1-\mathrm i\omega\tau_{Na})\delta N_{(a)}\\
			&-\frac{\mathrm i\chi_{Na}}{2}\epsilon_\perp^{ij}k_{\perp i}\Gamma_{s\parallel}\delta S^{zj},
		\end{aligned}
		\label{eq:Na-component-relaxation}
	\end{equation}
	\begin{equation}
		\begin{aligned}
			0={}&(1-\mathrm i\omega\tau_{\Omega\parallel})\delta\Omega_\parallel
			-\mathrm i\chi_{\Omega\parallel}k_z\Gamma_{s\perp}\delta S^{xy},\\
			0={}&(1-\mathrm i\omega\tau_{\Omega})\delta\Omega_i
			+\mathrm i\chi_{\Omega}k_{\perp i}\Gamma_{s\perp}\delta S^{xy}.
		\end{aligned}
	\end{equation}
	
	\subsection{Principal block form}
	\label{subsec:principal-block-matrix}
	
	For the block-matrix representation, we use the residual axial symmetry to choose the wave vector in the $x$--$z$ plane,
	\begin{equation}
		\begin{aligned}
			k_y&=0,\\
			k_x&\equiv k_\perp=k\sin\theta,\\
			k_z&\equiv k_\parallel=k\cos\theta .
		\end{aligned}
		\label{eq:kxz-plane-spinmhd}
	\end{equation}
	Equation~\eqref{eq:magnetic-gauss-linear} then becomes
	\begin{equation}
		\sin\theta\,\delta B_x+\cos\theta\,\delta B_z=0 .
		\label{eq:magnetic-gauss-angle}
	\end{equation}
	We solve this constraint without dividing by either trigonometric factor by introducing the
	divergence-free in-plane amplitude $\delta B_T$,
	\begin{equation}
		\delta B_x=\cos\theta\,\delta B_T,
		\qquad
		\delta B_z=-\sin\theta\,\delta B_T .
		\label{eq:BT-definition}
	\end{equation}
	This parametrization is regular for both parallel and transverse propagation.  The projected
	induction equation is
	\begin{equation}
		0=-\mathrm i\omega\delta B_T-\mathrm i Bk\delta u_x
		-\mathrm i k\delta\ell_x,
		\label{eq:BT-induction}
	\end{equation}
	where the last term is absent at ideal order.
	At ideal order the variables are
	\begin{equation}
		\begin{aligned}
			\mathbf X_{\rm ms}^{(0)}={}&(\delta\varepsilon,\delta u_x,\delta u_z,
			\delta B_T)^T,\\
			\mathbf X_{\rm A}^{(0)}={}&(\delta u_y,\delta B_y)^T,
		\end{aligned}
	\end{equation}
	\begin{equation}
		\begin{aligned}
			\mathbf X_{\rm sp}^{(0)}={}&(\delta S^{0z},\delta S^{0x},\delta S^{0y},\\
			&\delta S^{zx},\delta S^{zy},\delta S^{xy})^T .
		\end{aligned}
	\end{equation}
	For first order we use
	\begin{equation}
		\begin{aligned}
			\mathbf Y_{\rm ms}^{(1)T}=(&\delta\varepsilon,\delta u_x,\delta u_z,
			\delta B_T,\\
			&\delta\Pi_\parallel,\delta\Pi_\perp,\delta f_x,\delta\pi_{xx},\delta\ell_x,\\
			&\delta K,\delta q_x,\delta r_x,\delta S^{0z},\delta S^{0x},\delta S^{zx}),
		\end{aligned}
	\end{equation}
	\begin{equation}
		\begin{aligned}
			\mathbf Y_{\rm A}^{(1)T}=(&\delta u_y,\delta B_y,\delta f_y,
			\delta\pi_{xy},\delta m,\delta\ell_y,\\
			&\delta q_y,\delta r_y,\delta\phi,\delta S^{0y},\delta S^{zy},\delta S^{xy}).
		\end{aligned}
	\end{equation}
	For the minimal second-order system the spin-flux variables are added.  In the
	magnetosonic-type block,
	\begin{equation}
		\begin{aligned}
			\mathbf Y_{\rm ms}^{(2)T}=(&\mathbf Y_{\rm ms}^{(1)T},
			\delta\Lambda_\parallel,\delta\Lambda_{\perp x},\\
			&\delta\Phi_x,\delta W_x,\delta M_{(t)},\delta N_{(t)},\\
			&\delta M_{(s)xx},\delta N_{(s)xx}),
		\end{aligned}
		\label{eq:Y-ms-principal}
	\end{equation}
	and in the Alfv\'en-type block,
	\begin{equation}
		\begin{aligned}
			\mathbf Y_{\rm A}^{(2)T}=(&\mathbf Y_{\rm A}^{(1)T},
			\delta\Phi_y,\delta W_y,\\
			&\delta M_{(s)xy},\delta N_{(s)xy},\delta M_{(a)},\delta N_{(a)},\\
			&\delta\Omega_\parallel,\delta\Omega_x),
		\end{aligned}
		\label{eq:Y-A-principal}
	\end{equation}
	while $\Lambda_{\perp y}$ and $\Omega_y$ are diagonal spectators for $k_y=0$,
	\begin{equation}
		\begin{aligned}
			\mathbf Y_{\rm sp}^{(2)T}&=(\delta\Lambda_{\perp y},\delta\Omega_y),\\
			\mathbb M_{\rm sp}^{(2)}&=\operatorname{diag}\!\left(1-\mathrm i\omega\tau_{\Lambda\perp},\,1-\mathrm i\omega\tau_\Omega\right).
		\end{aligned}
		\label{eq:Y-sp-spectator}
	\end{equation}
	The full minimal second-order system therefore factorizes,
	\begin{equation}
		\begin{gathered}
			\mathbb M^{(2)}
			\begin{pmatrix}
				\mathbf Y_{\rm ms}^{(2)}\\
				\mathbf Y_{\rm A}^{(2)}\\
				\mathbf Y_{\rm sp}^{(2)}
			\end{pmatrix}=0,\\[2pt]
			\det\mathbb M^{(2)}
			=\det\mathbb M_{\rm ms}^{(2)}\det\mathbb M_{\rm A}^{(2)}\\[-1pt]
			{}\times(1-\mathrm i\omega\tau_{\Lambda\perp})
			(1-\mathrm i\omega\tau_\Omega).
		\end{gathered}
		\label{eq:principal-factorized-spinmhd}
	\end{equation}
	when the wave vector lies in the $x$--$z$ plane and the additional inter-block transport coefficients are omitted in the minimal ansatz.  The explicit nonzero entries are listed in
	Appendix~\ref{app:linear-matrix-elements}.

	\section{Ideal-order block analysis and asymptotic solutions}
	\label{sec:ideal-modular-analysis}
	
	For the unpolarized equilibrium of Sec.~\ref{sec:linearized-blocks}, ideal spin
	densities do not feed back into the fluid or magnetic equations.  The spectrum
	therefore consists of standard RMHD magnetosonic and Alfv\'en waves and six
	passive spin-density modes.  This ideal spin sector follows from the convective
	choice $S_{(0)}^{\lambda\mu\nu}=u^\lambda S^{\mu\nu}$; it is not a general
	property of an unpolarized spin fluid.  The kinetic spin current in
	Ref.~\cite{AmbrusRyblewskiSingh2022} instead supports transverse spin waves around
	an unpolarized background.  Antisymmetric stress and spin transport couple
	the sectors retained here beyond ideal order.

	We introduce the dimensionless Alfv\'en parameter
	\begin{equation}
		v_A^2\equiv \frac{B^2}{\mu_m h},
		\qquad
		h\equiv \epsilon+p_\perp,
	\end{equation}
	and write
	\begin{equation}
		k_\perp=k\sin\theta,
		\qquad
		k_\parallel=k\cos\theta .
	\end{equation}
	
	\begin{table*}[htbp]
	\caption{Dimensionless baseline parameters used in the plotted comparisons, grouped into (a) equilibrium and ordinary transport, (b) spin response and transport, and (c) transient relaxation and spin flux.  The enthalpy density $h$ has energy dimension four; a quantity $Q$ of energy dimension $d$ is listed as $Qh^{-d/4}$.  The two dissipative orders share the coefficients in (a) and (b).  In the last row, $X$ labels only the split spin-flux channels in Eq.~\eqref{eq:spinflux}; $\chi_X$ are transport amplitudes, distinct from the spin-density susceptibilities.  The magnetosonic matrices are reduced with Eq.~\eqref{eq:BT-definition} before diagonalization.}
	\label{tab:numerical-parameters}
	\centering
	\small
	\setlength{\tabcolsep}{6pt}
	\renewcommand{\arraystretch}{1.10}
	\begin{tabular}{p{0.32\textwidth}p{0.12\textwidth}p{0.32\textwidth}p{0.12\textwidth}}
	\toprule
	\multicolumn{4}{l}{(a) Equilibrium and ordinary transport}\\
	\midrule
	Quantity & Value & Quantity & Value\\
	\midrule
	$T_0 h^{-1/4}$ & $1$ & $B h^{-1/2}$ & $0.22$\\
	$\beta_\varepsilon h^{5/4}$ & $-0.3367003367$ & $c_s^2$ & $1/3$\\
	$v_A$ & $0.10$ & $\rho_\parallel h^{1/4}$ & $0.13$\\
	$\rho_\perp h^{1/4}$ & $0.12$ & $\zeta_\parallel h^{-3/4}$ & $0.12$\\
	$\zeta_\perp h^{-3/4}$ & $0.132$ & $\zeta_\times h^{-3/4}$ & $0.10$\\
	$\zeta_\times' h^{-3/4}$ & $0.10$ & $\eta_f h^{-3/4}$ & $0.165$\\
	$\eta_\perp h^{-3/4}$ & $0.33$ & & \\
	\bottomrule
	\end{tabular}
	\par\medskip
	\begin{tabular}{p{0.32\textwidth}p{0.12\textwidth}p{0.32\textwidth}p{0.12\textwidth}}
	\toprule
	\multicolumn{4}{l}{(b) Spin response and transport}\\
	\midrule
	Quantity & Value & Quantity & Value\\
	\midrule
	$\Gamma_{b\parallel}h^{3/4}$ & $0.80$ & $\Gamma_{b\perp}h^{3/4}$ & $0.85$\\
	$\Gamma_{s\parallel}h^{3/4}$ & $0.75$ & $\Gamma_{s\perp}h^{3/4}$ & $0.90$\\
	$\lambda_K h^{-1}$ & $2\times10^{-3}$ & $\lambda_q h^{-1}$ & $10^{-3}$\\
	$\lambda_r h^{-1}$ & $0.021$ & $\lambda_\phi h^{-1}$ & $0.014$\\
	$\kappa_{\ell q}h^{-3/8}$ & $0$ & $\kappa_{q\ell}h^{-3/8}$ & $0$\\
	$\kappa_{fr}h^{-1}$ & $0$ & $\kappa_{rf}h^{-1}$ & $0$\\
	$\kappa_{m\phi}h^{-3/8}$ & $0$ & $\kappa_{\phi m}h^{-3/8}$ & $0$\\
	\bottomrule
	\end{tabular}
	\par\medskip
	\begin{tabular}{p{0.32\textwidth}p{0.12\textwidth}p{0.32\textwidth}p{0.12\textwidth}}
	\toprule
	\multicolumn{4}{l}{(c) Relaxation times and spin flux (second order)}\\
	\midrule
	Quantity & Value & Quantity & Value\\
	\midrule
	$\tau_{\Pi\parallel}h^{1/4}$ & $8$ & $\tau_{\Pi\perp}h^{1/4}$ & $8$\\
	$\tau_K h^{1/4}$ & $30$ & $\tau_q h^{1/4}$ & $30$\\
	$\tau_f h^{1/4}$ & $8$ & $\tau_r h^{1/4}$ & $30$\\
	$\tau_\ell h^{1/4}$ & $8$ & $\tau_\pi h^{1/4}$ & $8$\\
	$\tau_m h^{1/4}$ & $8$ & $\tau_\phi h^{1/4}$ & $30$\\
	All $\chi_X h^{-1/2}$ & $2\times10^{-3}$ & All $\tau_X h^{1/4}$ & $50$\\
	\bottomrule
	\end{tabular}
\end{table*}
	The ideal determinant can then be expressed, up to an overall nonzero normalization, as
	\begin{equation}
		\det\mathbb M_{\rm ideal}
		\propto
		\omega^6
		\left(\omega^2-v_A^2 k^2\cos^2\theta\right)
		\left[\omega^4-A k^2\omega^2+C k^4\right],
		\label{eq:ideal-det-factorized}
	\end{equation}
	where
	\begin{equation}
		A=c_s^2+v_A^2-c_s^2v_A^2\sin^2\theta,
		\qquad
		C=c_s^2v_A^2\cos^2\theta .
	\end{equation}
	The first quadratic factor gives the Alfv\'en branches,
	\begin{equation}
		\omega_{A,\pm}=\pm v_A k\cos\theta .
		\label{eq:ideal-alfven-branches}
	\end{equation}
	The quartic factor gives the fast and slow magnetosonic branches,
	\begin{equation}
		\omega_{f,\pm}=\pm v_f(\theta) k,
		\qquad
		\omega_{s,\pm}=\pm v_s(\theta) k,
		\label{eq:ideal-ms-branches}
	\end{equation}
	with
	\begin{equation}
		v_{f,s}^2(\theta)
		=
		\frac12
		\left[
		A
		\pm
		\sqrt{A^2-4c_s^2v_A^2\cos^2\theta}
		\right].
		\label{eq:ideal-fast-slow-speeds}
	\end{equation}
	The upper sign corresponds to the fast branch and the lower sign to the slow branch.  In
	addition, Eq.~\eqref{eq:ideal-det-factorized} contains six zero-frequency roots from the
	passive spin-density block,
	\begin{equation}
		\omega_{{\rm spin},a}=0,
		\qquad a=1,\ldots,6 .
		\label{eq:ideal-zero-spin-roots}
	\end{equation}
	These roots are nondissipative at ideal order; they acquire nontrivial structure only after
	first- or second-order spin transport is included.
	
	The small-$k$ expansion is obtained by substituting
	\begin{equation}
		\omega(k)=\omega_0+\omega_1 k+\omega_2 k^2+\omega_3 k^3+\cdots
	\end{equation}
	into the determinant and solving order by order.  The leading equation gives
	\begin{equation}
		\omega_0=0,
	\end{equation}
	for all ideal branches.  The next order gives
	\begin{equation}
		\omega_1\in
		\left\{
		0,
		\pm v_A\cos\theta,
		\pm v_s(\theta),
		\pm v_f(\theta)
		\right\},
	\end{equation}
	while all higher coefficients vanish in the ideal theory,
	\begin{equation}
		\omega_{n\ge 2}=0 .
	\end{equation}
	Thus the long-wavelength expansion reproduces the exact ideal dispersion relations.
	
\begin{figure*}[htbp]
		\centering
		\includegraphics[width=0.48\textwidth]{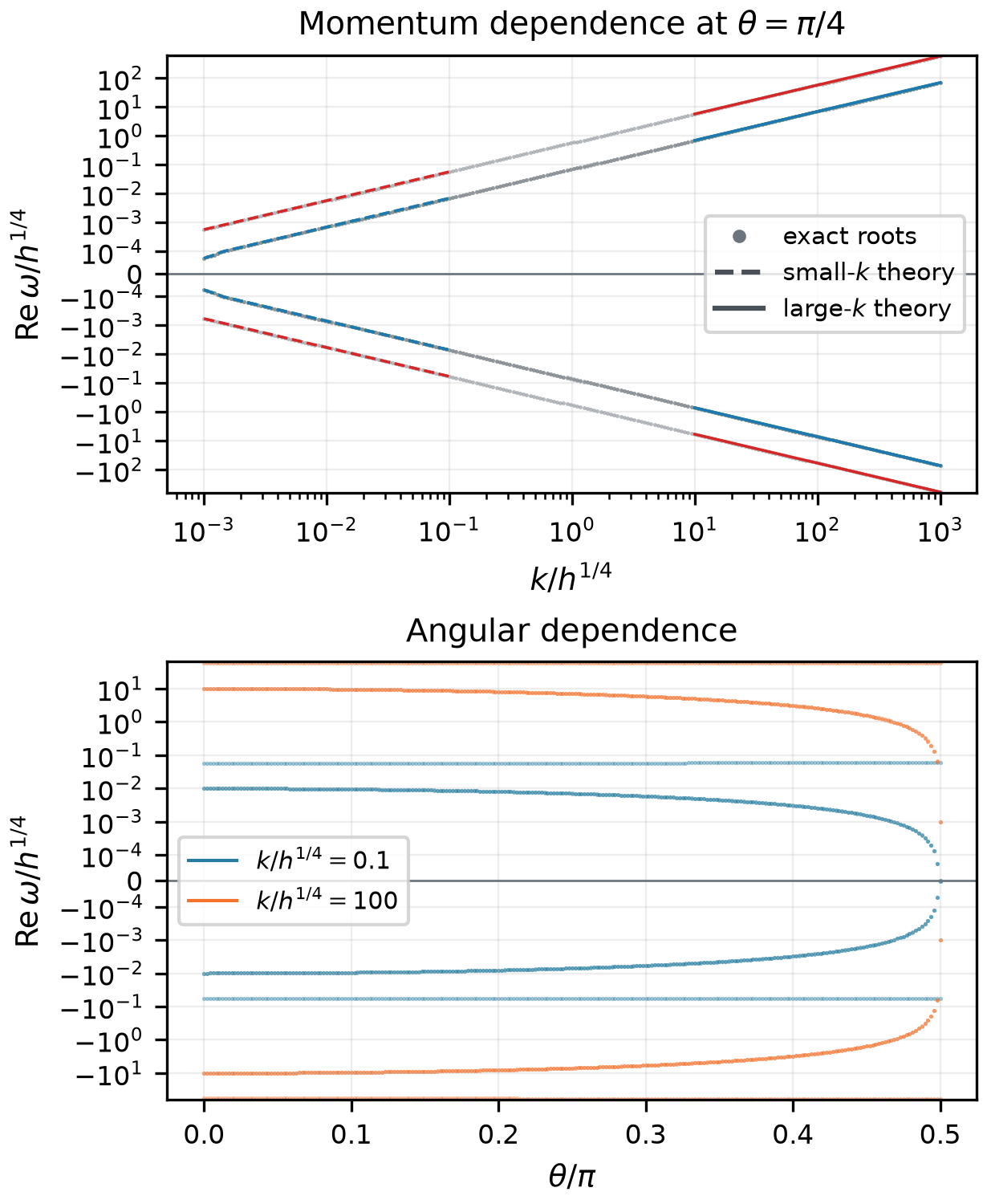}
		\caption{Real ideal-order spectrum.  The upper panel shows the exact roots and the coincident small- and large-$k$ solutions at $\theta=\pi/4$.  The lower panel shows all propagating branches over the angular interval at $k/h^{1/4}=0.1$ and $100$, using a common legend for both magnetosonic pairs and the Alfv\'en pair.  The slow magnetosonic and Alfv\'en branches vanish as $\theta\to\pi/2$, while the fast branch remains finite.}
		\label{fig:ideal-spin-mhd-2x2}
	\end{figure*}

	For the large-$k$ analysis one may use a singular expansion of the form
	\begin{equation}
		\omega(k)=v k+\omega^{(0)}+\frac{\omega^{(-1)}}{k}+\cdots .
	\end{equation}
	The propagating branches of Eq.~\eqref{eq:ideal-det-factorized} have the linear balance $\omega\sim vk$, with
	\begin{equation}
		v\in
		\left\{
		\pm v_A\cos\theta,
		\pm v_s(\theta),
		\pm v_f(\theta)
		\right\},
	\end{equation}
	whereas the six passive branches remain exactly at $\omega=0$.  Since the ideal determinant is homogeneous in $\omega$ and $k$, the subleading large-$k$
	coefficients vanish,
	\begin{equation}
		\omega^{(0)}=\omega^{(-1)}=\cdots=0 .
	\end{equation}
	Figure~\ref{fig:ideal-spin-mhd-2x2} compares the exact ideal roots with the
	coincident small- and large-$k$ solutions.
	
	The angular dependence follows from the magnetic restoring force.  The slow magnetosonic and
	Alfv\'en speeds vanish for transverse propagation, whereas the fast magnetosonic speed remains
	finite.  The magnetic Gauss constraint removes the additional longitudinal magnetic zero mode.

	\section{First-order block analysis and asymptotic solutions}
	\label{sec:first-order-modular-analysis}
	
	The first-order system provides the comparison with the transient theory.  Its
	constitutive relations reproduce the Navier--Stokes entropy production.  In the present
	closure, however, instantaneous response leads to nonhyperbolic ultraviolet behavior.
	The same mechanism underlies the acausality of first-order relativistic hydrodynamics
	and the instabilities identified in
	first-order spin hydrodynamics \cite{HiscockLindblom1985,XieWangYangPu2023}.  The mode
	analysis below identifies how the antisymmetric spin channels enter this pathology before
	finite relaxation times are introduced.  Because the magnetic field distinguishes the scalar
	$K$ channel from the transverse $q^\mu$ and $\phi^{\mu\nu}$ responses, the ultraviolet
	balance is both mode dependent and angle dependent rather than a universal isotropic limit.
	
	The Schur reduction and the coefficient functions required for the asymptotic expansion are
	given in Appendix~\ref{app:first-order-asymptotic-coefficients}.  Both the small- and
	large-$k$ expansions and the numerical spectra use the full first-order parameter set in
	Table~\ref{tab:numerical-parameters}.  The resulting spectrum contains the MHD-like and
	spin-rotation branches together with the nonhydrodynamic boost-sector poles and singular
	ultraviolet balances of the instantaneous closure.
	
	We use the ideal speeds and coefficients $A,C$ defined in Sec.~\ref{sec:ideal-modular-analysis}.
	
	\subsection{Small-$k$ expansion}
	
	At $k=0$ the first-order magnetosonic determinant factorizes as
	\begin{equation}
		\begin{aligned}
			\det \mathbb M_{\rm ms}^{(1)}(\omega,0)
			={}&\frac{1}{8}\,\omega^4
			\left(4\Gamma_{s\parallel}\lambda_r-\mathrm i\omega\right)
			\\
			&\times\left(\beta_0\lambda_K\omega^2-4\mathrm i h_\parallel\omega
			+8\Gamma_{b\parallel}h_\parallel\lambda_K\right)
			\\
			&\times\left(\beta_0\lambda_q\omega^2-2\mathrm i h\omega
			+8\Gamma_{b\perp}h\lambda_q\right).
		\end{aligned}
		\label{eq:first-order-ms-det-k0-main}
	\end{equation}
	Thus the magnetosonic small-$k$ spectrum takes the form
	\begin{equation}
		\begin{aligned}
			\omega_{{\rm ms},f,\pm}
			&=\pm v_f(\theta)k
			+\Omega_{{\rm ms},f,\pm}^{(2)}k^2
			+\mathcal O(k^3),
			\\
			\omega_{{\rm ms},s,\pm}
			&=\pm v_s(\theta)k
			+\Omega_{{\rm ms},s,\pm}^{(2)}k^2
			+\mathcal O(k^3),
			\\
			\omega_{{\rm ms},r}
			&=-4\mathrm i\Gamma_{s\parallel}\lambda_r
			+\Omega_{{\rm ms},r}^{(2)}k^2
			+\mathcal O(k^3),
			\\
			\omega_{{\rm ms},K,\pm}
			&=\omega_{K,\pm}
			+\Omega_{{\rm ms},K,\pm}^{(2)}k^2
			+\mathcal O(k^3),
			\\
			\omega_{{\rm ms},q,\pm}
			&=\omega_{q,\pm}
			+\Omega_{{\rm ms},q,\pm}^{(2)}k^2
			+\mathcal O(k^3).
		\end{aligned}
		\label{eq:first-order-ms-smallk-main-solutions}
	\end{equation}
	Here the boost-sector roots at $k=0$ are
	\begin{equation*}
		\begin{aligned}
			\omega_{K,\pm}
			&=\frac{2\mathrm i}{\beta_0\lambda_K}
			\left[h_\parallel\pm\sqrt{h_\parallel^2
				+2\beta_0\Gamma_{b\parallel}h_\parallel\lambda_K^2}\right],
			\\
			\omega_{q,\pm}
			&=\frac{\mathrm i}{\beta_0\lambda_q}
			\left[h\pm\sqrt{h^2
				+8\beta_0\Gamma_{b\perp}h\lambda_q^2}\right].
		\end{aligned}
	\end{equation*}
	The coefficients $\Omega_{{\rm ms},f/s,\pm}^{(2)}$ and
	$\Omega_{{\rm ms},j}^{(2)}$ for the nonhydrodynamic roots are fixed by the coefficient
	functions in Appendix~\ref{app:first-order-asymptotic-coefficients}.  Under the positivity assumptions adopted here, each quadratic boost-sector factor contains one pole in the upper half of the complex-frequency plane and one in the lower half.  Setting $\lambda_K=\lambda_q=0$ before forming the reduced determinant removes the positive-imaginary branches throughout the sampled momentum and angular ranges.  The $K$ and $q^\mu$ spin channels are therefore the principal source of the first-order instability in this parameter slice.  Equation~\eqref{eq:first-order-ms-smallk-main-solutions} also shows that the propagating
	magnetosonic branches retain their ideal phase velocities at order $k$, while dissipation first enters these branches at order $k^2$ and also generates gapped spin-sector roots.  
	
	The Alfv\'en block has the $k=0$ factorization
	\begin{equation}
		\begin{aligned}
			\det \mathbb M_{\rm A}^{(1)}(\omega,0)
			={}&\frac12\,\omega^2
			\left(4\Gamma_{s\parallel}\lambda_r-\mathrm i\omega\right)
			\\
			&\times\left(4\Gamma_{s\perp}\lambda_\phi-\mathrm i\omega\right)
			\\
			&\times\left(\beta_0\lambda_q\omega^2-2\mathrm i h\omega
			+8\Gamma_{b\perp}h\lambda_q\right).
		\end{aligned}
		\label{eq:first-order-A-det-k0-main}
	\end{equation}
	Accordingly, the Alfv\'en small-$k$ spectrum is
	\begin{equation}
		\begin{aligned}
			\omega_{{\rm A},\pm}
			&=\pm v_A\cos\theta\,k
			+\Omega_{{\rm A},\pm}^{(2)}k^2
			+\mathcal O(k^3),
			\\
			\omega_{{\rm A},r}
			&=-4\mathrm i\Gamma_{s\parallel}\lambda_r
			+\Omega_{{\rm A},r}^{(2)}k^2
			+\mathcal O(k^3),
			\\
			\omega_{{\rm A},\phi}
			&=-4\mathrm i\Gamma_{s\perp}\lambda_\phi
			+\Omega_{{\rm A},\phi}^{(2)}k^2
			+\mathcal O(k^3),
			\\
			\omega_{{\rm A},q,\pm}
			&=\omega_{q,\pm}
			+\Omega_{{\rm A},q,\pm}^{(2)}k^2
			+\mathcal O(k^3).
		\end{aligned}
		\label{eq:first-order-A-smallk-main-solutions}
	\end{equation}
	The first-order Navier--Stokes system has no independent spectator spin-flux determinant,
	because the spin-flux variables are not retained as first-order dynamical variables in the
	present power counting.  With $q^\mu$ retained, the Alfv\'en block includes the same
	upper-half-plane boost pole $\omega_{q,+}$ that appears in the magnetosonic block.
	
	\subsection{Large-$k$ expansion}
	
	The ultraviolet expansion is obtained by a Newton--Puiseux dominant-balance analysis of
	Schur-reduced determinants,
	\begin{equation}
		D_X^{(1)}(\omega,k,\theta)=\sum_{a,b}C_{a,b}^{X}(\theta)\omega^a k^b,
		\qquad X={\rm ms},{\rm A}.
	\end{equation}
	For each candidate scaling
	\begin{equation}
		\omega(k)=\Omega k^p+\cdots,
	\end{equation}
	the exponent $p$ is fixed by requiring at least two monomials
	$C_{a,b}^{X}\omega^a k^b$ to have the same largest weight $b+pa$.
	The leading coefficient $\Omega$ is then obtained from the corresponding balance polynomial.
	Subleading coefficients follow by substituting a Puiseux series into the same polynomial and matching successive powers, as specified in Appendix~\ref{app:first-order-largek-coefficients}.
	
	For the full first-order magnetosonic determinant, the generic
	large-$k$ spectrum is
	\begin{equation}
		\begin{aligned}
			\omega_{{\rm ms},2}
			&=\Omega_{{\rm ms},2}^{[2]}k^2+o(k^2),
			\\
			\omega_{{\rm ms},4/3,\alpha}
			&=\Omega_{{\rm ms},4/3,\alpha}^{[4/3]}k^{4/3}
			+o(k^{4/3}),\ \alpha=1,2,3,
			\\
			\omega_{{\rm ms},1,\alpha}
			&=\Omega_{{\rm ms},1,\alpha}^{[1]}k+o(k),\ \alpha=1,2,
			\\
			\omega_{{\rm ms},0,\alpha}
			&=\Omega_{{\rm ms},0,\alpha}^{[0]}+o(1),\ \alpha=1,2,
			\\
			\omega_{{\rm ms},-2}
			&=\Omega_{{\rm ms},-2}^{[-2]}k^{-2}+o(k^{-2}).
		\end{aligned}
		\label{eq:first-order-ms-largek-main-solutions}
	\end{equation}
	The $k^{4/3}$ family is the fractional Newton--Puiseux balance generated by the
	boost-spin sector.  It is absent if the $K$ and $q^\mu$ channels are set to zero before
	constructing the Newton polygon.  The longitudinal magnetic zero branch of the
	unconstrained determinant is absent after Eq.~\eqref{eq:magnetic-gauss-angle} is imposed.
	
	For the Alfv\'en block with the $q^\mu$ channel kept one obtains
	\begin{equation}
		\begin{aligned}
			\omega_{{\rm A},2}
			&=\Omega_{{\rm A},2}^{[2]}k^2+o(k^2),
			\\
			\omega_{{\rm A},1,\alpha}
			&=\Omega_{{\rm A},1,\alpha}^{[1]}k+o(k),\ \alpha=1,2,
			\\
			\omega_{{\rm A},0,\alpha}
			&=\Omega_{{\rm A},0,\alpha}^{[0]}+o(1),\ \alpha=1,2,3.
		\end{aligned}
		\label{eq:first-order-A-largek-main-solutions}
	\end{equation}
	Thus the generic first-order theory is nonhyperbolic: it contains parabolic and fractional ultraviolet branches in addition to finite nonhydrodynamic roots.  The numerical spectra below use this full determinant.
	\begin{figure*}[htbp]
		\centering
		\includegraphics[width=0.96\textwidth]{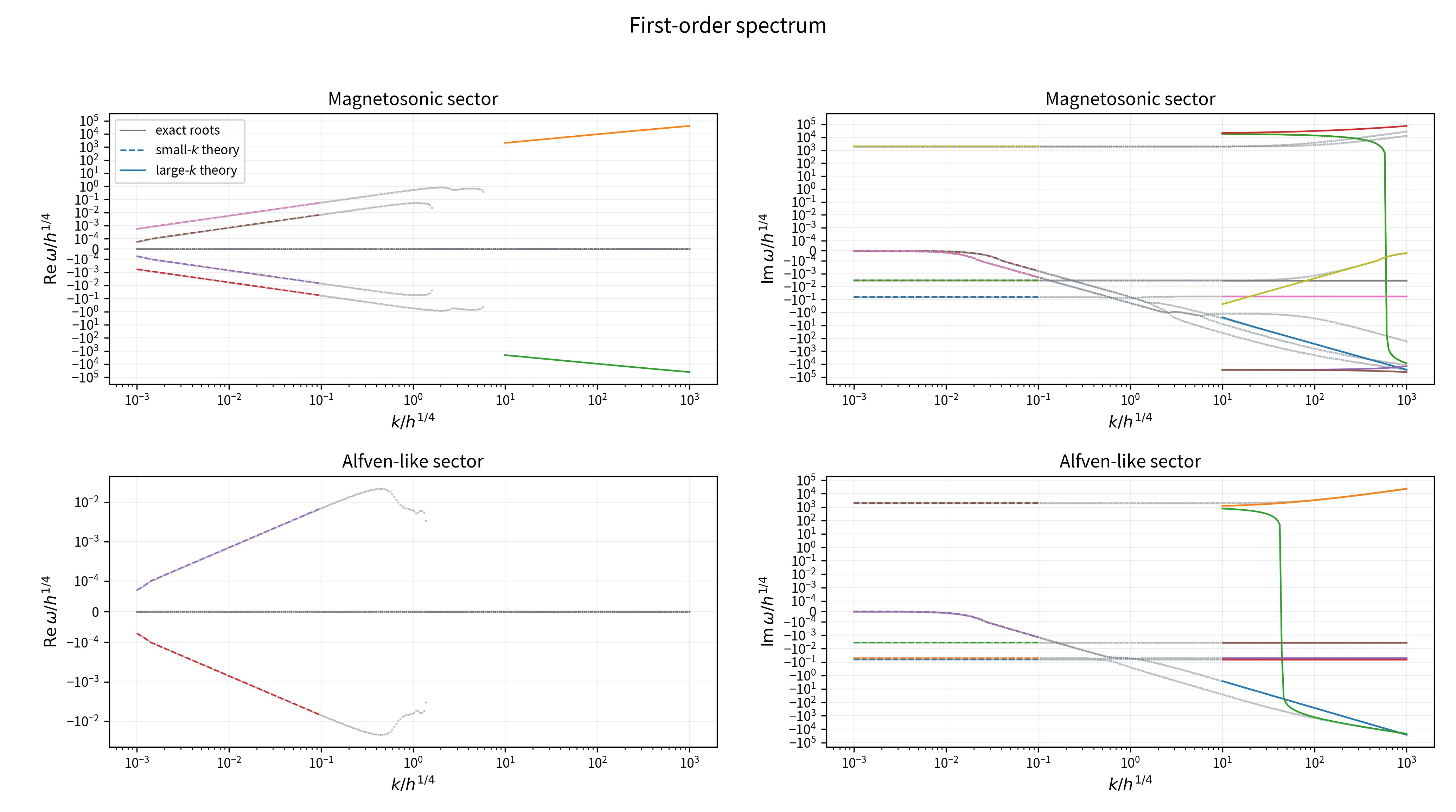}
		\caption{First-order spectrum at $\theta=\pi/4$: magnetosonic block (upper row) and Alfv\'en block (lower row).  Gray points are full-matrix roots; dashed curves retain the small-$k$ expansion through $k^2$.  Solid curves retain nonnegative large-$k$ powers through the constant term, or the leading term for a vanishing branch.  For the baseline $p=4/3$ family, keeping the first correction below 10\% of the leading term requires $k/h^{1/4}\gtrsim1.06\times10^6$, far above the plotted range.  The fractional-family expansion is discussed in Appendix~\ref{app:first-order-asymptotic-coefficients}.  Positive numerical imaginary parts signal instability.  The axes span $10^{-3}\leq k/h^{1/4}\leq10^3$; dashed and solid curves are displayed up to $10^{-1}$ and from $10^1$, respectively.}
		\label{fig:first-order-numeric-vs-theory}
	\end{figure*}
	\subsection{Comparison with numerical roots}
	
	Figure~\ref{fig:first-order-numeric-vs-theory} compares full-matrix roots with the characteristic-equation expansions.  In the infrared, the hydrodynamic pairs have ideal real slopes and quadratic attenuation, while the spin-conversion and boost poles remain gapped.  The small-$k$ coefficients follow from the coefficient equations in Appendix~\ref{app:second-order-asymptotic-extraction}.

	The ultraviolet balances in Eqs.~\eqref{eq:first-order-ms-largek-main-solutions} and \eqref{eq:first-order-A-largek-main-solutions} apply as $k\to\infty$.  The magnetosonic $p=4/3$ family has a large $k^{2/3}$ correction.  At the baseline parameters and $\theta=\pi/4$, its magnitude relative to the leading term is approximately $1038\,(k/h^{1/4})^{-2/3}$, still about $10.4$ at $k/h^{1/4}=10^3$.  Consequently, the nonzero real parts and changes of sign of some solid curves do not reproduce the numerical roots within this window.  The expansion converges at larger momentum.  The parabolic and fractional asymptotic families imply unbounded $|\omega|/k$.  Independently, the upper-half-plane numerical poles establish instability of the retained first-order closure \cite{XieWangYangPu2023,LuZhongFeng2026}.

\begin{figure*}[htbp]
		\centering
		\includegraphics[width=0.96\textwidth]{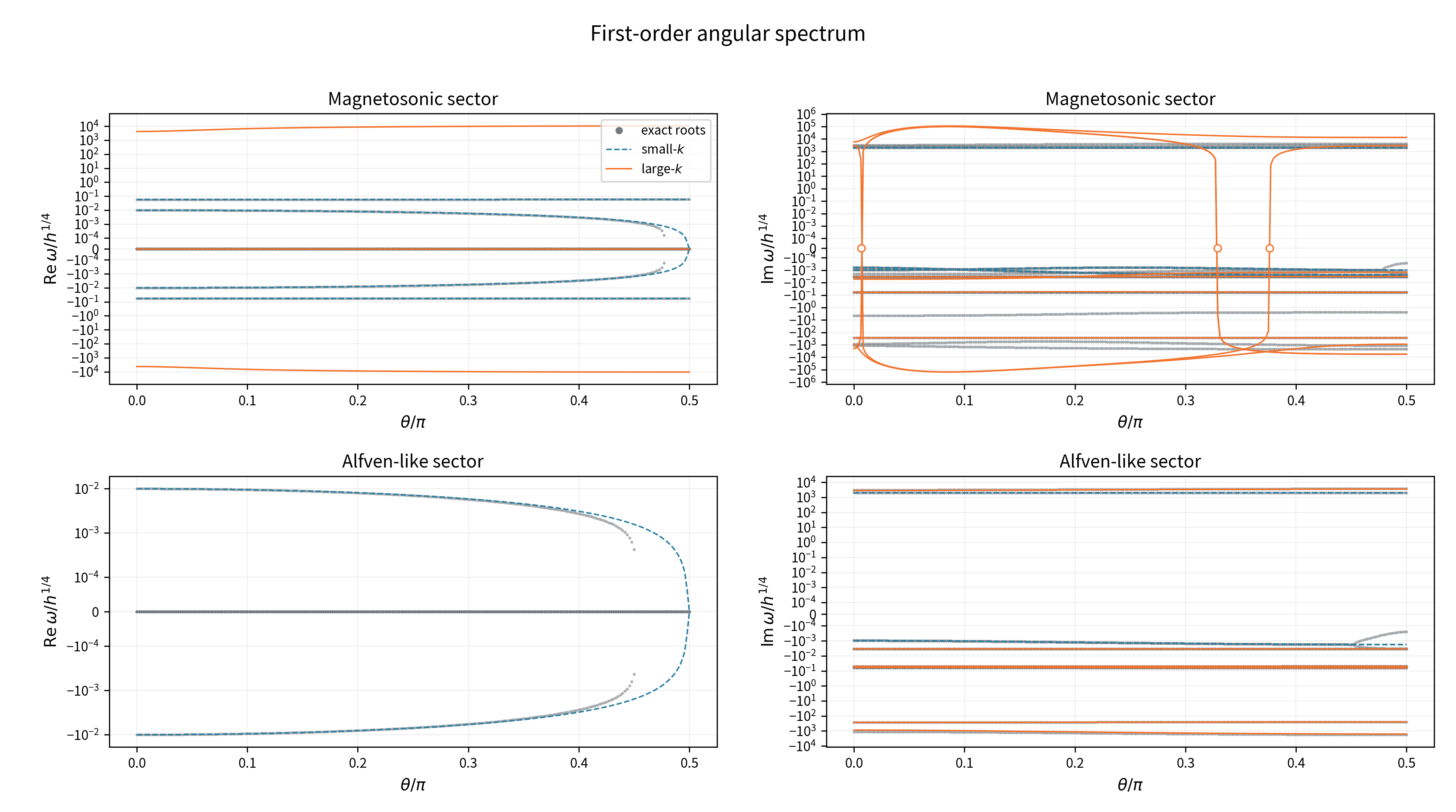}
		\caption{Angular first-order spectrum: magnetosonic block (upper row) and Alfv\'en block (lower row).  Gray points are full-matrix roots at $k/h^{1/4}=0.1$ and $100$; blue dashed and orange solid curves are the corresponding small- and large-$k$ truncations.  For the $p=4/3$ family, the $\theta=\pi/4$ benchmark needs $k/h^{1/4}\gtrsim1.06\times10^6$ to keep the first correction below 10\% of the leading term; this threshold is angle dependent (Appendix~\ref{app:first-order-asymptotic-coefficients}).  Open orange circles mark continuous zero crossings of the truncated imaginary parts, which appear steep on the symmetric-log scale.  The infrared expansion is nonuniform near transverse propagation.}
		\label{fig:first-order-theta-scan}
	\end{figure*}

	Figure~\ref{fig:first-order-theta-scan} compares the two fixed-momentum numerical spectra with the corresponding angular expansions.  At $k/h^{1/4}=0.1$, the fast wave retains a finite restoring force, whereas the slow and Alfv\'en pairs are suppressed as $k\cos\theta\to0$.  Close to transverse propagation, attenuation competes with the vanishing wave frequency, and the fixed-angle small-$k$ series is nonuniform.  Such competing momentum and angular limits, including conversion of wave pairs into diffusive branches, also occur in RMHD and spin MHD \cite{BiswasDashHaquePuRoy2020,FangHattoriHu2024,FangHattoriHu2025}.  At $k/h^{1/4}=100$, some magnetosonic ultraviolet families remain outside the validity range of the truncation, so their angular curves need not follow the numerical points.  The numerical spectra retain two growing magnetosonic roots and one growing Alfv\'en root at both plotted momenta throughout the sampled angular interval.

	\section{Minimal second-order analysis}
	\label{sec:second-order-modular-analysis}
	
	The minimal second-order theory replaces the algebraic first-order responses by finite-time
	relaxation equations for the viscous, magnetic, antisymmetric-stress, and spin-flux channels.
	This is an Israel--Stewart completion of the retained anisotropic spin-fluid channels
	\cite{BiswasSecondOrder2023,XieWangYangPu2023,DaherSecondOrder2024,LuZhongFeng2026}.  The resulting spectrum
	contains hydrodynamic waves and nonhydrodynamic relaxation poles, with diffusive balances
	in special angular limits.  Their angular dependence is measured relative to the magnetic field.  Each
	inequivalent $SO(2)$ channel has its own transient scale, and these scales enter the
	magnetosonic and Alfv\'en-like principal parts in different combinations.
	
	The numerical examples use the second-order parameter set in
	Table~\ref{tab:numerical-parameters}.  In this set the $K$ and $q^\mu$ channels are retained,
	and all tracked roots remain in the lower half of the frequency plane over the sampled
	domain.  Their zero-momentum gapped modes of the boost sector are stable precisely when
	$\tau_K-\beta_0\lambda_K/(4h_\parallel)>0$ and
	$\tau_q-\beta_0\lambda_q/(2h)>0$, for positive $\lambda_K\Gamma_{b\parallel}$
	and $\lambda_q\Gamma_{b\perp}$.  Appendix~\ref{app:zero-momentum-gaps}
	derives these conditions and relates their boundaries to the singularities of the time-derivative matrix.
	
	\subsection{Small-$k$ leading structure}
	
	At a fixed generic angle, the infrared spectrum contains four magnetosonic and two
	Alfv\'en hydrodynamic roots; the remaining modes are gapped.  Diffusive angular
	limits arise when a wave speed vanishes and require a separate dominant balance.
	
	First, the propagating hydrodynamic branches obey
	\begin{equation}
		\omega_{\rm prop,\pm}(k,\theta)
		=\pm v_{\rm eff}(\theta)k-\mathrm i\Gamma_{\rm eff}(\theta)k^2+O(k^3),
		\label{eq:second-order-smallk-propagating-structure}
	\end{equation}
	The ideal slopes are $v_f(\theta)$, $v_s(\theta)$, and $v_A\cos\theta$.
	The attenuation $\Gamma_{\rm eff}(\theta)$ follows from the order-$k^2$
	solvability condition.
	When the slow or Alfv\'en restoring force vanishes in a special angular limit,
	a diffusive branch may instead take the form
	\begin{equation}
		\omega_{\rm diff}(k,\theta)
		=-\mathrm iD_{\rm eff}(\theta)k^2+O(k^3).
		\label{eq:second-order-smallk-diffusive-structure}
	\end{equation}
	This is not an additional gapless family at a fixed generic angle.
	The nonhydrodynamic relaxation branches remain finite at $k=0$ and can be
	written as
	\begin{equation}
		\begin{aligned}
			\omega_{a,{\rm gap}}(k,\theta)
			={}&\omega_{a,0}+\omega_{a,1}(\theta)k\\
			&+\omega_{a,2}(\theta)k^2+O(k^3).
		\end{aligned}
		\label{eq:second-order-smallk-gapped-structure}
	\end{equation}
	For an isolated gap, the parity-even determinant implies $\omega_{a,1}=0$.  Repeated gaps must first be resolved by the projected coefficient equations; for the baseline set their first-order splittings also vanish.
	Finally, for the $k_y=0$ minimal reduction there are two exactly decoupled
	spectator spin-flux modes,
	\begin{equation}
		\begin{aligned}
			\omega_{\Lambda_\perp,{\rm spec}}
			&=-\frac{\mathrm i}{\tau_{\Lambda\perp}},
			&
			\omega_{\Omega,{\rm spec}}
			&=-\frac{\mathrm i}{\tau_\Omega}.
		\end{aligned}
		\label{eq:second-order-spectator-branches}
	\end{equation}
	These modes are included in the Alfv\'en panel of the figures below.
	
	The displayed small-$k$ curves retain
	\begin{equation}
		\omega(k)=\omega_0+\omega_1k+\omega_2k^2,
		\label{eq:second-order-smallk-theory-truncation}
	\end{equation}
	with coefficients obtained from successive solvability conditions of the affine matrix pencil.  The construction in Appendix~\ref{app:second-order-asymptotic-extraction} resolves repeated gaps and repeated first-order splittings before determining $\omega_2$.  Real and imaginary coefficients are parts of the same complex-frequency expansion.  The curves extend through $k/h^{1/4}=10^{-1}$, without assuming uniform validity near a vanishing wave speed.
	
	\subsection{Large-$k$ leading structure}
	
	At large momentum the minimal second-order determinant contains propagating ultraviolet families and finite relaxation families.  For the retained minimal matrix and parameter set, the principal-symbol balance gives no parabolic $p=2$ or fractional ultraviolet branches once the relaxation terms are included.  At fixed angle its affine pencil can be written exactly as
	\begin{equation}
		M^{(2)}(\omega,k,\theta)=C(\theta)+kK(\theta)+\omega W(\theta).
		\label{eq:second-order-affine-pencil}
	\end{equation}
	For a nonzero simple principal speed and a propagating branch $\omega=vk+\omega_0+O(k^{-1})$, the leading coefficient and first offset follow from
	\begin{align}
		[K+vW]r&=0,
		&
		\omega_0&=-\frac{\ell^\dagger C r}{\ell^\dagger W r},
		\qquad \ell^\dagger[K+vW]=0.
		\label{eq:second-order-principal-symbol-offset}
	\end{align}
	The propagating families obey
	\begin{equation}
		\omega_{\rm UV}(k,\theta)=v(\theta)k+\omega_0(\theta)+O(k^{-1}),
		\label{eq:second-order-largek-linear-structure}
	\end{equation}
	while the finite nonhydrodynamic families obey
	\begin{equation}
		\omega_{\rm rel}(k,\theta)=\Omega_0(\theta)+O(k^{-1}).
		\label{eq:second-order-largek-constant-structure}
	\end{equation}
	The finite leading values are obtained by projecting $C+\Omega_0W$ onto the left and right null spaces of $K$, as detailed in Appendix~\ref{app:second-order-asymptotic-extraction}.  At the angle used below this gives fourteen linear and nine finite branches in the magnetosonic block, and ten linear and ten finite branches in the Alfv\'en block.  Six linear branches in each block belong to the spin-density/flux principal subblocks, whose explicit factors are given in Appendix~\ref{app:spinflux-characteristics}.  All nonzero principal speeds $v$ are also real for the displayed parameter set.  Hence
	\begin{equation}
		\operatorname{Im}\omega(k,\theta)
		=\operatorname{Im}\omega_0(\theta)+O(k^{-1}),
		\label{eq:second-order-uv-damping-plateau}
	\end{equation}
	so the theoretical ultraviolet damping curves must approach horizontal plateaux rather than lines proportional to $k$.  For the retained matrix and parameter set, all propagating branches thus have finite ultraviolet slopes.
	\begin{figure*}[htbp]
		\centering
		\includegraphics[width=0.96\textwidth]{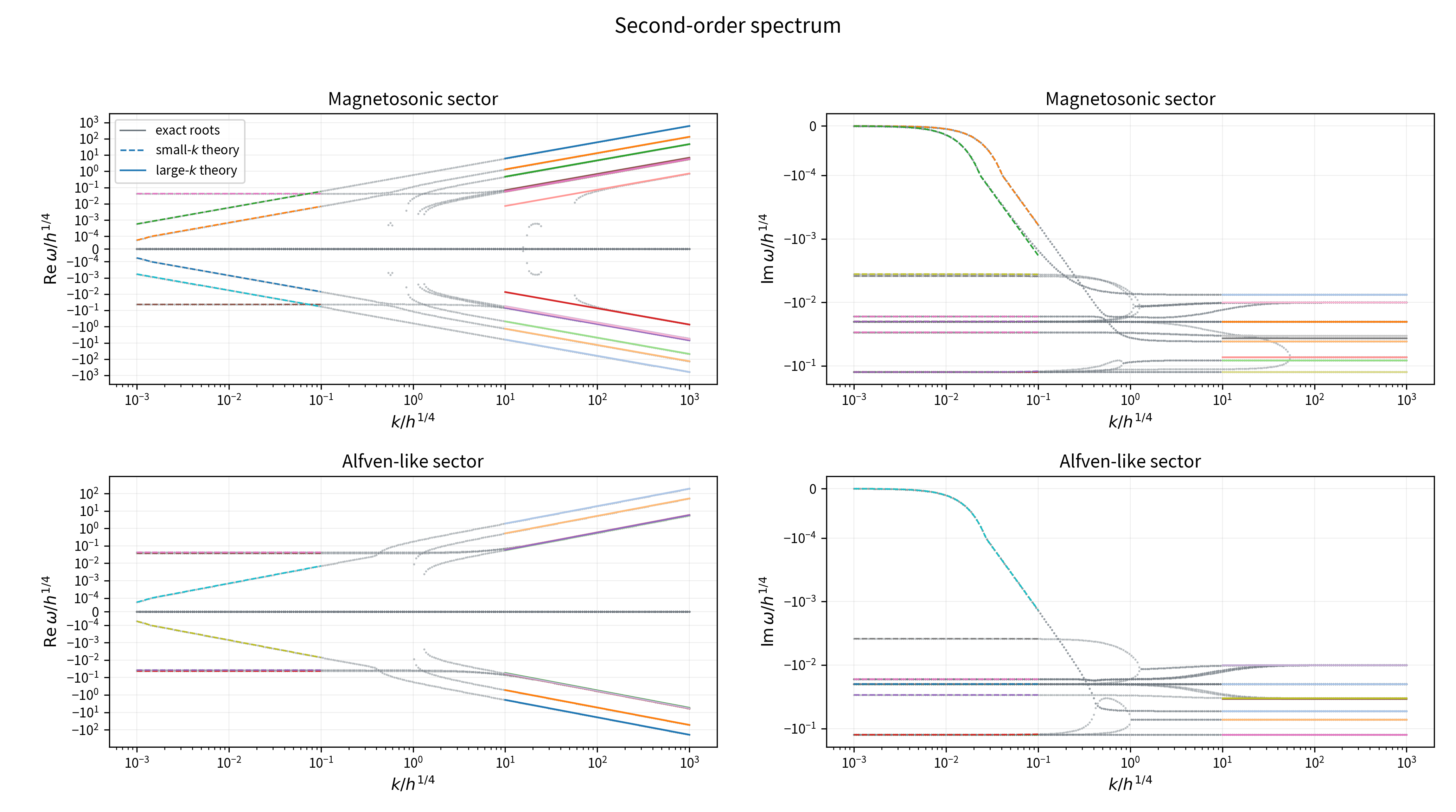}
		\caption{Minimal second-order spectrum at $\theta=\pi/4$: the 23-branch magnetosonic block (upper row) and the 20-branch Alfv\'en block with two spectator poles (lower row).  Gray points are full-matrix roots.  Dashed curves retain the small-$k$ coefficient expansion through $k^2$; solid curves use $vk+\omega_0$ or a finite pole from the projected principal pencil.  Their ultraviolet imaginary parts are horizontal, and the numerical damping converges to these plateaux.  All sampled numerical imaginary parts are negative for the baseline parameters.  Momentum and theory-display ranges are the same as in Fig.~\ref{fig:first-order-numeric-vs-theory}.}
		\label{fig:second-order-k-scan}
	\end{figure*}
	\begin{figure*}[htbp]
	\centering
	\includegraphics[width=0.96\textwidth]{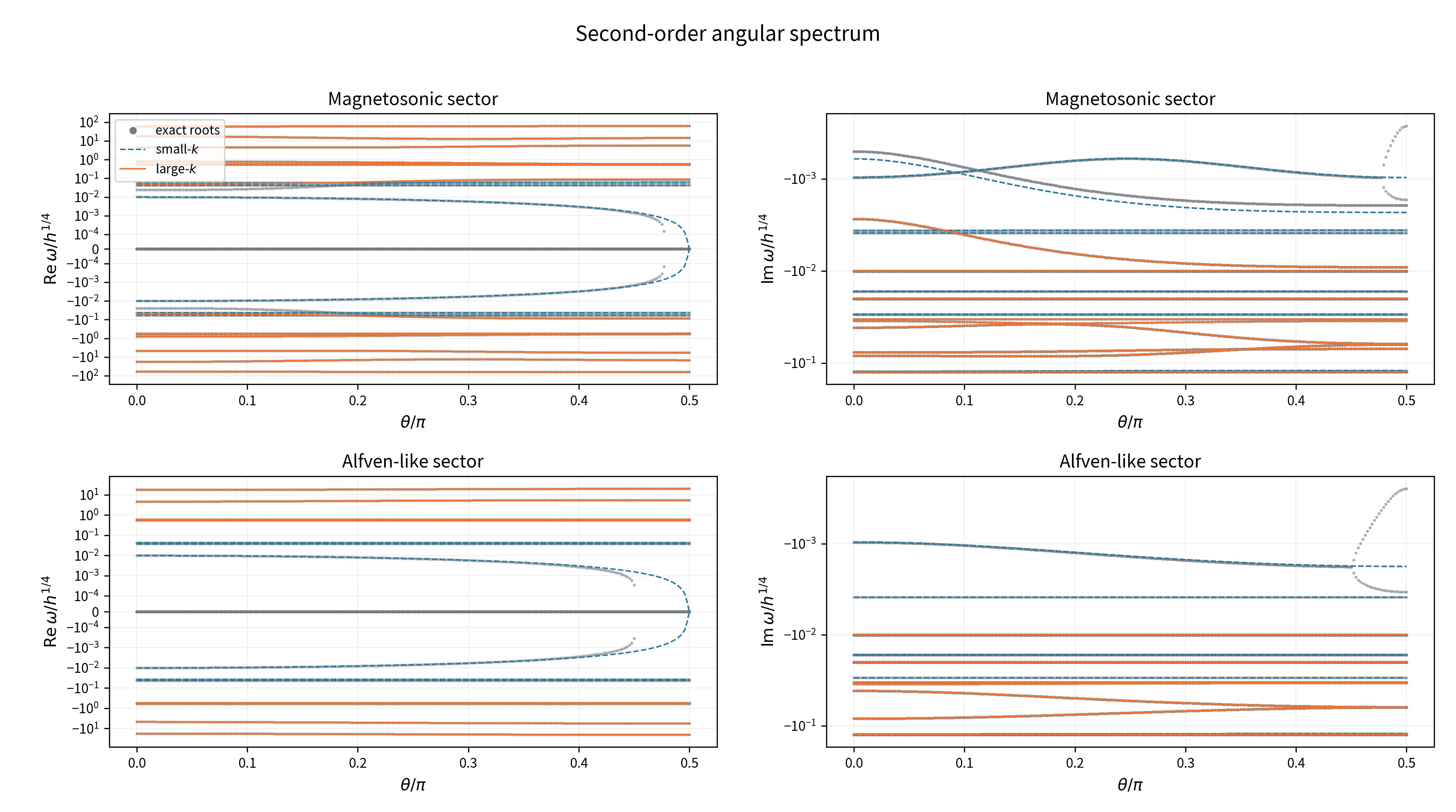}
	\caption{Angular minimal second-order spectrum: magnetosonic block (upper row) and Alfv\'en block with two spectator poles (lower row).  Gray points are full-matrix roots at $k/h^{1/4}=0.1$ and $100$.  Blue dashed curves retain the small-$k$ expansion through $k^2$; orange solid curves use the ultraviolet principal-pencil truncations.  All numerical imaginary parts remain negative.}
	\label{fig:second-order-theta-scan}
\end{figure*}
	\subsection{Comparison with numerical roots}
	
	Figure~\ref{fig:second-order-k-scan} compares the full minimal spectrum with the infrared solvability expansion and ultraviolet principal-symbol theory at $\theta=\pi/4$.  The low-momentum pairs have ideal real slopes and quadratic attenuation; the gapped roots depend jointly on relaxation times and spin response.

	At intermediate momentum, wave and relaxation roots approach one another and may become overdamped.  At sufficiently large momentum, the 14 magnetosonic and 10 Alfv\'en propagating branches approach finite real slopes, while the remaining 9 and 10 matrix roots approach finite poles.  The solid curves use the independently calculated complex offsets in Eq.~\eqref{eq:second-order-principal-symbol-offset}.  Their imaginary parts are constant in $k$, and the numerical damping rates converge to these plateaux.  All sampled numerical roots remain in the lower half-plane for the baseline set.  The finite-slope reorganization parallels the role of relaxation in second-order spin hydrodynamics and anisotropic RMHD \cite{DaherSecondOrder2024,LuZhongFeng2026,QiuSheHou2026}.

	Figure~\ref{fig:second-order-theta-scan} gives the angular comparison at $k/h^{1/4}=0.1$ and $100$.  The fast infrared pair retains a finite restoring force, whereas slow and Alfv\'en propagation is suppressed near the transverse direction.  There the damping and wave terms compete, and the fixed-angle small-$k$ series cannot uniformly resolve the conversion into diffusive roots \cite{FangHattoriHu2024,FangHattoriHu2025}.  The ultraviolet angular curves follow the characteristic speeds and projected finite poles, but finite-momentum corrections become important when a principal speed is small.  All numerical imaginary parts remain negative at both plotted momenta across the sampled angular interval.

	Related direction-dependent branch splitting occurs in the one-form effective theory of strong-field MHD \cite{VardhanEtAl2024,VardhanEtAl2025}.

	\section{Characteristic speeds and causality}
	\label{sec:numerical-causality-analysis}
	
	Signal propagation is governed by the large-$k$ characteristic slopes of the transient
	system, not by the ideal or hydrodynamic velocities alone.  For each propagating branch we
	define
	
	\begin{equation}
		v_g^{(a)}(k,\theta)
		\equiv
		\left|\frac{\partial\,\mathrm{Re}\,\omega_a(k,\theta)}{\partial k}\right|.
		\label{eq:causality-group-velocity}
	\end{equation}
	For $\omega_a=v_a(\theta)k+O(k^0)$, the ultraviolet limit is $|v_a(\theta)|$.
	For a real principal spectrum, a necessary causality condition is
	\begin{equation}
		v_{g,\mathrm{UV}}^{(a)}(\theta)\leq 1
		\qquad \text{for every propagating branch }a .
		\label{eq:causality-criterion}
	\end{equation}
	We rank the positive ultraviolet speeds within each
	block and retain the fastest and second-fastest magnetosonic modes and the fastest
	Alfv\'en-like mode.  This criterion is necessary but does not replace a proof of strong
	hyperbolicity.
	
\begin{figure*}[htbp]
		\centering
		\includegraphics[width=0.98\textwidth]{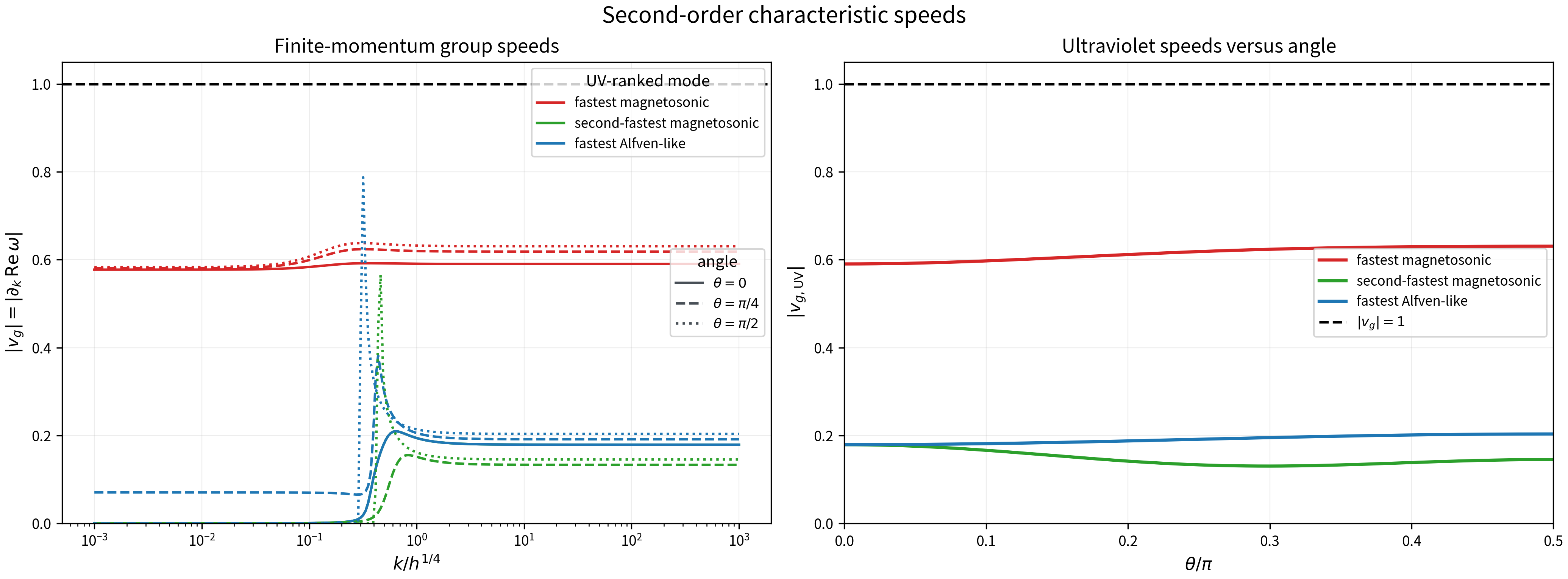}
		\caption{Characteristic speeds of the minimal second-order system.  The left panel follows the three branches ranked by ultraviolet speed through finite momentum for parallel, oblique, and transverse propagation; peaks mark mixing with transient modes.  The right panel gives their principal-symbol limits over the full angular interval.  The dashed horizontal line is $|v_g|=1$.}
		\label{fig:causality-group-velocity-k}
	\end{figure*}

	Figure~\ref{fig:causality-group-velocity-k} shows the relaxation crossover and the angular
	dependence of the ultraviolet limits.  The fastest magnetosonic speed is weakly anisotropic,
	whereas the other two speeds respond more strongly to the magnetic direction and transient
	mode mixing.  All three baseline ultraviolet characteristic speeds remain below the light cone.
	The dependence of causal bounds on propagation direction,
	background data, and relaxation parameters is also found in second-order spin hydrodynamics
	and anisotropic RMHD \cite{LuZhongFeng2026,QiuSheHou2026}.
	
For example, the fastest ultraviolet value in the Alfv\'en-like block at
$\theta=\pi/4$ is $v=0.1916034$.  This is a spin-coupled transverse-vector
relaxation mode, not the ideal Alfv\'en wave.  At exactly transverse propagation,
the minimal principal equations give
\begin{equation}
v_{\rm vec}^2\!\left(\frac\pi2\right)=
\frac{\eta_\perp/\tau_\pi+\beta_0\lambda_\phi/(2\tau_\phi)}
{h-\beta_0\lambda_q/(2\tau_q)}.
\label{eq:transverse-vector-uv-limit}
\end{equation}
For the baseline parameters this gives $v_{\rm vec}(\pi/2)=0.2036763$,
whereas $v_{A,\rm ideal}=v_A|\cos\theta|\to0$.
The vector branch retains the shear-relaxation contribution
$\eta_\perp/(h\tau_\pi)$ when spin transport is switched off; it is therefore
not a pure spin-density wave.  Here ``Alfv\'en-like'' names the coupled symmetry
block, rather than identifying each transient branch with the ideal Alfv\'en mode.

	The large-$k$ construction in
	Eqs.~\eqref{eq:second-order-affine-pencil}--\eqref{eq:second-order-principal-symbol-offset}
	reduces the ultraviolet problem to an angle- and parameter-dependent generalized eigenvalue
	equation.  We evaluate the principal spectra of the 23- and 20-branch blocks across the coefficient planes.
	
	The spin-sensitive ratios entering the principal matrix are
	\begin{equation}
		\xi_K\equiv \frac{\beta_0\lambda_K}{h_\parallel\tau_K},
		\qquad
		\xi_q\equiv \frac{\beta_0\lambda_q}{h\tau_q},
		\qquad
		\xi_\phi\equiv \frac{\beta_0\lambda_\phi}{h\tau_\phi} .
		\label{eq:causality-xi-definitions}
	\end{equation}
	For the baseline, $\xi_K=6.7340\times10^{-5}$ and
	$\xi_q=3.3333\times10^{-5}$: at fixed equilibrium data and relaxation times,
	\(\lambda_K\) and \(\lambda_q\) would have to increase by factors
	$5.94\times10^4$ and $6.00\times10^4$, respectively, to reach the singular
	values $\xi_K=4$ and $\xi_q=2$.  The scan covers this wider range of coefficients.
	These dimensionless ratios use the inertias appearing in the principal time matrix:
	\begin{equation}
		\begin{aligned}
			\det W_{(u_z,K)}&=-h_\parallel\tau_K\left(1-\frac{\xi_K}{4}\right),\\
			\det W_{(u_\perp,q)}&=h\tau_q\left(1-\frac{\xi_q}{2}\right).
		\end{aligned}
		\label{eq:principal-inertial-singularities}
	\end{equation}
	Combining the time-derivative terms in the longitudinal momentum and $K$ equations gives
	the effective inertia $h_\parallel-\beta_0\lambda_K/(4\tau_K)$.
	The spin correction is therefore applied to the longitudinal equilibrium
	inertia $h(1-v_A^2)$.
	Consequently, $\xi_K=4$ is a singular magnetosonic line and $\xi_q=2$ is a
	singular line in both blocks.  For the coefficients used in the scans, these lines give infinite
	generalized eigenvalues of the principal pencil.  Beyond the corresponding line, $\xi_K>4$ in the magnetosonic sector or
	$\xi_q>2$ in either sector, a principal pair becomes complex.  With the Fourier convention
	$e^{-\mathrm i\omega t}$, the conjugate principal pair contains an ultraviolet mode with
	$\operatorname{Im}\omega\simeq k\operatorname{Im}v>0$ and is therefore unstable as well as
	nonhyperbolic.

	These boundaries are not violations of the local entropy-production inequalities.  In the
	present scans all dissipative coefficients are nonnegative, the vector
	cross coefficients vanish, and the retained bulk matrix $\mathsf Z$ is positive definite,
	so Eqs.~\eqref{eq:scalar-positive-main}--\eqref{eq:at-positive-main} remain satisfied.
	Positive relaxation times also give the required entropy-density signs,
	$a_K=-\tau_K/\lambda_K<0$, $a_q=2\tau_q/\lambda_q>0$, and
	$a_\phi=-\tau_\phi/\lambda_\phi<0$ in the interior of the scan.  The second law supplies
	necessary local sign constraints, but it does not by itself guarantee an invertible principal
	time matrix, hyperbolicity, or linear stability.

	\begin{figure*}[htbp]
		\centering
		\includegraphics[width=0.98\textwidth]{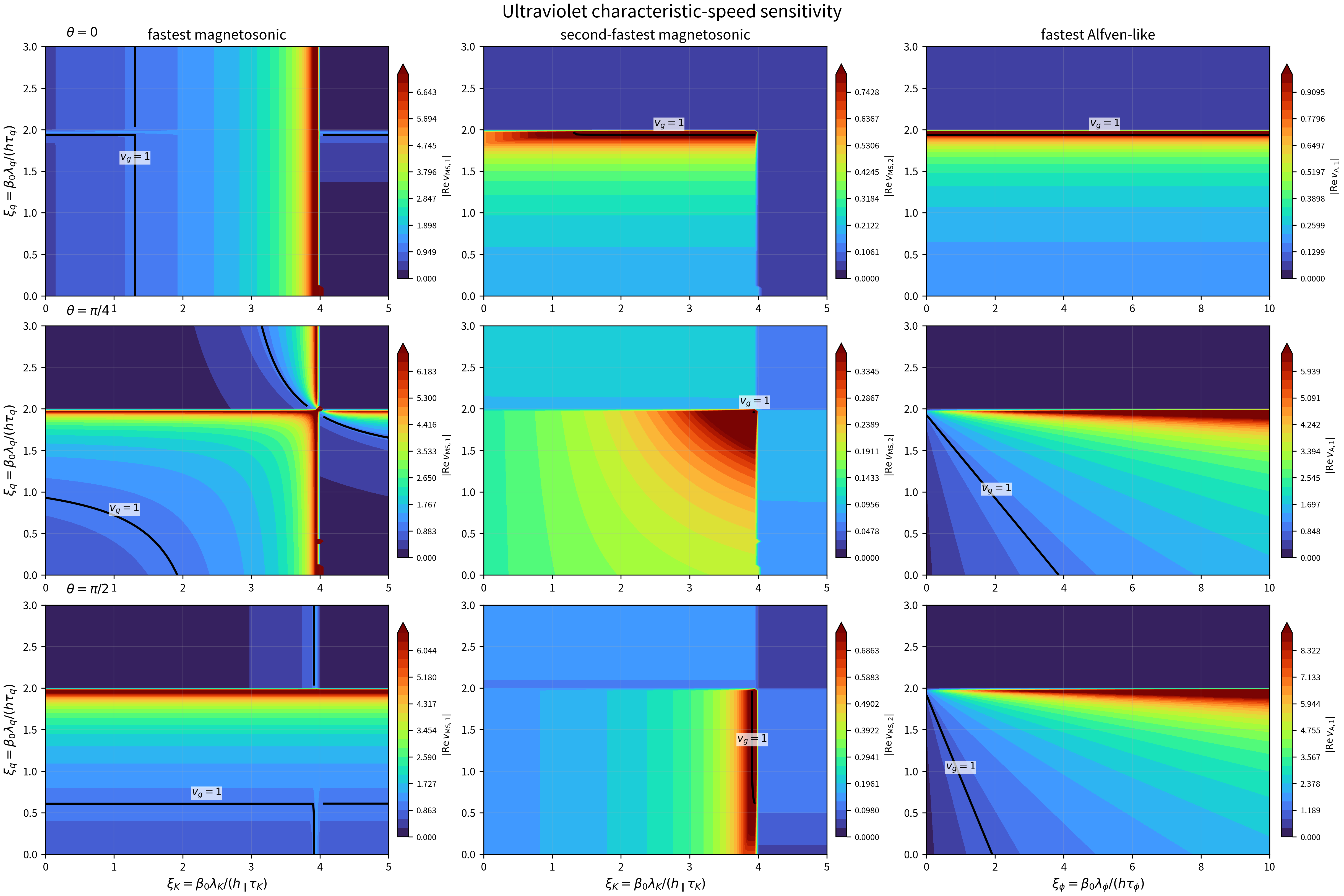}
		\caption{Ultraviolet characteristic-speed sensitivity.  Columns show the fastest magnetosonic, second-fastest magnetosonic, and fastest Alfv\'en-like ranked values of $|\mathrm{Re}\,v|$.  The magnetosonic columns scan $(\xi_K,\xi_q)$ and the Alfv\'en column scans $(\xi_\phi,\xi_q)$, with all other coefficients fixed.  Colors are retained in complex-spectrum regions and each panel has an independent, upper-saturated color scale.  The labeled black curves mark $|\mathrm{Re}\,v|=1$ and are not stability boundaries; the singular and unstable regions are identified in the text.}
		\label{fig:causality-dissipation-scan}
	\end{figure*}

	Figure~\ref{fig:causality-dissipation-scan} scans $(\xi_K,\xi_q)$ for the two magnetosonic
	speeds and $(\xi_\phi,\xi_q)$ for the Alfv\'en-like speed.  The rows show parallel, oblique,
	and transverse propagation.  The $K$ channel controls compressional propagation,
	$\phi^{\mu\nu}$ acts in the transverse antisymmetric Alfv\'en block, and $q^\mu$ enters both
	sectors through different transverse components.  Their competition produces curved
	light-cone boundaries.  Each scan point shows the ranked
	$|\mathrm{Re}\,v|$, including where the principal spectrum is complex.  The upper strip
	$\xi_q>2$ in all three columns and the right strip $\xi_K>4$ in the two magnetosonic
	columns therefore display the real part of an unstable principal spectrum, not causal
	propagation speeds.  On the singular lines themselves the plot shows the remaining finite
	ranks after the infinite generalized eigenvalues are omitted.  The upper $1.5\%$ of each
	panel is color-saturated for readability, so the large values adjacent to these lines should
	not be read from the color bar as unsaturated magnitudes.

	Below the inertial singularities, the sampled principal spectra are real:
	$0\leq\xi_K<4$, $0\leq\xi_q<2$ for the magnetosonic block and
	$0\leq\xi_q<2$ over the displayed $\xi_\phi$ interval for the Alfv\'en block.
	An independent finite-momentum scan over $10^{-3}\leq k/h^{1/4}\leq10^3$
	finds no growing roots in the sampled interiors at the three displayed angles.
	The baseline coefficients of Table~\ref{tab:numerical-parameters} lie well inside
	the nonsingular region and have no growing roots on the sampled finite-$k$ grid.
	
	Causality is therefore mode-, angle-, and parameter-dependent.  The black contours delimit
	the sampled subluminal real parts only where the principal spectrum is real and the time matrix
	is nonsingular.

	\section{Summary and outlook}
	\label{sec:summary}
	
	We have developed an Israel--Stewart entropy-current formulation of anisotropic
	spin hydrodynamics coupled to dynamical magnetic flux.  The magnetic direction
	resolves spin susceptibilities, antisymmetric stresses, and rank-three spin fluxes
	into distinct $SO(2)$ channels.  Entropy production constrains their transport
	matrices, and a restricted quadratic entropy current supplies finite relaxation
	times.  The spectral model assumes an unpolarized equilibrium, constant magnetic
	permeability, and no explicit spin-magnetization coupling.

	Imposing the magnetic Gauss constraint gives the usual ideal RMHD waves and six
	passive spin-density modes.  With the instantaneous $K$ and $q^\mu$ channels
	retained, the first-order spectrum develops unstable boost-sector poles and a
	fractional ultraviolet family $\omega\sim k^{4/3}$ in addition to parabolic,
	linear, finite, and vanishing branches.  These features belong to the truncated
	equations.  The fractional family has a large subleading correction, so its
	asymptotic hierarchy lies beyond the plotted momentum window.  The infrared
	wave expansion is also nonuniform near transverse propagation.

	For the baseline second-order model, finite relaxation replaces these superlinear
	ultraviolet families by branches with finite slopes or finite poles, together
	with two exact spin-flux spectators.  No growing root is found on the sampled
	baseline grid.  Positive effective inertias stabilize the zero-momentum gapped modes of the boost sector, and the spin-flux principal subblocks admit explicit characteristic speeds.
	The scans resolve the dependence on magnetic angle and transport ratios, with
	singular principal time matrices at $\xi_K=4$ or $\xi_q=2$.
	Complex ultraviolet speeds beyond these lines show that positive entropy
	production alone does not ensure stability or hyperbolicity.  The real,
	nonsingular scan interiors pass the sampled finite-momentum stability checks;
	the light-cone contours locate unit characteristic speeds in the real-spectrum regions.  Related asymptotic and numerical
	methods are used in second-order RMHD and spin hydrodynamics
	\cite{QiuSheHou2026,LuZhongFeng2026}.

	A complete second-order constitutive classification, analytic hyperbolicity and
	stability conditions across coefficient space, and microscopic determinations
	of the new transport coefficients remain open.  Extending the closure to a
	polarized equilibrium with explicit spin-magnetization feedback would connect
	the present mode analysis to a broader spin-MHD description.

	\section*{Acknowledgments}
	This work was supported in part by the National Key Research and Development Program of China under Contract No.~2022YFA1604900; the National Natural Science Foundation of China (NSFC) under Grant Nos.~12435009, 12275104, and 12605222; the Startup Research Fund of Henan Academy of Sciences (No.~231820058); the High-level Achievements Reward and Cultivation Projects (No.~20252320001); the Basic Research Fund of Henan Academy of Sciences (No.~20260620005); the Key Laboratory of Quark and Lepton Physics under Contract No.~QLPL2025P01; and the Fundamental Research Funds for the Central Universities at Central China Normal University (No.~2026CXGR050).
	
	\onecolumngrid
	\appendix

	\section{Tensor decompositions}
	\label{app:tensor-decompositions}
	
	\subsection{Decomposition of an antisymmetric rank-two tensor}
	
	Let $X^{\mu\nu}$ be an arbitrary antisymmetric rank-two tensor,
	\begin{equation}
		X^{\mu\nu}=-X^{\nu\mu}.
	\end{equation}
	Using
	\begin{equation}
		g^{\mu\nu}=u^\mu u^\nu-b^\mu b^\nu+\Xi^{\mu\nu},
	\end{equation}
	one may decompose $X^{\mu\nu}$ as
	\begin{equation}
		X^{\mu\nu}
		=
		x\,u^{[\mu}b^{\nu]}
		+
		2u^{[\mu}x_\Xi^{\nu]}
		+
		2b^{[\mu}y_\Xi^{\nu]}
		+
		x_{\rm A}^{\mu\nu},
	\end{equation}
	where
	\begin{equation}
		x=-2u_\mu b_\nu X^{\mu\nu},
		\qquad
		x_\Xi^\rho=u_\mu \Xi^\rho{}_{\nu} X^{\mu\nu},
		\qquad
		y_\Xi^\rho=-b_\mu \Xi^\rho{}_{\nu} X^{\mu\nu},
		\qquad
		x_{\rm A}^{\mu\nu}=\Xi^{\mu[\alpha}\Xi^{\beta]\nu}X_{\alpha\beta}.
	\end{equation}
	
	\subsection{Decomposition of a rank-three tensor antisymmetric in its last two indices}
	
	For a rank-three tensor $\phi^{\lambda\mu\nu}$ satisfying
	$\phi^{\lambda\mu\nu}=-\phi^{\lambda\nu\mu}$, we start from the identity
	\begin{equation}
		\phi^{\lambda\mu\nu}
		=
		g^\lambda{}_{\alpha}\phi^{\alpha\mu\nu}.
	\end{equation}
	Using
	$g^\lambda{}_{\alpha}=u^\lambda u_{\alpha}-b^\lambda b_{\alpha}+\Xi^\lambda{}_{\alpha}$,
	we obtain
	\begin{equation}
		\phi^{\lambda\mu\nu}
		=u^{\lambda}u_{\alpha}\phi^{\alpha\mu\nu}
		-b^{\lambda}b_{\alpha}\phi^{\alpha\mu\nu}
		+\Xi^{\lambda}{}_{\alpha}\phi^{\alpha\mu\nu}.
	\end{equation}
	The last two indices are decomposed by inserting the same identity twice,
	\begin{equation}
		\begin{aligned}
			\phi^{\alpha\mu\nu}
			&=g^{\mu}{}_{\beta}g^{\nu}{}_{\gamma}\phi^{\alpha\beta\gamma},
			\\
			\phi^{\alpha\mu\nu}
			&=\left(u^{\mu}u_{\beta}-b^{\mu}b_{\beta}+\Xi^{\mu}{}_{\beta}\right)
			\left(u^{\nu}u_{\gamma}-b^{\nu}b_{\gamma}+\Xi^{\nu}{}_{\gamma}\right)
			\phi^{\alpha\beta\gamma}.
		\end{aligned}
	\end{equation}
	After using the antisymmetry in the last two indices, this becomes
	\begin{equation}
		\begin{aligned}
			\phi^{\alpha\mu\nu}= {}&
			\left(u^{\nu}b^{\mu}-u^{\mu}b^{\nu}\right)u_{\beta}b_{\gamma}\phi^{\alpha\beta\gamma}
			\\
			&+
			\left(u^{\mu}\Xi^{\nu}{}_{\gamma}-u^{\nu}\Xi^{\mu}{}_{\gamma}\right)u_{\beta}\phi^{\alpha\beta\gamma}
			\\
			&+
			\left(b^{\nu}\Xi^{\mu}{}_{\gamma}-b^{\mu}\Xi^{\nu}{}_{\gamma}\right)b_{\beta}\phi^{\alpha\beta\gamma}
			\\
			&+
			\Xi^{\mu}{}_{\beta}\Xi^{\nu}{}_{\gamma}\phi^{\alpha\beta\gamma}.
		\end{aligned}
	\end{equation}
	Inserting the first-index decomposition and imposing the matching condition
	$u_\lambda\phi^{\lambda\mu\nu}=0$ gives the transverse correction below.
	With $A^{[\mu}B^{\nu]}=\tfrac12(A^\mu B^\nu-A^\nu B^\mu)$,
	\begin{equation}
		\begin{aligned}
			\phi^{\langle\lambda\rangle\mu\nu}= {}&
			2b^{\lambda}u^{[\mu}b^{\nu]}b_{\alpha}u_{\beta}b_{\gamma}\phi^{\alpha\beta\gamma}
			\\
			&-2u^{[\mu}b^{\nu]}\Xi^{\lambda}{}_{\alpha}u_{\beta}b_{\gamma}\phi^{\alpha\beta\gamma}
			\\
			&-2b^{\lambda}u^{[\mu}\Xi^{\nu]}{}_{\gamma}b_{\alpha}u_{\beta}\phi^{\alpha\beta\gamma}
			\\
			&+2u^{[\mu}\Xi^{\nu]}{}_{\gamma}\Xi^{\lambda}{}_{\alpha}u_{\beta}\phi^{\alpha\beta\gamma}
			\\
			&+2b^{\lambda}b^{[\mu}\Xi^{\nu]}{}_{\gamma}b_{\alpha}b_{\beta}\phi^{\alpha\beta\gamma}
			\\
			&-2b^{[\mu}\Xi^{\nu]}{}_{\gamma}\Xi^{\lambda}{}_{\alpha}b_{\beta}\phi^{\alpha\beta\gamma}
			\\
			&-b^{\lambda}\Xi^{[\mu}{}_{\beta}\Xi^{\nu]}{}_{\gamma}b_{\alpha}\phi^{\alpha\beta\gamma}
			+\Xi^{\lambda}{}_{\alpha}\Xi^{[\mu}{}_{\beta}\Xi^{\nu]}{}_{\gamma}\phi^{\alpha\beta\gamma}.
		\end{aligned}
		\label{app:eq:phi-final-decomp}
	\end{equation}
	
	To express Eq.~\eqref{app:eq:phi-final-decomp} in the notation of Eq.~\eqref{eq:spinflux}, we define the following projected amplitudes.
	The scalar and vector contractions are defined by
	\begin{equation}
		\begin{aligned}
			\Lambda_{\parallel}&\equiv
			b_{\alpha}u_{\beta}b_{\gamma}\phi^{\alpha\beta\gamma},
			\\
			\Lambda_{\perp}^{\lambda}&\equiv
			\Xi^{\lambda}{}_{\alpha}u_{\beta}b_{\gamma}\phi^{\alpha\beta\gamma},
			\\
			\Phi^{\nu}&\equiv
			\Xi^{\nu}{}_{\gamma}b_{\alpha}u_{\beta}\phi^{\alpha\beta\gamma},
			\\
			W^{\nu}&\equiv
			\Xi^{\nu}{}_{\gamma}b_{\alpha}b_{\beta}\phi^{\alpha\beta\gamma}.
		\end{aligned}
	\end{equation}
	The rank-two transverse contractions are
	\begin{equation}
		\begin{aligned}
			M^{\lambda\nu}&\equiv
			\Xi^{\lambda}{}_{\alpha}\Xi^{\nu}{}_{\gamma}u_{\beta}\phi^{\alpha\beta\gamma},
			\\
			M_{(t)}&\equiv
			\Xi_{\lambda\nu}M^{\lambda\nu}
			=\Xi_{\lambda\nu}\Xi^{\lambda}{}_{\alpha}\Xi^{\nu}{}_{\gamma}u_{\beta}\phi^{\alpha\beta\gamma},
			\\
			M_{(s)}^{\lambda\nu}&\equiv
			M^{(\lambda\nu)}-\frac{1}{2}\Xi^{\lambda\nu}M_{(t)},
			\\
			M_{(s)}^{\lambda\nu}&=
			\frac{1}{2}\left(
			\Xi^{\lambda}{}_{\alpha}\Xi^{\nu}{}_{\gamma}
			+\Xi^{\nu}{}_{\alpha}\Xi^{\lambda}{}_{\gamma}
			\right)u_{\beta}\phi^{\alpha\beta\gamma}
			-\frac{1}{2}\Xi^{\lambda\nu}
			\Xi_{\rho\sigma}\Xi^{\rho}{}_{\alpha}\Xi^{\sigma}{}_{\gamma}u_{\beta}\phi^{\alpha\beta\gamma}.
		\end{aligned}
		\label{eq:M-decomp}
	\end{equation}
	\begin{equation}
		M_{(a)}^{\lambda\nu}\equiv
		M^{[\lambda\nu]}
		=\Xi^{[\lambda}{}_{\alpha}\Xi^{\nu]}{}_{\gamma}u_{\beta}\phi^{\alpha\beta\gamma}.
	\end{equation}
	Similarly,
	\begin{equation}
		\begin{aligned}
			N^{\lambda\nu}&\equiv
			\Xi^{\lambda}{}_{\alpha}\Xi^{\nu}{}_{\gamma}b_{\beta}\phi^{\alpha\beta\gamma},
			\\
			N_{(t)}&\equiv
			\Xi_{\lambda\nu}N^{\lambda\nu}
			=\Xi_{\lambda\nu}\Xi^{\lambda}{}_{\alpha}\Xi^{\nu}{}_{\gamma}b_{\beta}\phi^{\alpha\beta\gamma},
			\\
			N_{(s)}^{\lambda\nu}&\equiv
			N^{(\lambda\nu)}-\frac{1}{2}\Xi^{\lambda\nu}N_{(t)},
			\\
			N_{(s)}^{\lambda\nu}&=
			\frac{1}{2}\left(
			\Xi^{\lambda}{}_{\alpha}\Xi^{\nu}{}_{\gamma}
			+\Xi^{\nu}{}_{\alpha}\Xi^{\lambda}{}_{\gamma}
			\right)b_{\beta}\phi^{\alpha\beta\gamma}
			-\frac{1}{2}\Xi^{\lambda\nu}
			\Xi_{\rho\sigma}\Xi^{\rho}{}_{\alpha}\Xi^{\sigma}{}_{\gamma}b_{\beta}\phi^{\alpha\beta\gamma}.
		\end{aligned}
		\label{eq:N-decomp}
	\end{equation}
	\begin{equation}
		N_{(a)}^{\lambda\nu}\equiv
		N^{[\lambda\nu]}
		=\Xi^{[\lambda}{}_{\alpha}\Xi^{\nu]}{}_{\gamma}b_{\beta}\phi^{\alpha\beta\gamma}.
	\end{equation}
	Finally,
	\begin{equation}
		\begin{aligned}
			\Omega_{\parallel}^{\mu\nu}&\equiv
			\Xi^{[\mu}{}_{\beta}\Xi^{\nu]}{}_{\gamma}b_{\alpha}\phi^{\alpha\beta\gamma},
			\\
			\Omega^{\lambda\mu\nu}&\equiv
			\Xi^{\lambda}{}_{\alpha}\Xi^{[\mu}{}_{\beta}\Xi^{\nu]}{}_{\gamma}\phi^{\alpha\beta\gamma}.
		\end{aligned}
	\end{equation}

	\section{Matching conditions}
	\label{app:matching-conditions}
	The matching is made at nonzero magnetic field and nondegenerate directional enthalpies.
	
	The nonequilibrium scalar shifts in the energy density and magnetic-field magnitude are
	absorbed into the local equilibrium variables,
	\begin{equation}
		\epsilon \rightarrow \epsilon'=\epsilon+\delta\epsilon+\mathcal O(\partial^2),
		\qquad
		B\rightarrow B'=B+\delta B+\mathcal O(\partial^2).
	\end{equation}
	Accordingly, $\delta\epsilon$ and $\delta B$ are not retained as independent first-order
	dissipative variables.
	
	Under first-order redefinitions of the basis vectors,
	\begin{equation}
		u^\mu\rightarrow u'^\mu=u^\mu+\delta u^\mu,
		\qquad
		b^\mu\rightarrow b'^\mu=b^\mu+\delta b^\mu,
		\qquad
		\delta u^\mu,\delta b^\mu=\mathcal O(\partial),
	\end{equation}
	the first-order vectors in $T^{\mu\nu}_{(1)}$ and $\tilde F^{\mu\nu}_{(1)}$ transform.  Before matching, we denote the corresponding flow-transverse components by $2u^{(\mu}h^{\nu)}$ in $T^{\mu\nu}_{(1)}$ and $2u^{[\mu}g^{\nu]}$ in $\tilde F^{\mu\nu}_{(1)}$.  The component of $g^\mu$ along $b^\mu$ is absorbed into $\delta B$.  A Landau-type frame choice adapted to RMHD removes the remaining components,
	\begin{equation}
		h^\mu=0,
		\qquad
		g^\mu=0.
	\end{equation}
	After this matching choice, the independent first-order structures are precisely those
	kept in Eqs.~\eqref{eq:T1}, \eqref{eq:F1}, and \eqref{eq:S1}.
	Under the phenomenological matching adopted here, the equilibrium spin density is identified with the $u^\lambda$-projected component, while the explicit first-order spin correction is fully contained in $\phi^{\langle\lambda\rangle\mu\nu}$.

	\section{First-order entropy-production matrices}
	\label{app:first-order-entropy-matrices}
	
	This appendix gives the explicit quadratic forms obtained by inserting the first-order
	constitutive relations into the Navier--Stokes entropy production in
	Eq.~\eqref{eq:sNSdiv-full}.  They give the positivity conditions in
	Eqs.~\eqref{eq:scalar-positive-main}--\eqref{eq:at-positive-main}.
	
	For the scalar sector,
	\begin{equation}
		\begin{aligned}
			\partial_\mu s^\mu_{\rm NS}\Big|_{\rm scalar}
			={}&
			\beta
			\begin{pmatrix}
				\theta_\perp & \theta_\parallel
			\end{pmatrix}
			\begin{pmatrix}
				\zeta_\perp & \dfrac{\zeta_\times+\zeta_\times'}{2} \\
				\dfrac{\zeta_\times+\zeta_\times'}{2} & \zeta_\parallel
			\end{pmatrix}
			\begin{pmatrix}
				\theta_\perp \\
				\theta_\parallel
			\end{pmatrix}
			\\
			&+
			\lambda_K
			\left[
			u^\mu b^\nu
			\Big(
			\partial_{[\mu}\beta_{\nu]}+2\beta\omega_{\mu\nu}
			\Big)
			\right]^2 .
		\end{aligned}
		\label{eq:first-order-scalar-matrix}
	\end{equation}
	Therefore
	\begin{equation}
		\zeta_\perp\ge 0,
		\qquad
		\zeta_\parallel\ge 0,
		\qquad
		4\zeta_\perp\zeta_\parallel-(\zeta_\times+\zeta_\times')^2\ge 0,
		\qquad
		\lambda_K\ge 0 .
	\end{equation}
	
	For the axial-vector sector, substitution gives
	\begin{equation}
		\partial_\mu s^\mu_{\rm NS}\Big|_{\rm axial}
		=
		-
		\begin{pmatrix}
			b^{(\alpha}\Xi^{\beta)\mu}\partial_\alpha\beta_\beta
			&
			b^{\alpha}\Xi^{\mu\beta}
			\Big(\partial_{[\alpha}\beta_{\beta]}+2\beta\omega_{\alpha\beta}\Big)
		\end{pmatrix}
		\mathbb K_{\rm AV}
		\begin{pmatrix}
			b^{(\rho}\Xi^{\sigma)}{}_{\mu}\partial_\rho\beta_\sigma
			\\[1ex]
			b^{\rho}\Xi_{\mu}{}^{\sigma}
			\Big(\partial_{[\rho}\beta_{\sigma]}+2\beta\omega_{\rho\sigma}\Big)
		\end{pmatrix},
		\label{eq:first-order-axial-matrix}
	\end{equation}
	where
	\begin{equation}
		\mathbb K_{\rm AV}
		=
		\begin{pmatrix}
			4\eta_\parallel & \kappa_{fr}+\kappa_{rf}
			\\[1ex]
			\kappa_{fr}+\kappa_{rf} & 2\lambda_r
		\end{pmatrix}.
		\label{eq:KA-matrix}
	\end{equation}
	Since the two displayed vector forces are transverse spacelike vectors, the second
	law requires $\mathbb K_{\rm AV}\ge0$, namely
	\begin{equation}
		\eta_\parallel\ge 0,
		\qquad
		\lambda_r\ge 0,
		\qquad
		8\eta_\parallel\lambda_r-(\kappa_{fr}+\kappa_{rf})^2\ge 0 .
		\label{eq:KA-positive}
	\end{equation}
	
	For the polar-vector sector,
	\begin{equation}
		\partial_\mu s^\mu_{\rm NS}\Big|_{\rm polar}
		=
		-
		\begin{pmatrix}
			b^{[\alpha}\Xi^{\beta]\mu}\partial_\alpha\mathcal H_\beta
			&
			u^{\alpha}\Xi^{\mu\beta}
			\Big(\partial_{[\alpha}\beta_{\beta]}+2\beta\omega_{\alpha\beta}\Big)
		\end{pmatrix}
		\mathbb K_{\rm PV}
		\begin{pmatrix}
			b^{[\rho}\Xi^{\sigma]}{}_{\mu}\partial_\rho\mathcal H_\sigma
			\\[1ex]
			u^{\rho}\Xi_{\mu}{}^{\sigma}
			\Big(\partial_{[\rho}\beta_{\sigma]}+2\beta\omega_{\rho\sigma}\Big)
		\end{pmatrix},
		\label{eq:first-order-polar-matrix}
	\end{equation}
	where
	\begin{equation}
		\mathbb K_{\rm PV}
		=
		\begin{pmatrix}
			4\rho_\parallel/\beta_0 & (\kappa_{\ell q}+\kappa_{q\ell})/\sqrt{\beta_0}
			\\[1ex]
			(\kappa_{\ell q}+\kappa_{q\ell})/\sqrt{\beta_0} & 2\lambda_q
		\end{pmatrix}.
		\label{eq:KP-matrix}
	\end{equation}
	The second law requires $\mathbb K_{\rm PV}\ge0$, i.e.
	\begin{equation}
		\rho_\parallel\ge0,
		\qquad
		\lambda_q\ge0,
		\qquad
		8\rho_\parallel\lambda_q-(\kappa_{\ell q}+\kappa_{q\ell})^2\ge0 .
		\label{eq:KP-positive}
	\end{equation}
	
	For the transverse symmetric-traceless tensor sector,
	\begin{equation}
		\partial_\mu s^\mu_{\rm NS}\Big|_{\rm ST}
		=
		2\beta\eta_\perp
		\left(\Xi^{\mu\nu\alpha\beta}\partial_\alpha u_\beta\right)
		\left(\Xi_{\mu\nu}{}^{\rho\sigma}\partial_\rho u_\sigma\right),
		\label{eq:first-order-ST-matrix}
	\end{equation}
	and hence
	\begin{equation}
		\eta_\perp\ge0 .
		\label{eq:KST-positive}
	\end{equation}
	
	For the transverse antisymmetric tensor sector,
	\begin{equation}
		\partial_\mu s^\mu_{\rm NS}\Big|_{\rm AT}
		=
		\begin{pmatrix}
			\Xi^{\mu[\alpha}\Xi^{\beta]\nu}\partial_\alpha\mathcal H_\beta
			&
			\Xi^\mu{}_{\rho}\Xi^\nu{}_{\sigma}
			\Big(\partial^{[\rho}\beta^{\sigma]}+2\beta\omega^{\rho\sigma}\Big)
		\end{pmatrix}
		\mathbb K_{\rm AT}
		\begin{pmatrix}
			\Xi_{\mu[\gamma}\Xi_{\delta]\nu}\partial^\gamma\mathcal H^\delta
			\\[1ex]
			\Xi_{\mu}{}^{\gamma}\Xi_{\nu}{}^{\delta}
			\Big(\partial_{[\gamma}\beta_{\delta]}+2\beta\omega_{\gamma\delta}\Big)
		\end{pmatrix},
		\label{eq:first-order-AT-matrix}
	\end{equation}
	where
	\begin{equation}
		\mathbb K_{\rm AT}
		=
		\begin{pmatrix}
			2\rho_\perp/\beta_0 & 0
			\\[1ex]
			0 & \lambda_\phi
		\end{pmatrix}.
		\label{eq:KAT-matrix}
	\end{equation}
	The positivity condition $\mathbb K_{\rm AT}\ge0$ is therefore
	\begin{equation}
		\rho_\perp\ge0,
		\qquad
		\lambda_\phi\ge0 .
		\label{eq:KAT-positive}
	\end{equation}

	\section{Structure of the second-order entropy-current contributions}
	\label{app:second-order-entropy-contributions}
	
	This appendix gives the source terms in the grouped divergence
	Eq.~\eqref{eq:dQ-grouped}, using the entropy-current split in
	Eq.~\eqref{eq:Q-split} and the quadratic forms in
	Eqs.~\eqref{eq:Qu} and \eqref{eq:Qperp}.

	The bulk contribution from $u^\mu Q_u$ is
	\begin{equation}
		\mathcal A^{(a)}_\Pi
		=A_\Pi D\Pi
		+\frac12(D A_\Pi+\theta A_\Pi)\Pi .
		\label{app:eq:bulk-entropy-source}
	\end{equation}
	The remaining diagonal pieces are denoted by the superscript $(a)$:
	\begin{equation}
		\begin{aligned}
			\mathcal A_K^{(a)}&=a_K DK+\frac12(Da_K+a_K\theta)K,
			&
			\mathcal A_{\Lambda\parallel}^{(a)}&=a_{\Lambda\parallel}D\Lambda_\parallel+\frac12(Da_{\Lambda\parallel}+a_{\Lambda\parallel}\theta)\Lambda_\parallel,\\
			\mathcal A_{Mt}^{(a)}&=a_{Mt}DM_{(t)}+\frac12(Da_{Mt}+a_{Mt}\theta)M_{(t)},
			&
			\mathcal A_{Nt}^{(a)}&=a_{Nt}DN_{(t)}+\frac12(Da_{Nt}+a_{Nt}\theta)N_{(t)}.
		\end{aligned}
		\label{app:eq:diag-scalar-new}
	\end{equation}
	For vector-like channels,
	\begin{equation}
		\begin{aligned}
			\mathcal B_f^{(a)\mu}&=\frac{a_f}{2}\Xi^\mu{}_{\nu}Df^\nu+\frac14(Da_f+a_f\theta)f^\mu,
			&
			\mathcal B_\ell^{(a)\mu}&=\frac{a_\ell}{2}\Xi^\mu{}_{\nu}D\ell^\nu+\frac14(Da_\ell+a_\ell\theta)\ell^\mu,\\
			\mathcal B_q^{(a)\mu}&=\frac{a_q}{2}\Xi^\mu{}_{\nu}Dq^\nu+\frac14(Da_q+a_q\theta)q^\mu,
			&
			\mathcal B_r^{(a)\mu}&=\frac{a_r}{2}\Xi^\mu{}_{\nu}Dr^\nu+\frac14(Da_r+a_r\theta)r^\mu,\\
			\mathcal B_{\Lambda\perp}^{(a)\mu}&=\frac{a_{\Lambda\perp}}{2}\Xi^\mu{}_{\nu}D\Lambda_\perp^\nu+\frac14(Da_{\Lambda\perp}+a_{\Lambda\perp}\theta)\Lambda_\perp^\mu,
			&
			\mathcal B_\Phi^{(a)\mu}&=\frac{a_\Phi}{2}\Xi^\mu{}_{\nu}D\Phi^\nu+\frac14(Da_\Phi+a_\Phi\theta)\Phi^\mu,\\
			\mathcal B_W^{(a)\mu}&=\frac{a_W}{2}\Xi^\mu{}_{\nu}DW^\nu+\frac14(Da_W+a_W\theta)W^\mu.
		\end{aligned}
		\label{app:eq:diag-vector-new}
	\end{equation}
	For two-index tensor channels,
	\begin{equation}
		\begin{aligned}
			\mathcal C_\pi^{(a)\mu\nu}&=\frac{a_\pi}{\beta}\Xi^{\mu\nu}{}_{\alpha\beta}D\pi_\perp^{\alpha\beta}
			+\frac{1}{2\beta}(Da_\pi+a_\pi\theta)\pi_\perp^{\mu\nu},\\
			\mathcal C_m^{(a)\mu\nu}&=a_m\Xi^{[\mu}{}_{\alpha}\Xi^{\nu]}{}_{\beta}Dm^{\alpha\beta}
			+\frac12(Da_m+a_m\theta)m^{\mu\nu},\\
			\mathcal C_\phi^{(a)\mu\nu}&=a_\phi\Xi^{[\mu}{}_{\alpha}\Xi^{\nu]}{}_{\beta}D\phi^{\alpha\beta}
			+\frac12(Da_\phi+a_\phi\theta)\phi^{\mu\nu},\\
			\mathcal D_{Ms}^{(a)\mu\nu}&=\frac{a_{Ms}}{2}\Xi^{\mu\nu}{}_{\alpha\beta}DM_{(s)}^{\alpha\beta}
			+\frac14(Da_{Ms}+a_{Ms}\theta)M_{(s)}^{\mu\nu},\\
			\mathcal D_{Ns}^{(a)\mu\nu}&=\frac{a_{Ns}}{2}\Xi^{\mu\nu}{}_{\alpha\beta}DN_{(s)}^{\alpha\beta}
			+\frac14(Da_{Ns}+a_{Ns}\theta)N_{(s)}^{\mu\nu},\\
			\mathcal E_{Ma}^{(a)\mu\nu}&=\frac{a_{Ma}}{2}\Xi^{[\mu}{}_{\alpha}\Xi^{\nu]}{}_{\beta}DM_{(a)}^{\alpha\beta}
			+\frac14(Da_{Ma}+a_{Ma}\theta)M_{(a)}^{\mu\nu},\\
			\mathcal E_{Na}^{(a)\mu\nu}&=\frac{a_{Na}}{2}\Xi^{[\mu}{}_{\alpha}\Xi^{\nu]}{}_{\beta}DN_{(a)}^{\alpha\beta}
			+\frac14(Da_{Na}+a_{Na}\theta)N_{(a)}^{\mu\nu},\\
			\mathcal E_{\Omega\parallel}^{(a)\mu\nu}&=a_{\Omega\parallel}\Xi^{[\mu}{}_{\alpha}\Xi^{\nu]}{}_{\beta}D\Omega_\parallel^{\alpha\beta}
			+\frac12(Da_{\Omega\parallel}+a_{\Omega\parallel}\theta)\Omega_\parallel^{\mu\nu}.
		\end{aligned}
		\label{app:eq:diag-tensor-new}
	\end{equation}
	Finally, the fully transverse spin-flux tensor enters Eq.~\eqref{eq:dQ-grouped} with unit prefactor, so its source is
	\begin{equation}
		\mathcal G_\Omega^{(a)\lambda\mu\nu}
		=a_\Omega\Xi^\lambda{}_{\rho}\Xi^{[\mu}{}_{\alpha}\Xi^{\nu]}{}_{\beta}D\Omega^{\rho\alpha\beta}
		+\frac12(Da_\Omega+a_\Omega\theta)\Omega^{\lambda\mu\nu}.
		\label{app:eq:diag-Omega-new}
	\end{equation}
	
	The contribution generated by $Q_\perp^\mu$ follows by differentiating Eq.~\eqref{eq:Qperp}.  It contains gradients of the coefficients $b_i$ and $c_i$, divergences of the coupled currents, and derivatives of the projectors $u^\mu$, $b^\mu$, and $\Xi^{\mu\nu}$.  These terms enter the hatted sources in the main text and generally produce cross-gradient couplings in the linearized relaxation equations.  The minimal model sets $b_i=c_i=0$, so they do not enter the matrices analyzed here.  Their componentwise expressions are therefore needed only for a nonminimal second-order closure.
	
	The comoving-derivative terms above define the relaxation times through
	Eqs.~\eqref{eq:ordinary-source-relaxation-defs} and
	\eqref{eq:spinflux-source-relaxation-defs}; the remaining terms form the hatted sources.

	\section{Nonzero entries of the linearized block matrices}

	\label{app:linear-matrix-elements}
	
	This appendix lists the nonzero entries of the minimal linear systems in the
	$x$--$z$ plane,
	$k_x=k_\perp$, $k_y=0$, $k_z=k_\parallel$.  Entries are listed explicitly or
	defined by the substitutions below; all others vanish.  We write
	\begin{equation}
		L_X\equiv 1-\mathrm i\omega\tau_X .
	\end{equation}
	Here $\varepsilon$ is the matter-energy perturbation, $h=\epsilon+p_\perp$,
	$h_\parallel=h(1-v_A^2)$, and $\mu_m$ is the magnetic permeability.
	The spin-response abbreviations are
	\begin{equation}
	\Gamma_{b\parallel}=\frac{\beta_0}{\chi_b^\parallel},\qquad
	\Gamma_{b\perp}=\frac{\beta_0}{\chi_b^\perp},\qquad
	\Gamma_{s\parallel}=\frac{\beta_0}{\chi_s^\parallel},\qquad
	\Gamma_{s\perp}=\frac{\beta_0}{\chi_s^\perp}.
	\end{equation}
	
	\subsection*{Ideal-order matrix entries}
	
	For the magnetosonic ordering
	\begin{equation}
		(\varepsilon,u_x,u_z,B_T),
	\end{equation}
	the nonzero entries are
	\begin{equation}
		\begin{aligned}
			&M_{E,\varepsilon}=-\mathrm i\omega,
			&&M_{E,u_x}=-\mathrm ih k_\perp,
			&&M_{E,u_z}= -\mathrm ih k_\parallel+\mathrm i\frac{B^2}{\mu_m}k_\parallel,
			&&M_{E,B_T}=-\mathrm i\omega\frac{B}{\mu_m}\sin\theta,
		\end{aligned}
	\end{equation}
	\begin{equation}
		\begin{aligned}
			&M_{P_z,\varepsilon}=-\mathrm i c_s^2 k_\parallel,
			&&M_{P_z,u_z}=-\mathrm i\omega h_\parallel,
		\end{aligned}
	\end{equation}
	\begin{equation}
		\begin{aligned}
			&M_{P_x,\varepsilon}=\mathrm i c_s^2 k_\perp,
			&&M_{P_x,u_x}=\mathrm i\omega h,
			&&M_{P_x,B_T}=\mathrm i\frac{B}{\mu_m}k,
		\end{aligned}
	\end{equation}
	\begin{equation}
		M_{B_T,u_x}=-\mathrm iB k,
		\qquad
		M_{B_T,B_T}=-\mathrm i\omega .
	\end{equation}
	For the Alfv\'en ordering $(u_y,B_y)$,
	\begin{equation}
		M_{P_y,u_y}=\mathrm i\omega h,
		\qquad
		M_{P_y,B_y}=\mathrm i\frac{B}{\mu_m}k_\parallel,
		\qquad
		M_{B_y,u_y}=-\mathrm iB k_\parallel,
		\qquad
		M_{B_y,B_y}=-\mathrm i\omega .
	\end{equation}
	The ideal spin rows are diagonal with entries $-\mathrm i\omega$.
	
	\subsection*{First-order matrix entries}
	
	For first order the magnetosonic ordering is
	\begin{equation}
		(\varepsilon,u_x,u_z,B_T,\Pi_\parallel,\Pi_\perp,f_x,\pi_{xx},\ell_x,
		K,q_x,r_x,S^{0z},S^{0x},S^{zx}).
	\end{equation}
	The conservation rows are the ideal rows supplemented by
	\begin{equation}
		M_{E,q_x}=\mathrm i k_\perp,
		\qquad
		M_{E,K}=-\frac{\mathrm i}{2}k_\parallel,
	\end{equation}
	\begin{equation}
		M_{P_z,\Pi_\parallel}=-\mathrm i k_\parallel,
		\qquad
		M_{P_z,f_x}=\mathrm i k_\perp,
		\qquad
		M_{P_z,K}=\frac{\mathrm i}{2}\omega,
		\qquad
		M_{P_z,r_x}=-\mathrm i k_\perp,
	\end{equation}
	\begin{equation}
		M_{P_x,\Pi_\perp}=\mathrm i k_\perp,
		\qquad
		M_{P_x,f_x}=-\mathrm i k_\parallel,
		\qquad
		M_{P_x,\pi_{xx}}=\mathrm i k_\perp,
		\qquad
		M_{P_x,q_x}=\mathrm i\omega,
		\qquad
		M_{P_x,r_x}=-\mathrm i k_\parallel,
	\end{equation}
	\begin{equation}
		M_{B_T,\ell_x}=-\mathrm i k .
	\end{equation}
	The first-order spin-balance rows are
	\begin{equation}
		M_{S^{0z},S^{0z}}=-\mathrm i\omega,
		\quad M_{S^{0z},K}=1,
		\qquad
		M_{S^{0x},S^{0x}}=-\mathrm i\omega,
		\quad M_{S^{0x},q_x}=-2,
	\end{equation}
	\begin{equation}
		M_{S^{zx},S^{zx}}=-\mathrm i\omega,
		\quad M_{S^{zx},r_x}=-2 .
	\end{equation}
	The algebraic constitutive rows have
	\begin{equation}
		M_{\Pi_\perp,\Pi_\perp}=1,
		\quad M_{\Pi_\perp,u_x}=-\mathrm i\zeta_\perp k_\perp,
		\quad M_{\Pi_\perp,u_z}=-\mathrm i\zeta_\times k_\parallel,
	\end{equation}
	\begin{equation}
		M_{\Pi_\parallel,\Pi_\parallel}=1,
		\quad M_{\Pi_\parallel,u_z}=-\mathrm i\zeta_\parallel k_\parallel,
		\quad M_{\Pi_\parallel,u_x}=-\mathrm i\zeta_\times' k_\perp,
	\end{equation}
	\begin{equation}
		M_{K,K}=1,
		\quad M_{K,\varepsilon}=\frac{\mathrm i}{2}\lambda_K\beta_\varepsilon k_\parallel,
		\quad M_{K,u_z}=\frac{\mathrm i}{2}\lambda_K\beta_0\omega,
		\quad M_{K,S^{0z}}=-2\lambda_K\Gamma_{b\parallel},
	\end{equation}
	\begin{equation}
		\begin{aligned}
			&M_{f_x,f_x}=1,
			&&M_{f_x,u_x}=\mathrm i\eta_f k_\parallel+\frac{\mathrm i}{2}\kappa_{fr}\beta_0k_\parallel,
			&&M_{f_x,u_z}=\mathrm i\eta_f k_\perp-\frac{\mathrm i}{2}\kappa_{fr}\beta_0k_\perp,
			&&M_{f_x,S^{zx}}=2\kappa_{fr}\Gamma_{s\parallel},
		\end{aligned}
	\end{equation}
	\begin{equation}
		\begin{aligned}
			&M_{r_x,r_x}=1,
			&&M_{r_x,u_x}=\frac{\mathrm i\kappa_{rf}}{2T_0}k_\parallel+\frac{\mathrm i}{2}\lambda_r\beta_0k_\parallel,
			&&M_{r_x,u_z}=\frac{\mathrm i\kappa_{rf}}{2T_0}k_\perp-\frac{\mathrm i}{2}\lambda_r\beta_0k_\perp,
			&&M_{r_x,S^{zx}}=2\lambda_r\Gamma_{s\parallel},
		\end{aligned}
	\end{equation}
	\begin{equation}
		\begin{aligned}
			&M_{\ell_x,\ell_x}=1,
			&&M_{\ell_x,B_T}=-\mathrm i\frac{\rho_\parallel}{\mu_m}k,
			&&M_{\ell_x,\varepsilon}=-\mathrm i
			\left(\frac{\rho_\parallel B}{\beta_0\mu_m}
			+\frac{\kappa_{\ell q}}{2\sqrt{\beta_0}}\right)\beta_\varepsilon k_\perp,
		\end{aligned}
	\end{equation}
	\begin{equation}
		M_{\ell_x,u_x}=-\frac{\mathrm i}{2}\kappa_{\ell q}\sqrt{\beta_0}\,\omega,
		\qquad
		M_{\ell_x,S^{0x}}=\frac{2\kappa_{\ell q}}{\sqrt{\beta_0}}\Gamma_{b\perp},
	\end{equation}
	\begin{equation}
		\begin{aligned}
			&M_{q_x,q_x}=1,
			&&M_{q_x,B_T}=-\frac{\mathrm i\sqrt{\beta_0}\kappa_{q\ell}}{2\mu_m}k,
			&&M_{q_x,\varepsilon}=-\frac{\mathrm i}{2}
			\left(\lambda_q+\frac{\kappa_{q\ell}B}{\sqrt{\beta_0}\mu_m}\right)
			\beta_\varepsilon k_\perp,
		\end{aligned}
	\end{equation}
	\begin{equation}
		M_{q_x,u_x}=-\frac{\mathrm i}{2}\lambda_q\beta_0\omega,
		\qquad
		M_{q_x,S^{0x}}=2\lambda_q\Gamma_{b\perp},
		\qquad
		M_{\pi_{xx},\pi_{xx}}=1,
		\qquad
		M_{\pi_{xx},u_x}=-\mathrm i\eta_\perp k_\perp .
	\end{equation}
	For the first-order Alfv\'en block the ordering is
	\begin{equation}
		(u_y,B_y,f_y,\pi_{xy},m,\ell_y,q_y,r_y,\phi,S^{0y},S^{zy},S^{xy}).
	\end{equation}
	The entries are obtained from the transverse equations with $x\to y$ and $k_y=0$, together
	with the antisymmetric rows
	\begin{equation}
		M_{S^{0y},S^{0y}}=-\mathrm i\omega,
		\quad M_{S^{0y},q_y}=-2,
		\qquad
		M_{S^{zy},S^{zy}}=-\mathrm i\omega,
		\quad M_{S^{zy},r_y}=-2,
	\end{equation}
	\begin{equation}
		M_{S^{xy},S^{xy}}=-\mathrm i\omega,
		\qquad
		M_{S^{xy},\phi}=2 .
	\end{equation}
	
	\subsection*{Second-order minimal matrix entries}
	
	First impose the minimal closure: the vector cross coefficients
	$\kappa$ and the transverse entropy-flux coefficients $b_i,c_i$ vanish,
	$\zeta_\times'=\zeta_\times$, and
	$\tau_{\Pi_\perp}=\tau_{\Pi_\parallel}$.  For the ordinary dissipative rows,
	replace the unit diagonal entry of this first-order matrix with $L_X$ for the corresponding current,
	e.g.
	\begin{equation}
		M_{K,K}=L_K,
		\qquad
		M_{q_x,q_x}=L_q,
		\qquad
		M_{\Pi_\perp,\Pi_\perp}=L_{\Pi_\perp},
		\qquad
		M_{\pi_{xx},\pi_{xx}}=L_\pi,
	\end{equation}
	and similarly for $\Pi_\parallel,f,\ell,r,m,\phi$.  The conservation rows are unchanged
	from first order.  The second-order spin-balance rows in the magnetosonic block
	are
	\begin{equation}
		M_{S^{0z},S^{0z}}=-\mathrm i\omega,
		\quad M_{S^{0z},K}=1,
		\quad M_{S^{0z},\Lambda_\parallel}=\mathrm ik_\parallel,
		\quad M_{S^{0z},\Lambda_{\perp x}}=-\mathrm ik_\perp,
	\end{equation}
	\begin{equation}
		M_{S^{0x},S^{0x}}=-\mathrm i\omega,
		\quad M_{S^{0x},q_x}=-2,
		\quad M_{S^{0x},\Phi_x}=-\mathrm ik_\parallel,
		\quad M_{S^{0x},M_{(t)}}=-\frac{\mathrm i}{2}k_\perp,
		\quad M_{S^{0x},M_{(s)xx}}=\mathrm ik_\perp,
	\end{equation}
	\begin{equation}
		M_{S^{zx},S^{zx}}=-\mathrm i\omega,
		\quad M_{S^{zx},r_x}=-2,
		\quad M_{S^{zx},W_x}=\mathrm ik_\parallel,
		\quad M_{S^{zx},N_{(t)}}=\frac{\mathrm i}{2}k_\perp,
		\quad M_{S^{zx},N_{(s)xx}}=-\mathrm ik_\perp .
	\end{equation}
	The Alfv\'en spin-balance rows are
	\begin{equation}
		M_{S^{0y},S^{0y}}=-\mathrm i\omega,
		\quad M_{S^{0y},q_y}=-2,
		\quad M_{S^{0y},\Phi_y}=-\mathrm ik_\parallel,
		\quad M_{S^{0y},M_{(s)xy}}=\mathrm ik_\perp,
		\quad M_{S^{0y},M_{(a)}}=\mathrm ik_\perp,
	\end{equation}
	\begin{equation}
		M_{S^{zy},S^{zy}}=-\mathrm i\omega,
		\quad M_{S^{zy},r_y}=-2,
		\quad M_{S^{zy},W_y}=\mathrm ik_\parallel,
		\quad M_{S^{zy},N_{(s)xy}}=-\mathrm ik_\perp,
		\quad M_{S^{zy},N_{(a)}}=-\mathrm ik_\perp,
	\end{equation}
	\begin{equation}
		M_{S^{xy},S^{xy}}=-\mathrm i\omega,
		\quad M_{S^{xy},\phi}=2,
		\quad M_{S^{xy},\Omega_\parallel}=-\mathrm ik_\parallel,
		\quad M_{S^{xy},\Omega_x}=\mathrm ik_\perp .
	\end{equation}
	The diagonal spin-flux relaxation rows are
	\begin{equation}
		M_{\Lambda_\parallel,\Lambda_\parallel}=L_{\Lambda\parallel},
		\qquad
		M_{\Lambda_\parallel,S^{0z}}=2\mathrm i\chi_{\Lambda\parallel}k_\parallel\Gamma_{b\parallel},
	\end{equation}
	\begin{equation}
		M_{\Lambda_{\perp x},\Lambda_{\perp x}}=L_{\Lambda\perp},
		\qquad
		M_{\Lambda_{\perp x},S^{0z}}=-\mathrm i\chi_{\Lambda\perp}k_\perp\Gamma_{b\parallel},
	\end{equation}
	\begin{equation}
		M_{\Phi_x,\Phi_x}=L_\Phi,
		\quad M_{\Phi_x,S^{0x}}=-\mathrm i\chi_\Phi k_\parallel\Gamma_{b\perp},
		\qquad
		M_{W_x,W_x}=L_W,
		\quad M_{W_x,S^{zx}}=\mathrm i\chi_W k_\parallel\Gamma_{s\parallel},
	\end{equation}
	\begin{equation}
		M_{M_{(t)},M_{(t)}}=L_{Mt},
		\quad M_{M_{(t)},S^{0x}}=-\mathrm i\chi_{Mt}k_\perp\Gamma_{b\perp},
		\qquad
		M_{N_{(t)},N_{(t)}}=L_{Nt},
		\quad M_{N_{(t)},S^{zx}}=\mathrm i\chi_{Nt}k_\perp\Gamma_{s\parallel},
	\end{equation}
	\begin{equation}
		M_{M_{(s)xx},M_{(s)xx}}=L_{Ms},
		\quad M_{M_{(s)xx},S^{0x}}=\frac{\mathrm i}{2}\chi_{Ms}k_\perp\Gamma_{b\perp},
		\qquad
		M_{N_{(s)xx},N_{(s)xx}}=L_{Ns},
		\quad M_{N_{(s)xx},S^{zx}}=-\frac{\mathrm i}{2}\chi_{Ns}k_\perp\Gamma_{s\parallel}.
	\end{equation}
	For the Alfv\'en spin-flux rows,
	\begin{equation}
		M_{\Phi_y,\Phi_y}=L_\Phi,
		\quad M_{\Phi_y,S^{0y}}=-\mathrm i\chi_\Phi k_\parallel\Gamma_{b\perp},
		\qquad
		M_{W_y,W_y}=L_W,
		\quad M_{W_y,S^{zy}}=\mathrm i\chi_W k_\parallel\Gamma_{s\parallel},
	\end{equation}
	\begin{equation}
		M_{M_{(s)xy},M_{(s)xy}}=L_{Ms},
		\quad M_{M_{(s)xy},S^{0y}}=\frac{\mathrm i}{2}\chi_{Ms}k_\perp\Gamma_{b\perp},
		\quad
		M_{N_{(s)xy},N_{(s)xy}}=L_{Ns},
		\quad M_{N_{(s)xy},S^{zy}}=-\frac{\mathrm i}{2}\chi_{Ns}k_\perp\Gamma_{s\parallel},
	\end{equation}
	\begin{equation}
		M_{M_{(a)},M_{(a)}}=L_{Ma},
		\quad M_{M_{(a)},S^{0y}}=\frac{\mathrm i\chi_{Ma}}{2}k_\perp\Gamma_{b\perp},
		\qquad
		M_{N_{(a)},N_{(a)}}=L_{Na},
		\quad M_{N_{(a)},S^{zy}}=-\frac{\mathrm i\chi_{Na}}{2}k_\perp\Gamma_{s\parallel},
	\end{equation}
	\begin{equation}
		M_{\Omega_\parallel,\Omega_\parallel}=L_{\Omega\parallel},
		\quad M_{\Omega_\parallel,S^{xy}}=-\mathrm i\chi_{\Omega\parallel}k_\parallel\Gamma_{s\perp},
		\qquad
		M_{\Omega_x,\Omega_x}=L_\Omega,
		\quad M_{\Omega_x,S^{xy}}=\mathrm i\chi_\Omega k_\perp\Gamma_{s\perp}.
	\end{equation}
	The spectator block in the $x$--$z$ plane is
	\begin{equation}
		M_{\Lambda_{\perp y},\Lambda_{\perp y}}=L_{\Lambda\perp},
		\qquad
		M_{\Omega_y,\Omega_y}=L_\Omega .
	\end{equation}

	\section{First-order asymptotic dispersion relations}
	\label{app:first-order-asymptotic-coefficients}
	
	This appendix defines the coefficients used in
	Sec.~\ref{sec:first-order-modular-analysis} directly from the Schur-reduced
	determinants of the first-order block matrices.  The first-order matrices contain algebraic Navier--Stokes currents.  We eliminate these
	auxiliary variables by the exact Schur complement
	\begin{equation}
		\mathbb M_{X,{\rm eff}}^{(1)}
		=
		\mathbb M_{dd}^{(1)}
		-
		\mathbb M_{da}^{(1)}
		\left(\mathbb M_{aa}^{(1)}\right)^{-1}
		\mathbb M_{ad}^{(1)},
		\qquad X={\rm ms},{\rm A},
		\label{eq:app-schur-compact}
	\end{equation}
	and define
	\begin{equation}
		D_X^{(1)}(\omega,k,\theta)
		=
		\det\mathbb M_{X,{\rm eff}}^{(1)}(\omega,k,\theta).
		\label{eq:app-reduced-det-compact}
	\end{equation}
	Since \(\det\mathbb M_{aa}^{(1)}\) is nonzero near the roots considered here, the Schur
	reduction changes only the overall nonzero determinant factor and leaves the dispersion
	relations unchanged.
	
	For any expression \(F(k)\), we denote by \([k^n]F\) the coefficient of \(k^n\) in its
	Taylor expansion.  Similarly, \([\omega^a k^b]F\) denotes the coefficient multiplying
	\(\omega^a k^b\).  The large-\(k\) coefficients appearing in
	Eqs.~\eqref{eq:first-order-ms-largek-main-solutions} and
	\eqref{eq:first-order-A-largek-main-solutions} are therefore defined by
	\begin{equation}
		C^X_{a,b}(\theta)\equiv [\omega^a k^b]D_X^{(1)}(\omega,k,\theta),
		\qquad X={\rm ms},{\rm A}.
		\label{eq:app-Cab-definition}
	\end{equation}
	This coefficient-extraction definition is the compact form of the expanded large-momentum
	polynomials.
	
	\subsection{Zeroth-order determinants and branch polynomials}
	
	At \(k=0\), the magnetosonic determinant factorizes as
	\begin{equation}
		\begin{aligned}
			D_{{\rm ms},0}(\omega)
			\equiv D_{\rm ms}^{(1)}(\omega,0,\theta)
			={}&
			\frac18\,
			\omega^4
			\left(4\Gamma_{s\parallel}\lambda_r-\mathrm i\omega\right)
			\left(\beta_0\lambda_K\omega^2-4\mathrm i h_\parallel\omega
			+8\Gamma_{b\parallel}h_\parallel\lambda_K\right)
			\\
			&\times
			\left(\beta_0\lambda_q\omega^2-2\mathrm i h\omega
			+8\Gamma_{b\perp}h\lambda_q\right).
		\end{aligned}
		\label{eq:app-ms-D0-compact}
	\end{equation}
	The Alfv\'en determinant factorizes as
	\begin{equation}
		\begin{aligned}
			D_{{\rm A},0}(\omega)
			\equiv D_{\rm A}^{(1)}(\omega,0,\theta)
			={}&
			\frac12\,
			\omega^2
			\left(4\Gamma_{s\parallel}\lambda_r-\mathrm i\omega\right)
			\left(4\Gamma_{s\perp}\lambda_\phi-\mathrm i\omega\right)
			\\
			&\times
			\left(\beta_0\lambda_q\omega^2-2\mathrm i h\omega
			+8\Gamma_{b\perp}h\lambda_q\right).
		\end{aligned}
		\label{eq:app-A-D0-compact}
	\end{equation}
	The gapped magnetosonic branches are specified by
	\begin{equation}
		P_r(w)=w+4\mathrm i\Gamma_{s\parallel}\lambda_r=0,
		\label{eq:app-Pr-ms}
	\end{equation}
	\begin{equation}
		P_K(w)=
		\beta_0\lambda_Kw^2
		-4\mathrm i h_\parallel w
		+8\Gamma_{b\parallel}h_\parallel\lambda_K=0,
		\label{eq:app-PK-ms}
	\end{equation}
	\begin{equation}
		P_q(w)=
		\beta_0\lambda_qw^2
		-2\mathrm i h w
		+8\Gamma_{b\perp}h\lambda_q=0.
		\label{eq:app-Pq-ms}
	\end{equation}
	The gapped Alfv\'en branches are specified by \(P_r(w)=0\), \(P_q(w)=0\), and
	\begin{equation}
		P_\phi(w)=w+4\mathrm i\Gamma_{s\perp}\lambda_\phi=0.
		\label{eq:app-Pphi-A}
	\end{equation}
	The coefficient expressions below do not apply branch-polynomial reduction to the
	quadratic \(K\) and \(q\) branches.  The symbol \(w\) denotes the corresponding
	leading frequency.
	
	\subsection{Hydrodynamic branches}
	The scalar coefficient formulas below apply to simple ideal roots at fixed angle;
	repeated or zero-speed roots are resolved by the projected equations in
	Appendix~\ref{app:second-order-asymptotic-extraction}.
	
	For the magnetosonic hydrodynamic roots we set
	\begin{equation}
		\omega=w k+\Omega^{(2)} k^2+O(k^3).
	\end{equation}
	The leading coefficient is proportional to
	\begin{equation}
		w^4-Aw^2+C,
		\qquad
		A=c_s^2+v_A^2-c_s^2v_A^2\sin^2\theta,
		\qquad
		C=c_s^2v_A^2c_\theta^2 .
		\label{eq:app-ms-hydro-leading}
	\end{equation}
	Here $v_A^2\equiv B^2/(\mu_m h)$ and $c_\theta\equiv\cos\theta$.
	Thus the physical magnetosonic roots are \(w=\pm v_f(\theta)\) and
	\(w=\pm v_s(\theta)\).  The additional zero factor of the unreduced determinant is removed
	by the magnetic Gauss constraint.
	Define
	\begin{equation}
		\mathcal D_H(w,\Omega^{(2)})
		\equiv
		[k^5]D_{\rm ms}^{(1)}(w k+\Omega^{(2)}k^2,k,\theta).
	\end{equation}
	This coefficient is linear in \(\Omega^{(2)}\), and we write
	\begin{equation}
		\mathcal D_H(w,\Omega^{(2)})
		=
		\mathcal A_H(w)\Omega^{(2)}+\mathcal B_H(w),
	\end{equation}
	where
	\begin{equation}
		\mathcal A_H(w)=
		\frac{\partial\mathcal D_H(w,\Omega^{(2)})}{\partial\Omega^{(2)}},
		\qquad
		\mathcal B_H(w)=\mathcal D_H(w,0).
		\label{eq:app-AH-BH-definition}
	\end{equation}
	Therefore
	\begin{equation}
		\Omega_{{\rm ms},f/s,\pm}^{(2)}
		=
		-\frac{\mathcal B_H(w)}{\mathcal A_H(w)}
		\bigg|_{w=\pm v_{f/s}(\theta)} .
		\label{eq:app-ms-hydro-Omega-compact}
	\end{equation}
	
	For the Alfv\'en hydrodynamic roots we again set
	\begin{equation}
		\omega=w k+\Omega^{(2)} k^2+O(k^3).
	\end{equation}
	The leading equation is
	\begin{equation}
		h\mu_m w^2-B^2c_\theta^2=0,
	\end{equation}
	so that \(w=\pm v_Ac_\theta\).  Define
	\begin{equation}
		\mathcal D_A^{\rm hyd}(w,\Omega^{(2)})
		\equiv
		[k^3]D_{\rm A}^{(1)}(w k+\Omega^{(2)}k^2,k,\theta),
	\end{equation}
	\begin{equation}
		\mathcal D_A^{\rm hyd}(w,\Omega^{(2)})
		=
		\mathcal A_{A,h}(w)\Omega^{(2)}+\mathcal B_{A,h}(w),
	\end{equation}
	with
	\begin{equation}
		\mathcal A_{A,h}(w)=
		\frac{\partial\mathcal D_A^{\rm hyd}(w,\Omega^{(2)})}{\partial\Omega^{(2)}},
		\qquad
		\mathcal B_{A,h}(w)=\mathcal D_A^{\rm hyd}(w,0).
	\end{equation}
	The Alfv\'en attenuation coefficients are therefore
	\begin{equation}
		\Omega_{{\rm A},\pm}^{(2)}
		=
		-\frac{\mathcal B_{A,h}(w)}{\mathcal A_{A,h}(w)}
		\bigg|_{w=\pm v_Ac_\theta} .
		\label{eq:app-A-hydro-Omega-compact}
	\end{equation}
	
	\subsection{Universal formula for gapped branches}
	
	For a simple nonzero root \(w\) of \(D_{X,0}(\omega)\), away from branch degeneracies, the first correction satisfies
	\(\omega_1=0\).  We therefore set
	\begin{equation}
		\omega=w+\Omega^{(2)}k^2+O(k^3),
	\end{equation}
	and expand the Schur-reduced determinant at fixed \(w\):
	\begin{equation}
		D_X^{(1)}(\omega,k,\theta)
		=
		D_{X,0}(\omega)+k^2D_{X,2}(\omega,\theta)+O(k^3),
		\qquad
		D_{X,2}(\omega,\theta)\equiv [k^2]D_X^{(1)}(\omega,k,\theta).
		\label{eq:app-D2-definition}
	\end{equation}
	Substitution gives the universal coefficient equation
	\begin{equation}
		\left[\partial_\omega D_{X,0}(w)\right]\Omega_X^{(2)}
		+
		D_{X,2}(w,\theta)=0,
	\end{equation}
	namely
	\begin{equation}
		\Omega_X^{(2)}(w)
		=
		-
		\frac{D_{X,2}(w,\theta)}{\partial_\omega D_{X,0}(w)}
		\qquad (X={\rm ms},{\rm A}).
		\label{eq:app-universal-gapped-Omega}
	\end{equation}
	For the magnetosonic block this formula gives
	\begin{equation}
		\Omega_{{\rm ms},r}^{(2)}=\Omega_{\rm ms}^{(2)}(w_r),
		\qquad
		\Omega_{{\rm ms},K,\pm}^{(2)}=\Omega_{\rm ms}^{(2)}(w_{K,\pm}),
		\qquad
		\Omega_{{\rm ms},q,\pm}^{(2)}=\Omega_{\rm ms}^{(2)}(w_{q,\pm}),
	\end{equation}
	where \(w_r\), \(w_{K,\pm}\), and \(w_{q,\pm}\) are determined by
	Eqs.~\eqref{eq:app-Pr-ms}--\eqref{eq:app-Pq-ms}.  For the Alfv\'en block,
	\begin{equation}
		\Omega_{{\rm A},r}^{(2)}=\Omega_{\rm A}^{(2)}(w_r),
		\qquad
		\Omega_{{\rm A},\phi}^{(2)}=\Omega_{\rm A}^{(2)}(w_\phi),
		\qquad
		\Omega_{{\rm A},q,\pm}^{(2)}=\Omega_{\rm A}^{(2)}(w_{q,\pm}),
	\end{equation}
	where \(w_r\), \(w_\phi\), and \(w_{q,\pm}\) are determined by
	Eqs.~\eqref{eq:app-Pr-ms}, \eqref{eq:app-Pphi-A}, and \eqref{eq:app-Pq-ms}.
	
	Equations~\eqref{eq:app-AH-BH-definition}, \eqref{eq:app-D2-definition}, and
	\eqref{eq:app-universal-gapped-Omega} define the attenuation and gap corrections directly
	from the reduced determinants fixed by Eq.~\eqref{eq:app-schur-compact} and the matrix entries
	in Appendix~\ref{app:linear-matrix-elements}.

		\subsection{Large-$k$ coefficient equations}
	\label{app:first-order-largek-coefficients}
	
	The large-momentum expansions in Eqs.~\eqref{eq:first-order-ms-largek-main-solutions}
	and~\eqref{eq:first-order-A-largek-main-solutions} are obtained from the Newton--Puiseux
	ansatz
	\begin{equation}
		D_X^{(1)}(\omega,k,\theta)=\sum_{a,b}C^X_{a,b}(\theta)\omega^a k^b,
		\qquad
		\omega=\Omega k^p+\cdots,
		\qquad X={\rm ms},{\rm A}.
		\label{eq:app-largek-det-poly}
	\end{equation}
	The exponent $p$ is selected by a dominant balance of the weights $b+pa$.  The following balance equations retain the $K$ and $q^\mu$ channels and follow from coefficient extraction at each dominant weight.
	
	\subsubsection{Magnetosonic block}
	
	The constrained magnetosonic Newton polygon has five nontrivial slopes,
	\begin{equation}
		p\in\left\{2,\frac43,1,0,-2\right\},
	\end{equation}
	with the following leading balance polynomials:
	\begin{align}
		&p=2:
		&&C_{9,0}^{\rm ms}\Omega+C_{8,2}^{\rm ms}=0,
		\label{eq:app-ms-p2-generic}
		\\
		&p=\frac43:
		&&C_{8,2}^{\rm ms}\Omega^3+C_{5,6}^{\rm ms}=0,
		\label{eq:app-ms-p43-generic}
		\\
		&p=1:
		&&C_{5,6}^{\rm ms}\Omega^2+C_{3,8}^{\rm ms}=0,
		\label{eq:app-ms-p1-generic}
		\\
		&p=0:
		&&C_{3,8}^{\rm ms}\Omega^2+C_{2,8}^{\rm ms}\Omega+C_{1,8}^{\rm ms}=0,
		\label{eq:app-ms-p0-generic}
		\\
		&p=-2:
		&&C_{1,8}^{\rm ms}\Omega+C_{0,6}^{\rm ms}=0 .
		\label{eq:app-ms-pm2-generic}
	\end{align}
	Equations~\eqref{eq:app-ms-p2-generic}--\eqref{eq:app-ms-pm2-generic} generate,
	respectively, one $p=2$ branch, three $p=4/3$ branches, two $p=1$ branches, two finite
	$p=0$ branches, and one $p=-2$ branch.  This accounts for all nine asymptotic families of
	the constrained generic magnetosonic determinant.  If the $K$ and $q^\mu$ channels are set
	to zero before the Newton polygon is constructed, its support changes, but the magnetic
	constraint must still be imposed independently.
	
	\subsubsection{Alfv\'en block}
	
	For the generic Alfv\'en determinant with the $q^\mu$ channel retained the dominant
	balances are
	\begin{equation}
		p\in\{2,1,0\}.
	\end{equation}
	The leading coefficient equations are
	\begin{align}
		&p=2:
		&&C_{6,0}^{\rm A}\Omega+C_{5,2}^{\rm A}=0,
		\label{eq:app-A-p2-generic}
		\\
		&p=1:
		&&C_{5,2}^{\rm A}\Omega^2+C_{3,4}^{\rm A}=0,
		\label{eq:app-A-p1-generic}
		\\
		&p=0:
		&&C_{3,4}^{\rm A}\Omega^3+C_{2,4}^{\rm A}\Omega^2
		+C_{1,4}^{\rm A}\Omega+C_{0,4}^{\rm A}=0.
		\label{eq:app-A-p0-generic}
	\end{align}
	These equations produce one $p=2$ branch, two $p=1$ branches, and three finite branches.
	No fractional Alfv\'en exponent appears in the generic first-order balance.

	\subsubsection{Subleading Newton--Puiseux coefficients}

	For a dominant exponent $p=P/q$ in lowest terms, define
	$w_p=\max_{C_{a,b}\ne0}(b+pa)$ and $\epsilon=k^{-1/q}$.  The rescaled
	characteristic polynomial is
	\begin{equation}
		\mathcal F_p(Z,\epsilon)
		=\sum_{a,b}C_{a,b}Z^a\epsilon^{q(w_p-b-pa)},
		\qquad \mathcal P_p(Z)=\mathcal F_p(Z,0).
		\label{eq:app-puiseux-rescaled-polynomial}
	\end{equation}
	All exponents of $\epsilon$ are nonnegative integers.  Choose a nonzero simple root
	$\Omega_0$ of $\mathcal P_p$ and expand
	\begin{equation}
		\omega(k)=k^p\sum_{j=0}^{N}\Omega_j k^{-j/q}
		+O\!\left(k^{p-(N+1)/q}\right).
		\label{eq:app-puiseux-subleading-series}
	\end{equation}
	The coefficient equations are triangular:
	\begin{equation}
		\Omega_j=
		-\frac{[\epsilon^j]\mathcal F_p\!\left(
			\sum_{\ell=0}^{j-1}\Omega_\ell\epsilon^\ell,\epsilon\right)}
		{\mathcal P_p'(\Omega_0)},\qquad j\geq1.
		\label{eq:app-puiseux-recursion}
	\end{equation}
	They determine the subleading terms directly from the characteristic coefficients.
	For the generic magnetosonic fractional family, the nonzero displayed terms are
	$\Omega_0 k^{4/3}+\Omega_2 k^{2/3}+\Omega_4$; the intervening odd-$j$ coefficients
	vanish.  The figures retain all nonnegative powers through the constant term,
	or only the leading term when $p<0$.  Multiple leading roots
	require a further dominant balance; the simple-root recursion is used at the
	sampled angles and baseline coefficients considered here.
	
For the baseline coefficients at $\theta=\pi/4$, the fractional family has
\begin{equation}
\frac{|\Omega_2\hat k^{2/3}|}{|\Omega_0\hat k^{4/3}|}
\simeq1038.0333\,\hat k^{-2/3},\qquad \hat k=k/h^{1/4}.
\label{eq:first-order-fractional-hierarchy}
\end{equation}
The first correction is as large as the leading term at
$\hat k\simeq3.34\times10^4$ and falls below 10\% only at
$\hat k\gtrsim1.06\times10^6$.  Thus the $p=4/3$ family lies outside this
asymptotic hierarchy window throughout the displayed $\hat k\leq10^3$ range.
The threshold varies with angle and transport coefficients.

\section{Infrared and second-order ultraviolet dispersion relations}
	\label{app:second-order-asymptotic-extraction}

	The infrared construction applies to the full physical block matrices of both dissipative
	orders, before eliminating algebraic currents.  At a fixed angle they are affine pencils,
	\begin{equation}
		M^{(n)}(\omega,k)=C+kK+\omega W,\qquad n=1,2.
		\label{eq:app-ir-affine-pencil}
	\end{equation}
	The first-order time matrix may be singular; only finite eigenvalues are retained.
	The second-order magnetosonic and Alfv\'en matrices have 23 and 20 finite branches,
	respectively.  The two exact spectator modes are appended to the Alfv\'en set.
	The integer-power expansions below assume semisimple eigenvalues at each projection step.

	\subsection{Successive infrared solvability conditions}

	Let $\omega_0$ be a finite semisimple eigenvalue of the zero-momentum pencil and set
	$A_0=C+\omega_0W$.  The columns of $R$ and $L$ span its right and left null spaces.
	For a regular finite eigenspace, $E=L^\dagger WR$ is nonsingular.  Substituting
	\begin{equation}
		\omega=\omega_0+\omega_1k+\omega_2k^2+O(k^3),\qquad
		x=Ra_0+kx_1+k^2x_2+\cdots
	\end{equation}
	into Eq.~\eqref{eq:app-ir-affine-pencil} gives at first order
	\begin{equation}
		(F+\omega_1E)a_0=0,\qquad F=L^\dagger KR.
		\label{eq:app-ir-first-projection}
	\end{equation}
	This finite projected characteristic equation determines the hydrodynamic slopes
	and any splitting of repeated gaps, independently of finite-momentum root curves.

	For a chosen semisimple $\omega_1$, let $U$ and $V$ span the right and left null spaces of
	$F+\omega_1E$, and define $R_1=RU$, $L_1=LV$, and $Q=K+\omega_1W$.
	A particular first-order eigenvector correction is fixed by the bordered system
	\begin{equation}
		\begin{pmatrix}
			A_0&WR\\ L^\dagger W&0
		\end{pmatrix}
		\begin{pmatrix}X\\Y\end{pmatrix}
		=
		\begin{pmatrix}-QR_1\\0\end{pmatrix}.
		\label{eq:app-ir-bordered-correction}
	\end{equation}
	Here $L^\dagger WX=0$ fixes the null-space freedom, and the first-order
	solvability condition gives $Y=0$.  Projecting the order-$k^2$ equation onto $L_1$
	then determines the second-order frequency coefficients:
	\begin{equation}
		\det\!\left[L_1^\dagger QX+\omega_2L_1^\dagger WR_1\right]=0.
		\label{eq:app-ir-second-projection}
	\end{equation}
	This second projection resolves repeated first-order splittings, including repeated
	relaxation gaps.  For the baseline coefficients all nonhydrodynamic $\omega_1$ vanish.
	The simple-root determinant formulas in Appendix~\ref{app:first-order-asymptotic-coefficients}
	are recovered from these solvability conditions.  Both real and imaginary parts of
	the displayed infrared theory are taken from the same series
	$\omega_0+\omega_1k+\omega_2k^2$.

	\subsection{Second-order ultraviolet projections}

	For every nonzero simple principal speed $v_a$, let $r_a$ and $\ell_a$ be right and
	left null vectors of $K+v_aW$.  Substituting
	\begin{equation}
		\omega_a(k,\theta)=v_a(\theta)k+\omega_{a,0}(\theta)+O(k^{-1})
	\end{equation}
	into the affine pencil gives
	\begin{align}
		[K+v_aW]r_a&=0,&
		\ell_a^\dagger[K+v_aW]&=0,&
		\omega_{a,0}&=-\frac{\ell_a^\dagger C r_a}{\ell_a^\dagger W r_a}.
		\label{eq:app-second-order-uv-solvability}
	\end{align}
	If a nonzero speed is semisimple but repeated, the offsets instead follow from
	$\det[L_v^\dagger(C+\omega_0W)R_v]=0$ within its left and right eigenspaces.
	For real $v_a$, the imaginary part approaches the independently determined constant
	$\operatorname{Im}\omega_{a,0}$.

	For a semisimple zero principal speed, let the columns of $R_0$ span
	$\ker K$ and those of $L_0$ span $\ker K^\dagger$, with $L_0^\dagger WR_0$ nonsingular.
	The finite leading poles are
	the generalized eigenvalues of
	\begin{equation}
		\det\!\left[L_0^\dagger(C+\omega_0W)R_0\right]=0.
		\label{eq:app-second-order-finite-projection}
	\end{equation}
	At the generic angle and parameter set of Fig.~\ref{fig:second-order-k-scan},
	the projections give fourteen linear and nine finite magnetosonic branches,
	and ten linear and ten finite Alfv\'en branches.  All nonzero principal speeds
	are real within numerical precision.  The imaginary theory curves are therefore
	horizontal, and the full numerical damping converges to their values at large
	momentum.  The two decoupled spectator modes are exact,
	\begin{equation}
		\omega_{\Lambda_\perp,\rm spec}=-\frac{\mathrm i}{\tau_{\Lambda\perp}},
		\qquad
		\omega_{\Omega,\rm spec}=-\frac{\mathrm i}{\tau_\Omega}.
	\end{equation}
	Together with the first-order recursions in
	Appendix~\ref{app:first-order-asymptotic-coefficients}, these coefficient equations
	and projected pencils determine the theoretical curves.  Their coefficients are
	evaluated with the dimensionless inputs in Table~\ref{tab:numerical-parameters}.
	The resulting dispersions are compared with the full matrix roots.

\subsection{Zero-momentum modes and effective inertia}
\label{app:zero-momentum-gaps}

At $k=0$, the triples $(u_z,K,S^{0z})$ and $(u_i,q_i,S^{0i})$ close independently.
After removing the conserved momentum root, the two nonzero frequencies in each sector obey
\begin{equation}
\begin{aligned}
A_K\omega^2+\mathrm i\omega-\nu_K&=0,
& A_K&=\tau_K-\frac{\beta_0\lambda_K}{4h_\parallel},
&\nu_K&=2\lambda_K\Gamma_{b\parallel},\\
A_q\omega^2+\mathrm i\omega-\nu_q&=0,
& A_q&=\tau_q-\frac{\beta_0\lambda_q}{2h},
&\nu_q&=4\lambda_q\Gamma_{b\perp}.
\end{aligned}
\label{eq:boost-gap-quadratics}
\end{equation}
For either channel with $\nu>0$, the roots
\begin{equation}
\omega_\pm=\frac{-\mathrm i\pm\sqrt{-1+4A\nu}}{2A}
\label{eq:boost-gap-roots}
\end{equation}
both lie in the lower half-plane if $A>0$; if $A<0$, exactly one lies in the upper half-plane.
For positive relaxation time, $A_K=\tau_K(1-\xi_K/4)$ and $A_q=\tau_q(1-\xi_q/2)$.
Their signs coincide with those of the effective inertias, since
$A_K=(\tau_K/h_\parallel)[h_\parallel-\beta_0\lambda_K/(4\tau_K)]$ and
$A_q=(\tau_q/h)[h-\beta_0\lambda_q/(2\tau_q)]$.
At $A=0$, one frequency remains at $-\mathrm i\nu$ while the other escapes to infinity,
identifying the same $\xi_K=4$ and $\xi_q=2$ lines as Eq.~\eqref{eq:principal-inertial-singularities}.
The first-order truncation sets the relaxation terms to zero while retaining the derivative
forces, giving $A_K,A_q<0$ for positive transport coefficients.
Thus its zero-momentum boost instability follows analytically in this closure.
Positive $A_K,A_q$ stabilize these modes; finite-$k$ stability still requires a separate check.
At fixed first-order coefficients and $\tau_K,\tau_q,\tau_\phi$, individual factor-two
variations of the other relaxation-time groups also yield no growing root on the sampled grid.

\subsection{Spin-flux characteristic subblocks}
\label{app:spinflux-characteristics}

In the unpolarized minimal system, the principal matrices $K$ and $W$ separate into
ordinary fluid--magnetic and spin-density/flux subblocks.  The constant matrix $C$
retains their finite-momentum coupling.  With the common coefficients
$\chi_X=\chi_{\rm sf}>0$ and $\tau_X=\tau_{\rm sf}>0$ used in Table~\ref{tab:numerical-parameters},
write $r_{\rm sf}=\chi_{\rm sf}/\tau_{\rm sf}$.  The spin principal determinants factor as follows, with the two exact spectators already excluded from the Alfv\'en subblock:
\begin{equation}
\begin{aligned}
\mathcal P_{{\rm ms},{\rm sf}}(v)&\propto
v^5(v^2-v_{b\parallel}^2)(v^2-v_{b\perp}^2)(v^2-v_{s\parallel}^2),\\
\mathcal P_{{\rm A},{\rm sf}}(v)&\propto
v^5(v^2-v_{b\perp}^2)(v^2-v_{s\parallel}^2)(v^2-v_{s\perp}^2),
\end{aligned}
\label{eq:spinflux-principal-factors}
\end{equation}
where
\begin{equation}
\begin{aligned}
v_{b\parallel}^2&=\Gamma_{b\parallel}r_{\rm sf}(1+\cos^2\theta),
&v_{b\perp}^2&=\Gamma_{b\perp}r_{\rm sf},\\
v_{s\parallel}^2&=\Gamma_{s\parallel}r_{\rm sf},
&v_{s\perp}^2&=\Gamma_{s\perp}r_{\rm sf}.
\end{aligned}
\label{eq:spinflux-principal-speeds}
\end{equation}
Eliminating $\Lambda_\parallel$ and $\Lambda_\perp$ from the $S^{0z}$ principal balance
gives $v^2=\Gamma_{b\parallel}r_{\rm sf}(2\cos^2\theta+\sin^2\theta)$;
the other density/flux groups are resolved in the same way.
The remaining principal flux combinations supply the zero-speed factors.
The $1/2$ antisymmetric projection in Eqs.~\eqref{eq:Ma-component-relaxation}
and \eqref{eq:Na-component-relaxation} is needed for the
$v_{b\perp}$ and $v_{s\parallel}$ expressions in the Alfv\'en block.
The angle independence of the last three speeds follows from the common-rate choice;
it is not imposed for general anisotropic transport.

\begin{figure}[t]
\centering
\includegraphics[width=0.94\textwidth]{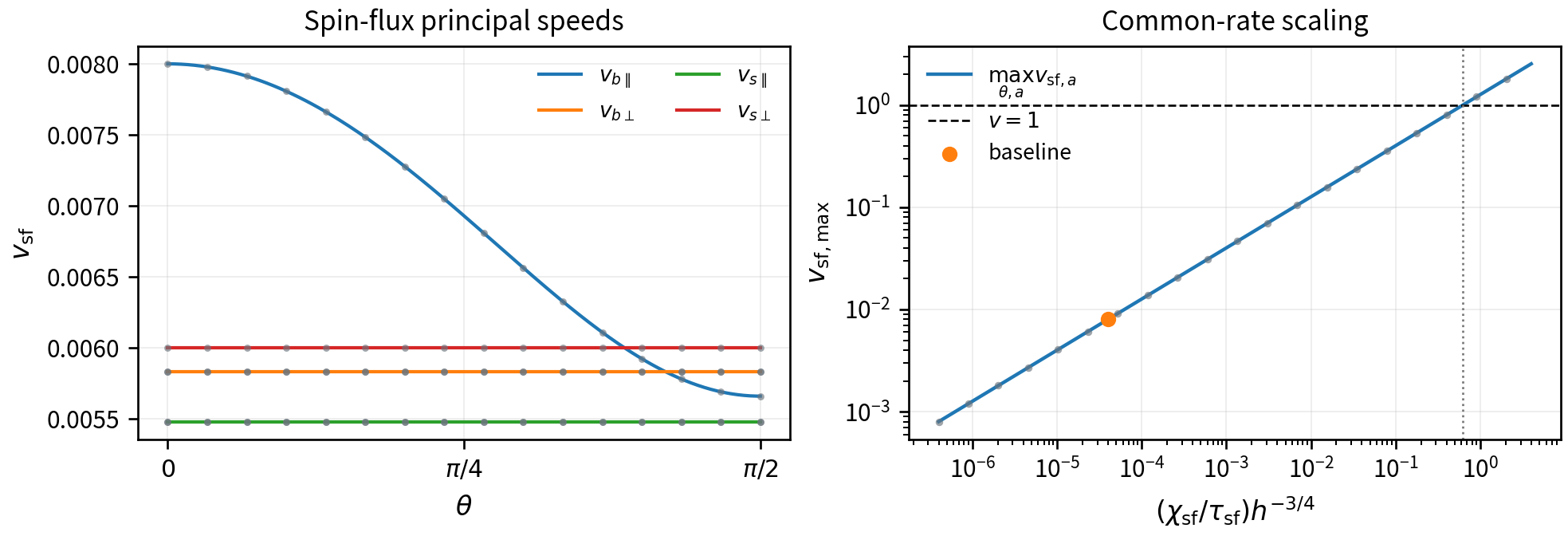}
\caption{Spin-flux characteristic identification.  Left: positive speeds from
Eq.~\eqref{eq:spinflux-principal-speeds} for the baseline coefficients.
Gray points are independently computed generalized eigenvalues of the spin principal subblocks;
the curves are analytic.  Right: the largest spin speed over all angles as the common
$\chi_{\rm sf}$ varies at fixed $\tau_{\rm sf}$ and fixed ordinary-sector coefficients.
The orange point is the baseline; the dashed horizontal and dotted vertical lines mark
$v=1$ and the spin bound in Eq.~\eqref{eq:spinflux-speed-bound}.
Above the critical common rate, the spin subblock has superluminal characteristics.}
\label{fig:spinflux-characteristics}
\end{figure}

For positive $\Gamma$ values, the spin speeds do not exceed unity at any angle exactly when
\begin{equation}
r_{\rm sf}\max\!\left(2\Gamma_{b\parallel},\Gamma_{b\perp},
\Gamma_{s\parallel},\Gamma_{s\perp}\right)\leq1.
\label{eq:spinflux-speed-bound}
\end{equation}
This is the spin-subblock part of the principal-speed criterion.
The baseline has $r_{\rm sf}h^{-3/4}=4\times10^{-5}$; for its $\Gamma$ values,
the bound is $r_{\rm sf}h^{-3/4}\leq0.625$.
Figure~\ref{fig:spinflux-characteristics} verifies the characteristic families and their square-root scaling.

\bibliography{spin_mhd_references}

@article{STAR2017,
  author = {L. Adamczyk and others},
  collaboration = {STAR Collaboration},
  title = {{Global $\Lambda$ hyperon polarization in nuclear collisions: Evidence for the most vortical fluid}},
  journal = {Nature},
  volume = {548},
  pages = {62},
  year = {2017},
}

@article{STAR2018,
  author = {J. Adam and others},
  collaboration = {STAR Collaboration},
  title = {{Global polarization of $\Lambda$ hyperons in Au+Au collisions at $\sqrt{s_{NN}}=200$ GeV}},
  journal = {Phys. Rev. C},
  volume = {98},
  pages = {014910},
  year = {2018},
}

@article{STAR2019Beam,
  author = {J. Adam and others},
  collaboration = {STAR Collaboration},
  title = {{Polarization of $\Lambda$ ($\bar\Lambda$) hyperons along the beam direction in Au+Au collisions at $\sqrt{s_{NN}}=200$ GeV}},
  journal = {Phys. Rev. Lett.},
  volume = {123},
  pages = {132301},
  year = {2019},
}

@article{ALICE2020,
  author = {S. Acharya and others},
  collaboration = {ALICE Collaboration},
  title = {{Global polarization of $\Lambda$ and $\bar\Lambda$ hyperons in Pb--Pb collisions at $\sqrt{s_{NN}}=2.76$ and $5.02$ TeV}},
  journal = {Phys. Rev. C},
  volume = {101},
  pages = {044611},
  year = {2020},
}

@article{ALICESpinOrbit2020,
  author = {S. Acharya and others},
  collaboration = {ALICE Collaboration},
  title = {{Evidence of spin-orbital angular momentum interactions in relativistic heavy-ion collisions}},
  journal = {Phys. Rev. Lett.},
  volume = {125},
  pages = {012301},
  year = {2020},
}

@article{STARXiOmega2021,
  author = {J. Adam and others},
  collaboration = {STAR Collaboration},
  title = {{Global polarization of $\Xi$ and $\Omega$ hyperons in Au+Au collisions at $\sqrt{s_{NN}}=200$ GeV}},
  journal = {Phys. Rev. Lett.},
  volume = {126},
  pages = {162301},
  year = {2021},
}

@article{FlorkowskiKumarRyblewski2019,
  author = {W. Florkowski and A. Kumar and R. Ryblewski},
  title = {{Relativistic hydrodynamics for spin-polarized fluids}},
  journal = {Prog. Part. Nucl. Phys.},
  volume = {108},
  pages = {103709},
  year = {2019},
}

@article{STAR3GeV2021,
  author = {M. S. Abdallah and others},
  collaboration = {STAR Collaboration},
  title = {{Global $\Lambda$-hyperon polarization in Au+Au collisions at $\sqrt{s_{NN}}=3$ GeV}},
  journal = {Phys. Rev. C},
  volume = {104},
  pages = {L061901},
  year = {2021},
}

@article{LiangWang2005,
  author = {Z.-T. Liang and X.-N. Wang},
  title = {{Globally polarized quark-gluon plasma in noncentral A+A collisions}},
  journal = {Phys. Rev. Lett.},
  volume = {94},
  pages = {102301},
  year = {2005},
  note = {Erratum: Phys. Rev. Lett. 96, 039901(E) (2006)},
}

@article{BetzGyulassyTorrieri2007,
  author = {B. Betz and M. Gyulassy and G. Torrieri},
  title = {{Polarization probes of vorticity in heavy ion collisions}},
  journal = {Phys. Rev. C},
  volume = {76},
  pages = {044901},
  year = {2007},
}

@article{BecattiniPiccininiRizzo2008,
  author = {F. Becattini and F. Piccinini and J. Rizzo},
  title = {{Angular momentum conservation in heavy ion collisions at very high energy}},
  journal = {Phys. Rev. C},
  volume = {77},
  pages = {024906},
  year = {2008},
}

@article{Barnett1915,
  author = {S. J. Barnett},
  title = {{Magnetization by rotation}},
  journal = {Phys. Rev.},
  volume = {6},
  pages = {239},
  year = {1915},
}

@article{EinsteinDeHaas1915,
  author = {A. Einstein and W. J. de Haas},
  title = {{Experimenteller Nachweis der Amp\`ereschen Molekularstr\"ome}},
  journal = {Verh. Dtsch. Phys. Ges.},
  volume = {17},
  pages = {152},
  year = {1915},
}

@article{HattoriEntropy2019,
  author = {K. Hattori and M. Hongo and X.-G. Huang and M. Matsuo and H. Taya},
  title = {{Fate of spin polarization in a relativistic fluid: An entropy-current analysis}},
  journal = {Phys. Lett. B},
  volume = {795},
  pages = {100},
  year = {2019},
}

@article{FukushimaPu2021,
  author = {K. Fukushima and S. Pu},
  title = {{Spin hydrodynamics and symmetric energy-momentum tensors: A current induced by the spin vorticity}},
  journal = {Phys. Lett. B},
  volume = {817},
  pages = {136346},
  year = {2021},
}

@article{LiStephanovYee2021,
  author = {S. Li and M. A. Stephanov and H.-U. Yee},
  title = {{Nondissipative second-order transport, spin, and pseudogauge transformations in hydrodynamics}},
  journal = {Phys. Rev. Lett.},
  volume = {127},
  pages = {082302},
  year = {2021},
}

@article{SheHuangHouLiao2022,
  author = {D. She and A. Huang and D. Hou and J. Liao},
  title = {{Relativistic viscous hydrodynamics with angular momentum}},
  journal = {Sci. Bull.},
  volume = {67},
  pages = {2265--2268},
  year = {2022},
  doi = {10.1016/j.scib.2022.10.020},
  eprint = {2105.04060},
  archivePrefix = {arXiv},
  primaryClass = {nucl-th},
}

@article{DaherCanonical2023,
  author = {A. Daher and A. Das and W. Florkowski and R. Ryblewski},
  title = {{Equivalence of canonical and phenomenological formulations of spin hydrodynamics}},
  journal = {Phys. Rev. C},
  volume = {108},
  pages = {024902},
  year = {2023},
}

@article{HongoEtAl2021,
  author = {M. Hongo and X.-G. Huang and M. Kaminski and M. A. Stephanov and H.-U. Yee},
  title = {{Relativistic spin hydrodynamics with torsion and linear response theory for spin relaxation}},
  journal = {J. High Energy Phys.},
  volume = {2021},
  number = {11},
  pages = {150},
  year = {2021},
}

@article{BecattiniDaherSheng2024,
  author = {F. Becattini and A. Daher and X.-L. Sheng},
  title = {{Entropy current and entropy production in relativistic spin hydrodynamics}},
  journal = {Phys. Lett. B},
  volume = {850},
  pages = {138533},
  year = {2024},
}

@article{DaherShengWagnerBecattini2025,
  author = {A. Daher and X.-L. Sheng and D. Wagner and F. Becattini},
  title = {{Dissipative currents and transport coefficients in relativistic spin hydrodynamics}},
  journal = {Phys. Rev. D},
  volume = {112},
  pages = {094020},
  year = {2025},
}

@article{WangFangPu2021,
  author = {D.-L. Wang and S. Fang and S. Pu},
  title = {{Analytic solutions of relativistic dissipative spin hydrodynamics with Bjorken expansion}},
  journal = {Phys. Rev. D},
  volume = {104},
  pages = {114043},
  year = {2021},
}

@article{GallegosGursoyYarom2021,
  author = {A. D. Gallegos and U. G\"ursoy and A. Yarom},
  title = {{Hydrodynamics of spin currents}},
  journal = {SciPost Phys.},
  volume = {11},
  pages = {041},
  year = {2021},
}

@article{FlorkowskiFrimanJaiswalSperanza2018,
  author = {W. Florkowski and B. Friman and A. Jaiswal and E. Speranza},
  title = {{Relativistic fluid dynamics with spin}},
  journal = {Phys. Rev. C},
  volume = {97},
  pages = {041901(R)},
  year = {2018},
}

@article{FlorkowskiFrimanJaiswalRyblewskiSperanza2018,
  author = {W. Florkowski and B. Friman and A. Jaiswal and R. Ryblewski and E. Speranza},
  title = {{Spin-dependent distribution functions for relativistic hydrodynamics of spin-$1/2$ particles}},
  journal = {Phys. Rev. D},
  volume = {97},
  pages = {116017},
  year = {2018},
}

@article{HidakaPuYang2018,
  author = {Y. Hidaka and S. Pu and D.-L. Yang},
  title = {{Nonlinear responses of chiral fluids from kinetic theory}},
  journal = {Phys. Rev. D},
  volume = {97},
  pages = {016004},
  year = {2018},
}

@article{FlorkowskiSperanzaBecattini2018,
  author = {W. Florkowski and E. Speranza and F. Becattini},
  title = {{Perfect-fluid hydrodynamics with constant acceleration along the stream lines and spin polarization}},
  journal = {Acta Phys. Pol. B},
  volume = {49},
  pages = {1409},
  year = {2018},
}

@article{WeickgenanntShengSperanzaWangRischke2019,
  author = {N. Weickgenannt and X.-L. Sheng and E. Speranza and Q. Wang and D. H. Rischke},
  title = {{Kinetic theory for massive spin-$1/2$ particles from the Wigner-function formalism}},
  journal = {Phys. Rev. D},
  volume = {100},
  pages = {056018},
  year = {2019},
}

@article{BhaduryEtAl2021PLB,
  author = {S. Bhadury and W. Florkowski and A. Jaiswal and A. Kumar and R. Ryblewski},
  title = {{Relativistic dissipative spin dynamics in the relaxation time approximation}},
  journal = {Phys. Lett. B},
  volume = {814},
  pages = {136096},
  year = {2021},
}

@article{WeickgenanntEtAl2021PRL,
  author = {N. Weickgenannt and E. Speranza and X.-L. Sheng and Q. Wang and D. H. Rischke},
  title = {{Generating spin polarization from vorticity through nonlocal collisions}},
  journal = {Phys. Rev. Lett.},
  volume = {127},
  pages = {052301},
  year = {2021},
}

@article{ShiGaleJeon2021,
  author = {S. Shi and C. Gale and S. Jeon},
  title = {{From chiral kinetic theory to relativistic viscous spin hydrodynamics}},
  journal = {Phys. Rev. C},
  volume = {103},
  pages = {044906},
  year = {2021},
}

@article{SperanzaWeickgenannt2021,
  author = {E. Speranza and N. Weickgenannt},
  title = {{Spin tensor and pseudo-gauges: From nuclear collisions to gravitational physics}},
  journal = {Eur. Phys. J. A},
  volume = {57},
  pages = {155},
  year = {2021},
}

@article{BhaduryEtAl2021PRD,
  author = {S. Bhadury and W. Florkowski and A. Jaiswal and A. Kumar and R. Ryblewski},
  title = {{Dissipative spin dynamics in relativistic matter}},
  journal = {Phys. Rev. D},
  volume = {103},
  pages = {014030},
  year = {2021},
}

@article{SinghSophysRyblewski2021,
  author = {R. Singh and G. Sophys and R. Ryblewski},
  title = {{Spin polarization dynamics in the Gubser-expanding background}},
  journal = {Phys. Rev. D},
  volume = {103},
  pages = {074024},
  year = {2021},
}

@article{BhaduryBhattJaiswalKumar2021,
  author = {S. Bhadury and J. R. Bhatt and A. Jaiswal and A. Kumar},
  title = {{New developments in relativistic fluid dynamics with spin}},
  journal = {Eur. Phys. J. Spec. Top.},
  volume = {230},
  pages = {655},
  year = {2021},
}

@article{PengZhangShengWang2021,
  author = {H.-H. Peng and J.-J. Zhang and X.-L. Sheng and Q. Wang},
  title = {{Ideal spin hydrodynamics from the Wigner function approach}},
  journal = {Chin. Phys. Lett.},
  volume = {38},
  pages = {116701},
  year = {2021},
}

@article{ShengEtAl2021,
  author = {X.-L. Sheng and N. Weickgenannt and E. Speranza and D. H. Rischke and Q. Wang},
  title = {{From Kadanoff--Baym to Boltzmann equations for massive spin-$1/2$ fermions}},
  journal = {Phys. Rev. D},
  volume = {104},
  pages = {016029},
  year = {2021},
}

@article{ShengWangRischke2022,
  author = {X.-L. Sheng and Q. Wang and D. H. Rischke},
  title = {{Lorentz-covariant kinetic theory for massive spin-$1/2$ particles}},
  journal = {Phys. Rev. D},
  volume = {106},
  pages = {L111901},
  year = {2022},
}

@article{HuChapmanEnskog2022,
  author = {J. Hu},
  title = {{Relativistic first-order spin hydrodynamics via the Chapman--Enskog expansion}},
  journal = {Phys. Rev. D},
  volume = {105},
  pages = {076009},
  year = {2022},
}

@article{HuLinearMode2022,
  author = {J. Hu},
  title = {{Linear mode analysis from spin transport equation}},
  journal = {Phys. Rev. D},
  volume = {106},
  pages = {036004},
  year = {2022},
}

@article{FangPuYang2022,
  author = {S. Fang and S. Pu and D.-L. Yang},
  title = {{Quantum kinetic theory for dynamical spin polarization from QED-type interaction}},
  journal = {Phys. Rev. D},
  volume = {106},
  pages = {016002},
  year = {2022},
}

@article{WangSpinEvolution2022,
  author = {Z. Wang},
  title = {{Spin evolution of a massive fermion in a QED plasma}},
  journal = {Phys. Rev. D},
  volume = {106},
  pages = {076011},
  year = {2022},
}

@article{MontenegroTintiTorrieri2017,
  author = {D. Montenegro and L. Tinti and G. Torrieri},
  title = {{Ideal relativistic fluid limit for a medium with polarization}},
  journal = {Phys. Rev. D},
  volume = {96},
  pages = {056012},
  year = {2017},
}

@article{MontenegroTintiTorrieriSound2017,
  author = {D. Montenegro and L. Tinti and G. Torrieri},
  title = {{Sound waves and vortices in a polarized relativistic fluid}},
  journal = {Phys. Rev. D},
  volume = {96},
  pages = {076016},
  year = {2017},
}

@article{MontenegroTorrieri2019,
  author = {D. Montenegro and G. Torrieri},
  title = {{Causality and dissipation in relativistic polarizable fluids}},
  journal = {Phys. Rev. D},
  volume = {100},
  pages = {056011},
  year = {2019},
}

@article{MontenegroTorrieri2020,
  author = {D. Montenegro and G. Torrieri},
  title = {{Linear response theory and effective action of relativistic hydrodynamics with spin}},
  journal = {Phys. Rev. D},
  volume = {102},
  pages = {036007},
  year = {2020},
}

@article{BecattiniPiccinini2008,
  author = {F. Becattini and F. Piccinini},
  title = {{The ideal relativistic spinning gas: Polarization and spectra}},
  journal = {Ann. Phys. (N.Y.)},
  volume = {323},
  pages = {2452},
  year = {2008},
}

@article{BecattiniTinti2010,
  author = {F. Becattini and L. Tinti},
  title = {{The ideal relativistic rotating gas as a perfect fluid with spin}},
  journal = {Ann. Phys. (N.Y.)},
  volume = {325},
  pages = {1566},
  year = {2010},
}

@article{BecattiniTinti2013,
  author = {F. Becattini and L. Tinti},
  title = {{Nonequilibrium thermodynamical inequivalence of quantum stress-energy and spin tensors}},
  journal = {Phys. Rev. D},
  volume = {87},
  pages = {025029},
  year = {2013},
}

@article{FlorkowskiHontarenko2025,
  author = {W. Florkowski and M. Hontarenko},
  title = {{Generalized thermodynamic relations for perfect spin hydrodynamics}},
  journal = {Phys. Rev. Lett.},
  volume = {134},
  pages = {082302},
  year = {2025},
}

@article{BecattiniFlorkowskiSperanza2019,
  author = {F. Becattini and W. Florkowski and E. Speranza},
  title = {{Spin tensor and its role in non-equilibrium thermodynamics}},
  journal = {Phys. Lett. B},
  volume = {789},
  pages = {419},
  year = {2019},
}

@article{HuKubo2021,
  author = {J. Hu},
  title = {{Kubo formulae for first-order spin hydrodynamics}},
  journal = {Phys. Rev. D},
  volume = {103},
  pages = {116015},
  year = {2021},
}

@article{Buzzegoli2022,
  author = {M. Buzzegoli},
  title = {{Pseudogauge dependence of the spin polarization and of the axial vortical effect}},
  journal = {Phys. Rev. C},
  volume = {105},
  pages = {044907},
  year = {2022},
}

@article{DasFlorkowskiRyblewskiSingh2021,
  author = {A. Das and W. Florkowski and R. Ryblewski and R. Singh},
  title = {{Pseudogauge dependence of quantum fluctuations of the energy in a hot relativistic gas of fermions}},
  journal = {Phys. Rev. D},
  volume = {103},
  pages = {L091502},
  year = {2021},
  doi = {10.1103/PhysRevD.103.L091502},
  eprint = {2103.01013},
  archivePrefix = {arXiv},
  primaryClass = {hep-ph},
}

@article{AmbrusRyblewskiSingh2022,
  author = {V. E. Ambrus and R. Ryblewski and R. Singh},
  title = {{Spin waves in spin hydrodynamics}},
  journal = {Phys. Rev. D},
  volume = {106},
  number = {1},
  pages = {014018},
  year = {2022},
  doi = {10.1103/PhysRevD.106.014018},
  eprint = {2202.03952},
  archivePrefix = {arXiv},
  primaryClass = {hep-ph},
}

@article{SinghShokriRyblewski2021,
  author = {R. Singh and M. Shokri and R. Ryblewski},
  title = {{Spin polarization dynamics in the Bjorken-expanding resistive MHD background}},
  journal = {Phys. Rev. D},
  volume = {103},
  number = {9},
  pages = {094034},
  year = {2021},
  doi = {10.1103/PhysRevD.103.094034},
  eprint = {2103.02592},
  archivePrefix = {arXiv},
  primaryClass = {hep-ph},
}

@article{BhambureEtAl2025DensityFrame,
  author = {J. Bhambure and A. Mazeliauskas and J.-F. Paquet and R. Singh and M. Singh and D. Teaney and F. Zhou},
  title = {{Relativistic viscous hydrodynamics in the density frame: Numerical tests and comparisons}},
  journal = {Phys. Rev. C},
  volume = {111},
  number = {6},
  pages = {064910},
  year = {2025},
  doi = {10.1103/PhysRevC.111.064910},
  eprint = {2412.10303},
  archivePrefix = {arXiv},
  primaryClass = {nucl-th},
}

@article{DaherDasRyblewski2023,
  author = {A. Daher and A. Das and R. Ryblewski},
  title = {{Stability studies of first-order spin-hydrodynamic frameworks}},
  journal = {Phys. Rev. D},
  volume = {107},
  pages = {054043},
  year = {2023},
}

@article{SarwarEtAl2023,
  author = {G. Sarwar and M. Hasanujjaman and J. R. Bhatt and H. Mishra and J.-e. Alam},
  title = {{Causality and stability of relativistic spin hydrodynamics}},
  journal = {Phys. Rev. D},
  volume = {107},
  pages = {054031},
  year = {2023},
}

@article{BiswasBoost2023,
  author = {R. Biswas and A. Daher and A. Das and W. Florkowski and R. Ryblewski},
  title = {{Boost-invariant spin hydrodynamics within the first order in derivative expansion}},
  journal = {Phys. Rev. D},
  volume = {107},
  pages = {094022},
  year = {2023},
}

@article{XieWangYangPu2023,
  author = {X.-Q. Xie and D.-L. Wang and C. Yang and S. Pu},
  title = {{Causality and stability analysis for the minimal causal spin hydrodynamics}},
  journal = {Phys. Rev. D},
  volume = {108},
  pages = {094031},
  year = {2023},
}

@article{Weickgenannt2023,
  author = {N. Weickgenannt},
  title = {{Linearly stable and causal relativistic first-order spin hydrodynamics}},
  journal = {Phys. Rev. D},
  volume = {108},
  pages = {076011},
  year = {2023},
}

@article{BiswasSecondOrder2023,
  author = {R. Biswas and A. Daher and A. Das and W. Florkowski and R. Ryblewski},
  title = {{Relativistic second-order spin hydrodynamics: An entropy-current analysis}},
  journal = {Phys. Rev. D},
  volume = {108},
  pages = {014024},
  year = {2023},
}

@article{DaherSecondOrder2024,
  author = {A. Daher and W. Florkowski and R. Ryblewski and F. Taghinavaz},
  title = {{Stability and causality of rest-frame modes in second-order spin hydrodynamics}},
  journal = {Phys. Rev. D},
  volume = {109},
  pages = {114001},
  year = {2024},
}

@article{DaherSpinEOS2024,
  author = {A. Daher and W. Florkowski and R. Ryblewski},
  title = {{Stability constraint for spin equation of state}},
  journal = {Phys. Rev. D},
  volume = {110},
  pages = {034029},
  year = {2024},
}

@article{RenYangWangPu2024,
  author = {X. Ren and C. Yang and D.-L. Wang and S. Pu},
  title = {{Thermodynamic stability in relativistic viscous and spin hydrodynamics}},
  journal = {Phys. Rev. D},
  volume = {110},
  pages = {034010},
  year = {2024},
}

@article{SheQiuHou2025,
  author = {D. She and Y.-W. Qiu and D. Hou},
  title = {{Relativistic second-order spin hydrodynamics: A Kubo-type formulation for the quark-gluon plasma}},
  journal = {Phys. Rev. D},
  volume = {111},
  pages = {036027},
  year = {2025},
  doi = {10.1103/PhysRevD.111.036027},
  eprint = {2410.15142},
  archivePrefix = {arXiv},
  primaryClass = {nucl-th},
}

@article{BhaduryNovelRTA2025,
  author = {S. Bhadury},
  title = {{Relativistic spin hydrodynamics from novel relaxation time approximation}},
  journal = {Phys. Rev. C},
  volume = {112},
  pages = {L021901},
  year = {2025},
}

@article{BhadurySpinMHD2022,
  author = {S. Bhadury and W. Florkowski and A. Jaiswal and A. Kumar and R. Ryblewski},
  title = {{Relativistic spin magnetohydrodynamics}},
  journal = {Phys. Rev. Lett.},
  volume = {129},
  pages = {192301},
  year = {2022},
}

@article{FangHattoriHu2025,
  author = {Z. Fang and K. Hattori and J. Hu},
  title = {{Anisotropic linear waves and breakdown of the momentum expansion in spin magnetohydrodynamics}},
  journal = {Phys. Rev. D},
  volume = {112},
  pages = {076016},
  year = {2025},
  doi = {10.1103/d86f-1vk8},
  eprint = {2409.07096},
  archivePrefix = {arXiv},
}

@article{KiamariSadooghiJafari2024,
  author = {M. Kiamari and N. Sadooghi and M. Sedighi Jafari},
  title = {{Relativistic magnetohydrodynamics of a spinful and vortical fluid: Entropy current analysis}},
  journal = {Phys. Rev. D},
  volume = {109},
  pages = {036024},
  year = {2024},
}

@article{HernandezKovtun2017,
  author = {J. Hernandez and P. Kovtun},
  title = {{Relativistic magnetohydrodynamics}},
  journal = {J. High Energy Phys.},
  volume = {2017},
  number = {05},
  pages = {001},
  year = {2017},
}

@article{GrozdanovHofmanIqbal2017,
  author = {S. Grozdanov and D. M. Hofman and N. Iqbal},
  title = {{Generalized global symmetries and dissipative magnetohydrodynamics}},
  journal = {Phys. Rev. D},
  volume = {95},
  pages = {096003},
  year = {2017},
}

@article{HongoHattori2021,
  author = {M. Hongo and K. Hattori},
  title = {{Revisiting relativistic magnetohydrodynamics from quantum electrodynamics}},
  journal = {J. High Energy Phys.},
  volume = {2021},
  number = {02},
  pages = {011},
  year = {2021},
}

@article{HattoriHongoHuang2022,
  author = {K. Hattori and M. Hongo and X.-G. Huang},
  title = {{New developments in relativistic magnetohydrodynamics}},
  journal = {Symmetry},
  volume = {14},
  pages = {1851},
  year = {2022},
}

@article{DenicolEtAl2018,
  author = {G. S. Denicol and others},
  title = {{Nonresistive dissipative magnetohydrodynamics from the Boltzmann equation in the 14-moment approximation}},
  journal = {Phys. Rev. D},
  volume = {98},
  pages = {076009},
  year = {2018},
}

@article{DenicolMolnarNiemiRischke2019,
  author = {G. S. Denicol and E. Moln\'ar and H. Niemi and D. H. Rischke},
  title = {{Resistive dissipative magnetohydrodynamics from the Boltzmann--Vlasov equation}},
  journal = {Phys. Rev. D},
  volume = {99},
  pages = {056017},
  year = {2019},
}

@article{PandaDashBiswasRoy2021,
  author = {A. K. Panda and A. Dash and R. Biswas and V. Roy},
  title = {{Relativistic non-resistive viscous magnetohydrodynamics from kinetic theory: A relaxation-time approach}},
  journal = {J. High Energy Phys.},
  volume = {2021},
  number = {03},
  pages = {216},
  year = {2021},
}

@article{BiswasDashHaquePuRoy2020,
  author = {R. Biswas and A. Dash and N. Haque and S. Pu and V. Roy},
  title = {{Causality and stability in relativistic viscous non-resistive magneto-fluid dynamics}},
  journal = {J. High Energy Phys.},
  volume = {2020},
  number = {10},
  pages = {171},
  year = {2020},
}

@article{FangHattoriHu2024,
  author = {Z. Fang and K. Hattori and J. Hu},
  title = {{Analytic solutions for the linearized first-order magnetohydrodynamics and implications for causality and stability}},
  journal = {Phys. Rev. D},
  volume = {110},
  pages = {056049},
  year = {2024},
}

@article{HoultKovtun2025,
  author = {R. E. Hoult and P. Kovtun},
  title = {{Causality in dissipative relativistic magnetohydrodynamics}},
  journal = {J. High Energy Phys.},
  volume = {2025},
  number = {04},
  pages = {009},
  year = {2025},
}

@article{MolnarRischke2025,
  author = {E. Moln\'ar and D. H. Rischke},
  title = {{Higher-order dissipative anisotropic magnetohydrodynamics from the Boltzmann--Vlasov equation}},
  journal = {Phys. Rev. D},
  volume = {111},
  pages = {036035},
  year = {2025},
}

@article{KushwahDenicol2024,
  author = {K. Kushwah and G. S. Denicol},
  title = {{Relativistic dissipative magnetohydrodynamics from the Boltzmann equation for a two-component gas}},
  journal = {Phys. Rev. D},
  volume = {109},
  pages = {096021},
  year = {2024},
}

@article{ArmasCamilloni2022,
  author = {J. Armas and F. Camilloni},
  title = {{A stable and causal model of magnetohydrodynamics}},
  journal = {J. Cosmol. Astropart. Phys.},
  volume = {2022},
  number = {10},
  pages = {039},
  year = {2022},
}

@article{ArmasJain2019,
  author = {J. Armas and A. Jain},
  title = {{Magnetohydrodynamics as superfluidity}},
  journal = {Phys. Rev. Lett.},
  volume = {122},
  pages = {141603},
  year = {2019},
}

@article{ArmasJain2020,
  author = {J. Armas and A. Jain},
  title = {{One-form superfluids and magnetohydrodynamics}},
  journal = {J. High Energy Phys.},
  volume = {2020},
  number = {01},
  pages = {041},
  year = {2020},
  eprint = {1811.04913},
  archivePrefix = {arXiv},
}

@article{GrozdanovPoovuttikul2019,
  author = {S. Grozdanov and N. Poovuttikul},
  title = {{Generalised global symmetries in holography: magnetohydrodynamic waves in a strongly interacting plasma}},
  journal = {J. High Energy Phys.},
  volume = {2019},
  number = {04},
  pages = {141},
  year = {2019},
  eprint = {1707.04182},
  archivePrefix = {arXiv},
}

@article{BenenowskiPoovuttikul2019,
  author = {B. Benenowski and N. Poovuttikul},
  title = {{Classification of magnetohydrodynamic transport at strong magnetic field}},
  journal = {arXiv:1911.05554},
  year = {2019},
}

@article{PoovuttikulRajagopal2021,
  author = {N. Poovuttikul and A. Rajagopal},
  title = {{Operator lifetime and the force-free electrodynamic limit of magnetised holographic plasma}},
  journal = {J. High Energy Phys.},
  volume = {2021},
  number = {09},
  pages = {091},
  year = {2021},
  eprint = {2101.12540},
  archivePrefix = {arXiv},
}

@article{VardhanEtAl2024,
  author = {S. Vardhan and S. Grozdanov and S. Leutheusser and H. Liu},
  title = {{Strong-field magnetohydrodynamics for neutron stars}},
  journal = {Phys. Rev. Research},
  volume = {6},
  pages = {L042050},
  year = {2024},
  eprint = {2207.01636},
  archivePrefix = {arXiv},
}

@article{VardhanEtAl2025,
  author = {S. Vardhan and S. Grozdanov and S. Leutheusser and H. Liu},
  title = {{Effective field theories of dissipative fluids with one-form symmetries}},
  journal = {J. High Energy Phys.},
  volume = {2025},
  number = {05},
  pages = {184},
  year = {2025},
  eprint = {2408.12868},
  archivePrefix = {arXiv},
}

@article{QiuSheHou2026,
  author = {Y. Qiu and D. She and D. Hou},
  title = {{Linear causality and stability constraints on relativistic second-order magnetohydrodynamics}},
  journal = {J. High Energy Phys.},
  year = {2026},
  volume = {2026},
  number = {10},
  pages = {155},
  doi = {10.1007/JHEP10(2026)155},
  eprint = {2606.00649},
  archivePrefix = {arXiv},
  primaryClass = {physics.flu-dyn},
}

@article{LuZhongFeng2026,
  author = {W. Lu and Y. Zhong and S.-Q. Feng},
  title = {{Causality and stability analysis of relativistic spin hydrodynamics: Insights from a nonvanishing spin density background}},
  journal = {Phys. Rev. D},
  volume = {114},
  pages = {034050},
  year = {2026},
  doi = {10.1103/rc82-5fjl},
  eprint = {2603.20770},
  archivePrefix = {arXiv},
}

@article{Muller1967,
  author = {I. M\"uller},
  title = {{Zum Paradoxon der W\"armeleitungstheorie}},
  journal = {Z. Phys.},
  volume = {198},
  pages = {329},
  year = {1967},
}

@article{Israel1976,
  author = {W. Israel},
  title = {{Nonstationary irreversible thermodynamics: A causal relativistic theory}},
  journal = {Ann. Phys. (N.Y.)},
  volume = {100},
  pages = {310},
  year = {1976},
}

@article{IsraelStewart1979,
  author = {W. Israel and J. M. Stewart},
  title = {{Transient relativistic thermodynamics and kinetic theory}},
  journal = {Ann. Phys. (N.Y.)},
  volume = {118},
  pages = {341},
  year = {1979},
}

@article{HiscockLindblom1983,
  author = {W. A. Hiscock and L. Lindblom},
  title = {{Stability and causality in dissipative relativistic fluids}},
  journal = {Ann. Phys. (N.Y.)},
  volume = {151},
  pages = {466},
  year = {1983},
}

@article{HiscockLindblom1985,
  author = {W. A. Hiscock and L. Lindblom},
  title = {{Generic instabilities in first-order dissipative relativistic fluid theories}},
  journal = {Phys. Rev. D},
  volume = {31},
  pages = {725},
  year = {1985},
}

@article{HiscockLindblom1987,
  author = {W. A. Hiscock and L. Lindblom},
  title = {{Linear plane waves in dissipative relativistic fluids}},
  journal = {Phys. Rev. D},
  volume = {35},
  pages = {3723},
  year = {1987},
}

@article{KoideDenicolMotaKodama2007,
  author = {T. Koide and G. S. Denicol and P. Mota and T. Kodama},
  title = {{Relativistic dissipative hydrodynamics: A minimal causal theory}},
  journal = {Phys. Rev. C},
  volume = {75},
  pages = {034909},
  year = {2007},
}

@article{DenicolKoideRischke2010,
  author = {G. S. Denicol and T. Koide and D. H. Rischke},
  title = {{Dissipative relativistic fluid dynamics: A new way to derive the equations of motion from kinetic theory}},
  journal = {Phys. Rev. Lett.},
  volume = {105},
  pages = {162501},
  year = {2010},
}

@article{PuKoideRischke2010,
  author = {S. Pu and T. Koide and D. H. Rischke},
  title = {{Does stability of relativistic dissipative fluid dynamics imply causality?}},
  journal = {Phys. Rev. D},
  volume = {81},
  pages = {114039},
  year = {2010},
}

@article{DenicolNiemiMolnarRischke2012,
  author = {G. S. Denicol and H. Niemi and E. Moln\'ar and D. H. Rischke},
  title = {{Derivation of transient relativistic fluid dynamics from the Boltzmann equation}},
  journal = {Phys. Rev. D},
  volume = {85},
  pages = {114047},
  year = {2012},
  note = {Erratum: Phys. Rev. D 91, 039902(E) (2015)},
}

@article{BaierEtAl2008,
  author = {R. Baier and P. Romatschke and D. T. Son and A. O. Starinets and M. A. Stephanov},
  title = {{Relativistic viscous hydrodynamics, conformal invariance, and holography}},
  journal = {J. High Energy Phys.},
  volume = {2008},
  number = {04},
  pages = {100},
  year = {2008},
}

@article{Kovtun2019,
  author = {P. Kovtun},
  title = {{First-order relativistic hydrodynamics is stable}},
  journal = {J. High Energy Phys.},
  volume = {2019},
  number = {10},
  pages = {034},
  year = {2019},
}

@article{BemficaDisconziNoronha2019,
  author = {F. S. Bemfica and M. M. Disconzi and J. Noronha},
  title = {{Nonlinear causality of general first-order relativistic viscous hydrodynamics}},
  journal = {Phys. Rev. D},
  volume = {100},
  pages = {104020},
  year = {2019},
}

@article{HoultKovtun2020,
  author = {R. E. Hoult and P. Kovtun},
  title = {{Stable and causal relativistic Navier--Stokes equations}},
  journal = {J. High Energy Phys.},
  volume = {2020},
  number = {06},
  pages = {067},
  year = {2020},
}

@article{BritoDenicol2020,
  author = {C. V. Brito and G. S. Denicol},
  title = {{Linear stability of Israel--Stewart theory in the presence of net-charge diffusion}},
  journal = {Phys. Rev. D},
  volume = {102},
  pages = {116009},
  year = {2020},
}

@article{FloerchingerGrossi2018,
  author = {S. Floerchinger and E. Grossi},
  title = {{Causality of fluid dynamics for high-energy nuclear collisions}},
  journal = {J. High Energy Phys.},
  volume = {2018},
  number = {08},
  pages = {186},
  year = {2018},
}

@article{RochaWagnerDenicolNoronhaRischke2024,
  author = {G. S. Rocha and D. Wagner and G. S. Denicol and J. Noronha and D. H. Rischke},
  title = {{Theories of relativistic dissipative fluid dynamics}},
  journal = {Entropy},
  volume = {26},
  pages = {189},
  year = {2024},
}

@article{FangHattoriHu2024FirstOrder,
  author = {Z. Fang and K. Hattori and J. Hu},
  title = {{First-order spin magnetohydrodynamics}},
  journal = {J. Subatomic Part. Cosmol.},
  volume = {1--2},
  pages = {100003},
  year = {2024},
  doi = {10.1016/j.jspc.2024.100003},
  eprint = {2410.00721},
  archivePrefix = {arXiv},
  primaryClass = {hep-ph},
}

@article{MoghaddamAlbericoSheKordAzadegan2020,
  author = {Haddadi Moghaddam, M. and Alberico, W. M. and She, Duan and Kord, A. F. and Azadegan, B.},
  title = {{Accelerating longitudinal expansion of resistive relativistic magnetohydrodynamics in heavy ion collisions}},
  journal = {Phys. Rev. D},
  volume = {102},
  pages = {014017},
  year = {2020},
  doi = {10.1103/PhysRevD.102.014017},
  eprint = {2002.09752},
  archivePrefix = {arXiv},
  primaryClass = {nucl-th},
}

@article{HuangSheShiHuangLiao2023,
  author = {Huang, Anping and She, Duan and Shi, Shuzhe and Huang, Mei and Liao, Jinfeng},
  title = {{Dynamical magnetic fields in heavy-ion collisions}},
  journal = {Phys. Rev. C},
  volume = {107},
  pages = {034901},
  year = {2023},
  doi = {10.1103/PhysRevC.107.034901},
  eprint = {2212.08579},
  archivePrefix = {arXiv},
  primaryClass = {hep-ph},
}

@article{SheQiuJiangHou2025,
  author = {She, Duan and Qiu, Yi-Wei and Jiang, Ze-Fang and Hou, Defu},
  title = {{Generalized relativistic second-order spin hydrodynamics from Zubarev's nonequilibrium statistical operator}},
  journal = {Phys. Rev. D},
  volume = {112},
  pages = {096013},
  year = {2025},
  doi = {10.1103/kbgm-vq87},
  eprint = {2505.01814},
  archivePrefix = {arXiv},
  primaryClass = {nucl-th},
}

@article{SheJiangHouYang2019,
  author = {She, Duan and Jiang, Ze Fang and Hou, Defu and Yang, C. B.},
  title = {{1+1 dimensional relativistic magnetohydrodynamics with longitudinal acceleration}},
  journal = {Phys. Rev. D},
  volume = {100},
  pages = {116014},
  year = {2019},
  doi = {10.1103/PhysRevD.100.116014},
  eprint = {1907.01250},
  archivePrefix = {arXiv},
  primaryClass = {hep-ph},
}

@article{JiangLiuHuZhengShe2025,
  author = {Jiang, Ze-Fang and Liu, Shuo-Yan and Hu, Tian-Yu and Zheng, Huang-Jing and She, Duan},
  title = {{1+1 dimensional relativistic viscous non-resistive magnetohydrodynamics with longitudinal boost invariance}},
  journal = {Chin. Phys. C},
  volume = {49},
  pages = {114104},
  year = {2025},
  doi = {10.1088/1674-1137/ade6d4},
  eprint = {2411.11398},
  archivePrefix = {arXiv},
  primaryClass = {nucl-th},
}

@article{JingSheJiang2026,
  author = {Jing, Jing and She, Duan and Jiang, Ze-Fang},
  title = {{Magnetization and magnetic field-induced correction: Implications for QGP thermal photon production in magnetohydrodynamic}},
  journal = {Phys. Rev. D},
  volume = {114},
  pages = {056002},
  year = {2026},
  doi = {10.1103/3rc1-gyx5},
  eprint = {2608.13364},
  archivePrefix = {arXiv},
  primaryClass = {hep-ph},
}

@article{JiangFanSheYeZhang2025,
  author = {Jiang, Ze-Fang and Fan, Xiang and She, Duan and Ye, Shasha and Zhang, Ben-Wei},
  title = {{Left-right splitting of elliptic flow in heavy ion collisions: TRENTo-3D initialization and CLVisc hydrodynamic simulations}},
  journal = {Phys. Rev. C},
  volume = {112},
  pages = {044906},
  year = {2025},
  doi = {10.1103/yyxf-dg63},
  eprint = {2505.14637},
  archivePrefix = {arXiv},
  primaryClass = {nucl-th},
}

@article{AbbasiDavody2016,
  author = {Abbasi, Navid and Davody, Ali},
  title = {{Dissipative charged fluid in a magnetic field}},
  journal = {Phys. Lett. B},
  volume = {756},
  pages = {161--165},
  year = {2016},
  doi = {10.1016/j.physletb.2016.02.071},
  eprint = {1508.06879},
  archivePrefix = {arXiv},
  primaryClass = {hep-th},
}

@article{AbbasiDavodyTahery2024,
  author = {Abbasi, Navid and Davody, Ali and Tahery, Sara},
  title = {{Correlation functions in stable first-order relativistic hydrodynamics}},
  journal = {Phys. Rev. D},
  volume = {109},
  pages = {036006},
  year = {2024},
  doi = {10.1103/PhysRevD.109.036006},
  eprint = {2212.14619},
  archivePrefix = {arXiv},
  primaryClass = {hep-th},
}
	
\end{document}